\documentclass[11pt]{article}
\usepackage{amssymb}
\usepackage{latexsym}
\usepackage{amsmath}
\usepackage{epsfig}
\usepackage{apalike}
\usepackage{graphicx}

\usepackage{xcolor, minitoc}

\newcommand{\IP}{\mathbb P}

\newcommand{\IE}{\mathbb E}
\newcommand{\IR}{\mathbb R}

\newtheorem{thm}{Theorem}[section]
\newtheorem{propn}[thm]{Proposition}
\newtheorem{lemma}[thm]{Lemma}

\newtheorem{defn}[thm]{Definition}
\newtheorem{remark}[thm]{Remark}
\newtheorem{notn}[thm]{Notation}
\newtheorem{corollary}[thm]{Corollary}

\def \hat{\widehat}

\begin{document}

\title{The infinitesimal model with dominance}
\author{ N H Barton\thanks{NHB supported in part by ERC Grants 250152 and 101055327}
\\Institute of Science and Technology
\\Am Campus I\\A-3400 Klosterneuberg\\Austria\\Email nick.barton@ist-austria.ac.at
\\~\\ A M Etheridge
\\Department of Statistics\\University of Oxford\\24--29 St Giles\\
Oxford OX1 3LB, UK\\~\\Amandine V\'eber\thanks{AV partly supported by the chaire Mod\'elisation Math\'ematique
et Biodiversit\'e of Veolia Environment - Ecole Polytechnique - Museum
National d'Histoire Naturelle - Fondation X}\\
Universit\'e Paris Cit\'e\\
CNRS\\
MAP5 F-75006 Paris\\
France}
\maketitle

\def\thefootnote{\fnsymbol{footnote}}
\def\@makefnmark{\hbox to\z@{$\m@th^{\@thefnmark}$\hss}}
\footnotesize\rm\noindent
\hspace*{16pt}\strut
\normalsize\rm

\begin{abstract}
The classical infinitesimal model is a simple and robust model for the inheritance of quantitative traits. In this model, a quantitative trait is expressed as the sum of a genetic and a non-genetic (environmental) component and the genetic component of offspring traits within a family follows a normal distribution around the average of the parents' trait values, and has a variance that is independent of the trait values of the parents.
Although the trait distribution across the whole population can be far from normal, the trait distributions within families are normally distributed with a variance-covariance matrix that is determined entirely by that in the ancestral population and the probabilities of identity determined by the pedigree. Moreover, conditioning on some of the trait values within the pedigree has predictable effects on the mean and variance within and between families. In previous work, Barton et al.~(2017), \nocite{barton/etheridge/veber:2017} we showed that when trait values are determined by the sum of a large number of Mendelian factors, each of small effect, one can justify the infinitesimal model as limit of Mendelian inheritance. It was also shown that under some forms of epistasis, trait values within a family are still normally distributed.

In this paper, we show that this extraordinary robustness of the infinitesimal model extends to include dominance. We define the model in terms of classical quantities of quantitative genetics, before justifying it as a limit of Mendelian inheritance as the number, $M$, of underlying loci tends to infinity. As in the additive case, the multivariate normal distribution of trait values across the pedigree can be expressed in terms of variance components in an ancestral population and probabilities of identity by descent determined by the pedigree. Now, with just first order dominance effects, we require two, three and four way identities. In this setting, it is natural to decompose trait values, not just into the additive and dominance components, but into a component that is shared by all individuals within the family and an independent `residual' for each offspring, which captures the randomness of Mendelian inheritance. In the additive case, the first term is just the mean of the parental trait values, but with dominance it is random. We show that, even if we condition on parental trait values, both the shared component and the residuals within each family will be asymptotically normally distributed as the number of loci tends to infinity, with an error of order $1/\sqrt{M}$.

We illustrate our results with some numerical examples.
\end{abstract}

\section{Introduction}\label{s:introduction}

In the classical infinitesimal model, a quantitative trait is expressed as the sum of a genetic and a non-genetic (environmental) component and the genetic component of offspring traits within a family follows a normal distribution around the average of the parents' trait values, and has a variance that is independent of the trait values of the parents. With inbreeding, the variance decreases in proportion to relatedness. When trait values are determined by the sum of a large number of Mendelian factors, each of small effect, as we show in Barton et al.~(2017), \nocite{barton/etheridge/veber:2017}one can justify the infinitesimal model as a limit of Mendelian inheritance. Crucially, the results of Barton et al.~(2017) \nocite{barton/etheridge/veber:2017} show that the evolutionary forces such as random drift and population structure are captured by the pedigree; conditioning on that pedigree, and trait values in the population in all generations before the present, the within family distributions in the present generation will be given by a multivariate normal, with variance determined by that in the ancestral population and probabilities of identity by descent that can be deduced from the pedigree. If some traits in the pedigree are unknown, then averaging with respect to the ancestral distribution, the multivariate normality is preserved. It was also shown that under some forms of epistasis, trait values within a family are still normally distributed, although the mean will no longer be a simple function of the traits in the parents (as there are epistatic components which cannot be observed directly).

We emphasize that as a result of selection, population structure, and so on, the trait distribution \emph{across} the population can be far from normal; the infinitesimal model as we define it only asserts that the \emph{within families} distributions of the genetic component of the trait are Gaussian, with a variance-covariance matrix that is determined entirely by that in an ancestral population and the probabilities of identity determined by the pedigree. Moreover, as a result of the multivariate normality, conditioning on some of the trait values within that pedigree has predictable effects on the mean and variance within and between families. In other words, knowing the traits values for some individuals in the population does not distort the multivariate normality of the distribution of the unobserved traits, and the mean and covariances of these traits may be derived explicitly (albeit after rather tedious calculations).

In this paper, we show that this extraordinary robustness of the infinitesimal model extends to include dominance. The distribution of the genetic part of the trait will once again be a multivariate normal distribution whose mean and variance is expressed in terms of the variance components in an ancestral population and probabilities of identity by descent determined by the pedigree, but now, with just first order dominance effects, the identities required will involve up to 4 genes. As with the case of epistasis, the mean is not a simple function of the trait values in the parents, and there is nontrivial covariance between families. One can think of the genetic component of the trait values within a family as consisting of two parts. Both are normally distributed. In the additive case, the first reduces to the mean of the trait values of the parents; with dominance it will be random (even if we condition on knowing the parental traits), but the same for all individuals in the family. What is at first sight surprising is that even if we condition on knowing the trait values of the parents, this shared quantity is normally distributed. Assuming there is no mutation to ease the presentation (the effect of mutation was studied in Barton et al.~(2017)), our first contribution is to show how to calculate its mean and variance from knowledge of variance components in the ancestral population and the pedigree, both with and without knowledge of the trait values of the parents. Knowing the trait values of the parents shifts the mean in a predictable way; the variance is independent of the parental trait values. The second part of the trait value, which is independent for each offspring in the family, is independent of the first; it encodes the randomness of Mendelian inheritance. It is a draw from a normally distributed random variable with mean zero and variance again determined by the pedigree and variance components in the ancestral population. It is not affected by conditioning on parental trait values. This segregation of the trait into a shared part and a residual part that is independent for each member of a family, is not the classical subdivision into additive and dominance components, but it arises naturally both in the formulation of the infinitesimal model and in its derivation as a limit of Mendelian inheritance for a large number of loci each of small effect. We give a more mathematical description of it in Eq.~\eqref{decomposition}.

Our work can be seen as an extension of that of Abney et al.~(2000), who \nocite{abney/mcpeek/ober:2000} establish sufficient conditions for a Central Limit Theorem to be applied to the vector of trait values in the presence of dominance and inbreeding. Our second contribution in this work is to  establish the magnitude of the error in that normal approximation, verify that in conditioning on the trait values of the parents of an individual we are not (unless those traits are very extreme or the pedigree is very inbred) leaving the domain where the normal approximation is valid, and write down the effect of knowing those parental trait values on the distribution of the individual's own trait. A careful statement of our results can be found in Theorems~\ref{convergence of residuals} and~\ref{shared parent theorem}. The notation we shall need is rather involved, but in a nutshell, we shall write the trait $\widetilde{Z}^i$ of a given diploid individual $i$ in generation $t$ as the sum over $M$ loci of per-locus allelic effects that are functions of the allelic states $\chi_l^1,\chi_l^2$ of the two genes of individual $i$ at locus $l$, plus an environmental contribution $E^i$ (that we shall assume to be Gaussian):
\begin{equation}\label{def trait}
\widetilde{Z}^i = \bar{z}_0 +\sum_{l=1}^M \frac{1}{\sqrt{M}}\big(\eta_l(\chi_l^1) + \eta_l(\chi_l^2) + \phi_l(\chi_l^1,\chi_l^2)\big) + E^i.
\end{equation}
Here, $\bar{z}_0$ is the average trait value in the ancestral population (itself a sum of average allelic effects) and the sum encodes the contribution of all loci to the deviation from this average (each per-locus deviation being of order $1/\sqrt{M}$, see Barton et al.~(2017) and Section~\ref{setting out the model} below for a justification). In this sum, the term $\eta_l(\chi_l^1) + \eta_l(\chi_l^2)$ models the additive part of the contribution of locus $l$ and $\phi_l(\chi_l^1,\chi_l^2)$ models the part due to dominance. Assuming Mendelian inheritance and no linkage between the $M$ loci, at each locus the allelic state $\chi_l^1$ is a copy of the allelic state of one of the two genes in the `first' parent of $i$, chosen at random, and $\chi_l^2$ is a copy of the allelic state of one of the two genes in the `second' parent of $i$, again chosen uniformly at random. Writing $\chi_l^{i[1],1},\chi_l^{i[1],2}$ for the alleles at locus $l$ in the first parent and $\chi_l^{i[2],1},\chi_l^{i[2],2}$ for the alleles in the second parent, we can then write the sum over all loci in~\eqref{def trait} as the sum of an \emph{average} parental contribution (shared by all offspring of these parents), and a residual term of mean zero that encodes the stochasticity of Mendelian inheritance (the \emph{actual} genetic contribution of the parents minus their average contribution). To avoid introducing even more notation, here we simply write $R_A^i$ and $R_D^i$ for the parts of the residual due to the additive terms and to the dominance terms respectively. Explicit formulae are given in \eqref{remainder R1}-\eqref{remainder S2}. Doing so, we obtain
\begin{align}
\widetilde{Z}^i =&\, \bar{z}_0 + \frac{1}{\sqrt{M}}\sum_{l=1}^M\bigg\{\frac{\eta_l(\chi_l^{i[1],1}) + \eta_l(\chi_l^{i[1],2})}{2} + \frac{\eta_l(\chi_l^{i[2],1}) + \eta_l(\chi_l^{i[2],2})}{2}\bigg\} \nonumber\\
& + \frac{1}{\sqrt{M}}\sum_{l=1}^M \frac{\phi_l(\chi_l^{i[1],1},\chi_l^{i[2],1})+\phi_l(\chi_l^{i[1],2},\chi_l^{i[2],1})+\phi_l(\chi_l^{i[1],1},\chi_l^{i[2],2})+\phi_l(\chi_l^{i[1],2},\chi_l^{i[2],2})}{4}\nonumber\\
& + R_A^i + R_D^i + E^i \nonumber\\
=: &\, \bar{z}_0 + {\cal A}^i + {\cal D}^i + R_A^i + R_D^i + E^i.\label{decomposition}
\end{align}
The genetic component of the trait can thus be seen either as the sum of an additive part (${\cal A}^i + R_A^i$) and a dominance part (${\cal D}^i + R_D^i$), or as the sum of a shared part (${\cal A}^i + {\cal D}^i$) and a residual part ($R_A^i + R_D^i$). Following the same strategy as in Barton et al.~(2017), in Theorem~\ref{convergence of residuals} we show that even conditionally on (\emph{i.e.}, knowing) the parental traits $\widetilde{Z}^{i[1]}$ and $\widetilde{Z}^{i[2]}$, as $M$ tends to infinity the residual part converges in distribution to a Gaussian distribution with mean $0$ and a variance depending only on variance components in the ancestral population and on the probability of identity by descent between two parental genes (which is fully determined by the pedigree). Crucially, the limiting variance does not depend on the parental traits. This convergence happens at a rate proportional to $1/\sqrt{M}$. Turning to the shared part, we use a different approach to prove that conditional on $\widetilde{Z}^{i[1]}$ and $\widetilde{Z}^{i[2]}$, ${\cal A}^i+{\cal D}^i$ also converges to a Gaussian distribution as $M$ tends to infinity. Again, the nonzero mean and the variance of the limiting normal distribution can be fully described, the variance is independent of the parental traits and the convergence happens at a rate proportional to $1/\sqrt{M}$. This is the content of Theorem~\ref{shared parent theorem}, in the special (and most difficult) case when individual~$i$ was produced by selfing. For both the shared and the residual parts, the rate of convergence deteriorates when the pedigree is too inbred (leading to probabilities of identity by descent close to $1$ between some pairs of parental genes), or when some traits in the population are too extreme (as knowing the trait value then gives us too much information about the unobserved underlying allelic states).

Our derivation of the infinitesimal model as the limit of a finite-locus model has two interesting corollaries. First, as mentioned above, we obtain that the error made by approximating the trait distribution within a family by a Gaussian distribution increases by a quantity of order $1/\sqrt{M}$ in each generation. Consequently, for very large $M$, we expect the infinitesimal model with dominance to be valid for a time of the order of $\sqrt{M}$ generations, provided the population is not too inbred and no too extreme traits appear in the meantime. Second, the set of technical lemmas that are key to the proofs of these results, presented in Appendix~\ref{key lemmas}, show that the infinitesimal model leaves essentially no signature on the allele frequencies at any given locus: even knowing the ancestral traits, the distribution of the allelic state at a single locus in a given individual is barely distorted by selection acting on the trait and the result is that, at the population level, the allelic distribution evolves in an essentially neutral way. In particular, its variance only depends on the variance of the allele distribution in the ancestral population and on identities by descent, that are not changed by knowledge of the trait values.

The rest of the paper is organised as follows. In Section~\ref{s:identity}, we define the identity coefficients (that is, probabilities of identity by descent) that we shall need to formulate the model precisely. We show how to compute them knowing the population pedigree in Appendix~\ref{appendix:id coef} and provide the corresponding \emph{Mathematica} code in Supplementary Material \cite{barton:2023}. In Section~\ref{setting out the model}, we spell out the model in terms of quantities that are familiar from classical quantitative genetics, and we explore its accuracy numerically in Section~\ref{s:numerics}.
Finally, in Section~\ref{mendelian inheritance} we derive this extension of the infinitesimal model as a limit of a model of Mendelian inheritance on the pedigree. The calculations are somewhat involved, and almost all will be relegated to the appendices. We must modify the strategy of Barton et al.~(2017), which, although valid for the part of the trait value which is independent for each individual within the family, does not suffice for proving normality of the part of the trait value that is shared by all individuals within a family.\nocite{barton/etheridge/veber:2017} To prove that this is normally distributed requires a new approach, based on an extension of Stein's method of exchangeable pairs. To keep the expressions in our calculations manageable, we satisfy ourselves with presenting the details only in the case in which we condition on knowing the trait values of the parents of an individual, in contrast to the additive case of Barton et al.~(2017), in which we conditioned on knowing all the trait values in the pedigree right back to the ancestral generation. Our approach could readily be extended to conditioning on knowledge of more trait values, which amounts to conditioning a multivariate normal on some of its marginals. In Appendix~\ref{appendix: accumulation of information} we present the new ideas that are required to control the way in which errors in the infinitesimal approximation accumulate from knowledge of trait values of more distant relatives in the presence of dominance.

Just as in the additive case, the key will be to show that because many different combinations of allelic states
are consistent with the same trait value, knowledge of the pedigree, and the trait values of the parents of an individual in that pedigree, actually gives very little information about the allelic state at a particular locus in that individual, or about correlations between two specific loci. An important consequence of this is that, in practice, it is going to be hard to observe signals of polygenic adaptation, because even a large shift in a trait
caused by strong selection does not yield a prediction about alleles at a particular locus.

\section{Identity coefficients}\label{s:identity}

In the case of an additive trait, the infinitesimal model can be expressed in terms of the variance in the ancestral population (that is, the base population which we shall call \emph{generation zero}) and two-way identity coefficients at a single locus. Recall that two genes at a given locus are identical by descent if their allelic states are identical and were inherited from a common ancestor. Since we assume that individuals are diploid, we need to specify which genes we consider when defining the identity coefficients.

For two distinct individuals $i$ and $j$ in the same generation, we define $F_{ij}$ to be the probability of identity by descent between two genes (at a given locus), one taken uniformly at random among the two genes of individual $i$ and one taken at random among the two genes of individual $j$. When $i=j$, $F_{ii}$ is defined to be the probability of identity by descent of the two \emph{distinct} genes in the diploid individual $i$.

The definition naturally extends to subsets of three or four genes taken from two distinct individuals (again, at a given locus), for which we shall talk about three- and four-way identities. These quantities will be required to state our results below.

\begin{figure}
\centerline{\includegraphics[width=4in]{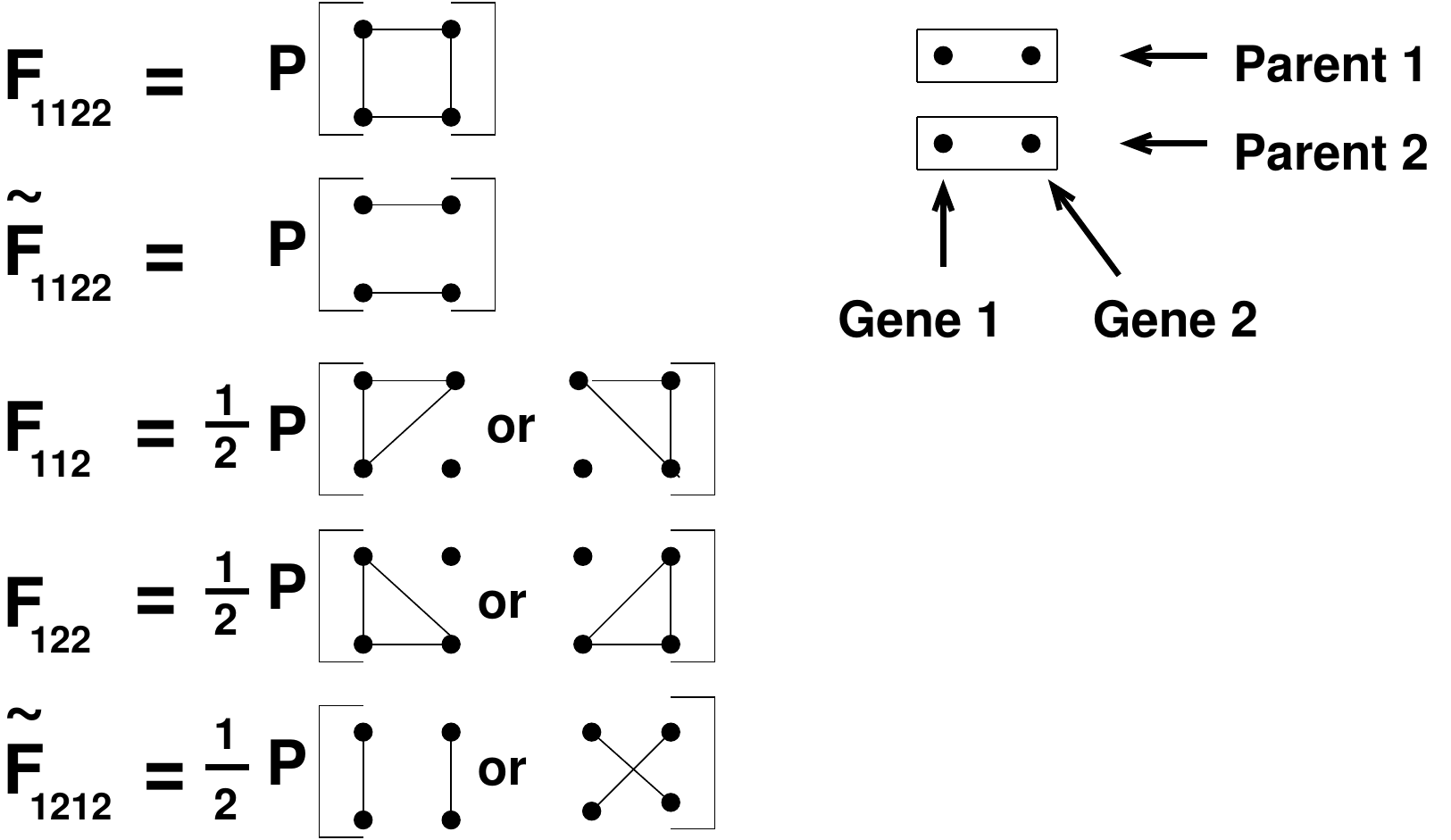}}
\caption{Three- and four-way identities. Lines indicate identity by descent between genes. See the main text for further explanation.}
\label{picture for identities}
\end{figure}

We use $F_{122}$ for the probability that the two genes in individual $2$ are identical by descent \emph{and} they are identical by descent with a gene chosen at random from individual $1$. We write $F_{1122}$ for the probability that all four genes across individuals $1$ and $2$ are identical by descent; this corresponds to the quantity $\delta$ in Walsh \& Lynch~(2018), \nocite{walsh/lynch:2018} Chapter 11. We need an expression for the probability that each gene in individual~$1$ is identical by descent with a \emph{different} gene in individual~$2$ and all four are not identical. We shall denote this by $\widetilde{F}_{1212}$. This is denoted by $(\Delta -\delta)$ in Walsh \& Lynch~(2018). \nocite{walsh/lynch:2018} Finally we need the probability that the two genes in individual~$1$ are identical, as are the two genes in individual $2$, but the four genes are not \emph{all} identical, which we shall denote
by $\widetilde{F}_{1122}$. We illustrate the three- and four-way identities in Figure~\ref{picture for identities}. During the course of our mathematical derivations, it will be convenient to express all two-, three-, and four-way identities in terms of the nine possible four-way identities \nocite{walsh/lynch:2018} (Walsh \& Lynch, 2018; Figure~11.5). This is illustrated in Figure~\ref{picture for all identities}.

In Appendix~\ref{appendix:id coef}, we discuss how to compute these identity coefficients given a pedigree. From now on we simply write `identity' instead of `identity by descent'.

\section{The infinitesimal model with dominance}
\label{setting out the model}
For ease of exposition, in this section we leave aside the environmental component of the trait value and we focus on its genetic component, which we denote by $Z$ (so that in the notation of~\eqref{def trait}, $\widetilde{Z} = Z+E$). We first introduce the different quantities that are involved in this component of the trait value in a rigorous way, most of which were already hinted at in the introduction, and then we compute the mean and variance of the shared and residual parts of $Z$ with and without knowledge of the parental traits.

The population is diploid and trait values are determined by the allelic states at $M$ \emph{unlinked} loci. Each locus thus corresponds to a pair of genes. We assume that in generation zero (\emph{i.e.}, in the `ancestral' population), the individuals that found the pedigree are unrelated and sampled from an ancestral population in which all loci are in linkage equilibrium and are in Hardy-Weinberg equilibrium (that is, in the ancestral population the two allelic states at each locus in a given individual are sampled independently of each other and therefore the probability that an individual carries a given pair of alleles is given by the product of the probabilities of each allele being sampled).

In order to define the various quantities that enter into our model, we introduce notation to express the trait as a sum of effects over loci. However, we emphasize that once these components, all of which are familiar from classical quantitative genetics, have been calculated for the ancestral population, the model can be defined without reference to the effects of individual loci.

To adhere to the notation of Barton et al.~(2017), \nocite{barton/etheridge/veber:2017} we use $\chi_l^1$, $\chi_l^2$ for the allelic states of the two genes at locus $l$ in a given individual in the pedigree. When we talk about the distribution of the allelic state of a single gene, we drop the superscript $1$ or $2$ and simply write $\chi_l$.
We write $\bar{z}_0$ for the mean trait value in the ancestral population and express the trait value of an individual as $\bar{z}_0$ plus a sum of allelic effects. The influence of each locus will scale as $1/\sqrt{M}$,
where $M$ is the total number of loci (assumed large). We write $\eta_l(\chi_l)$ to denote the (order one) scaled additive effect of the allele $\chi_l$ and $\phi_l(\chi_l^1,\chi_l^2)$ for the scaled dominance component (where $\phi_l$ is assumed to be a symmetric function of the two allelic states $\chi_l^1$ and $\chi_l^2$). That is, the total contribution of locus $l$ to the trait value will be of the form
$$
\frac{1}{\sqrt{M}}\big( \eta_l(\chi_l^1)+\eta_l(\chi_l^2)\big) + \frac{1}{\sqrt{M}}\, \phi_l(\chi_l^1,\chi_l^2).
$$

We shall assume that both $\eta_l$ and $\phi_l$ are uniformly bounded (\emph{i.e.}, they will all take their values in some finite interval $[-B,B]$.). We also suppose that dominance effects are sufficiently `balanced' that inbreeding depression is finite at least in the ancestral population. More precisely, let $\hat{\chi}_l$ denote an allele sampled at random from the distribution of alleles at locus $l$ in the ancestral population, then $\iota$ defined by
\begin{equation}\label{def iota}
\iota=\frac{1}{\sqrt{M}}\sum_{l=1}^M\IE[\phi_l(\hat{\chi}_l,\hat{\chi}_l)]
\end{equation}
is bounded (as a function of $M$). This condition is crucial to our result. It is not obvious that it can hold, as the number of terms in the sum grows linearly with $M$ while the scaling factor $1/\sqrt{M}$ decreases much more slowly. Such a uniform bound is possible for instance if we consider a situation in which the contributions of the different loci compensate each other in a `random-walk-like' way, \emph{i.e.}, each expectation is either positive or negative (by the same amount, say), and the number of positive and negative expectations differ by at most $\mathcal{O}(\sqrt{M})$. An example is presented at the beginning of Section~\ref{s:numerics}. Note however that the quantity $\iota$ may be bounded uniformly in $M$ for many other reasons. For simplicity, we do not consider higher order dominance components (that is $D\times D$ -- or more complex -- components) here.

\begin{remark} Note that $\hat{\chi}_l$ is the random variable describing a draw from the distribution of allelic states at locus~$l$ in the ancestral population (generation $0$), while we use $\chi_l$ to denote the allelic state at locus~$l$ in a given individual in the pedigree (living in generation $t$, say). \emph{A priori}, the law of $\chi_l$ is a biased version of the law of $\hat{\chi}_l$, obtained after letting selection and drift act over $t$ generations, but in Appendix~\ref{key lemmas} we shall show that, in effect, this distortion is very small for each given locus, and $\hat{\chi}_l$ and $\chi_l$ have the same distribution up to a small error even if we condition on knowing the parental (or ancestral) trait values.
\end{remark}

For an individual in the ancestral population, its allelic states at locus $l$, which we denote by $\widehat{\chi}^1_l, \widehat{\chi}^2_l$, are independent draws from a distribution $\widehat{\nu}_l$ on possible allelic states that we assume is known. It is convenient to normalise so that $\IE[\eta_l(\widehat{\chi}_l)]=0$,
$\IE[\phi_l(\widehat{\chi}_l^1,\widehat{\chi}_l^2)]=0$, and for any value $x'$ of the allelic state at locus $l$, the conditional expectations $\IE[\phi_l(\widehat{\chi}_l,x')]=0 =\IE[\phi_l(x',\widehat{\chi}_l)]$. We explain in Section~\ref{mendelian inheritance} why these assumptions do not result in a loss of generality. The genetic component of the trait value takes the form (compare with Eq.~\eqref{def trait}, the expression for the observed trait including environmental noise)
\begin{equation}
\label{expression for genetic component of trait}
Z=\bar{z}_0+\frac{1}{\sqrt{M}} \sum_{l=1}^M\left(\eta_l(\chi_l^1)
+\eta_l(\chi_l^2)
+\phi_l(\chi_l^1,\chi_l^2)\right).
\end{equation}

Let us write $i[1]$ and $i[2]$ for the parents of the individual labelled $i$. As advertised in the introduction, the genetic component of an offspring's trait value has two contributions. The first one is shared by all its siblings, and is a random quantity which is characteristic of the family. The second contribution is unique to the individual and independent of the first one. In our proofs, we shall investigate these two parts separately. We shall use the notation $Z^i=({\cal A}^i+{\cal D}^i)+(R_A^i+R_D^i)$, where the shared part has been further subdivided into the contribution ${\cal A}^i$ from the additive component, and the contribution ${\cal D}^i$ from the dominance component. The residuals $R_A^i$ and $R_D^i$ are determined by Mendelian inheritance and correspond to the contributions from the additive and dominance components respectively. Explicit expressions for these quantities are in Eq.~(\ref{remainder R1})-(\ref{defn of D}) below. In this notation, the additive part of the trait value is ${\cal A}^i+R_A^i$ and the dominance deviation is ${\cal D}^i+R_D^i$.

\subsection*{Trait values for a given pedigree}

We now define the infinitesimal model in terms of classical quantities of quantitative genetics that can be expressed in terms of expectations in the ancestral population and identities determined by the pedigree. We use the notation of Walsh \& Lynch~(2018), \nocite{walsh/lynch:2018} which we recall in Table~\ref{QG coefficients}.
\begin{table}
\centering
\begin{tabular}{ll}\\
\hline\\
Additive variance & $\sigma_A^2=\frac{2}{M}\sum_{l=1}^M
\IE[\eta_l(\widehat{\chi}_l)^2]$\\
\\
Dominance variance & $\sigma_D^2=\frac{1}{M}\sum_{l=1}^M
\IE[\phi_l(\widehat{\chi}_l^1,\widehat{\chi}_l^2)^2]$\\
\\
Inbreeding depression & $\iota =\frac{1}{\sqrt{M}}\sum_{l=1}^M
\IE[\phi_l(\widehat{\chi}_l,\widehat{\chi}_l)]$\\
\\
Sum of squared locus-specific & $\iota^*=\frac{1}{M}\sum_{l=1}^M
\IE[\phi_l(\widehat{\chi}_l,\widehat{\chi}_l)]^2$\\
inbreeding depressions & \\
\\
Variance of dominance effects & $\sigma_{DI}^2=\frac{1}{M}\sum_{l=1}^M
\left(\IE[\phi_l(\widehat{\chi}_l,\widehat{\chi}_l)^2]-
\IE[\phi_l(\widehat{\chi}_l,\widehat{\chi}_l)]^2\right)$\\
in inbred individuals & \\
\\
Covariance of additive and & $\sigma_{ADI}=\frac{2}{M}\sum_{l=1}^M
\IE\left[\eta_l(\widehat{\chi}_l)
\phi_l(\widehat{\chi}_l,\widehat{\chi}_l)\right]$
\\
dominance effects in inbred &
\\
individuals & \\
\\
\hline\hline \\
Additive part of the & ${\cal A}^i$ -- defined in \eqref{defn of A} \\
shared component & \\
\\
Dominance part of the & ${\cal D}^i$ -- defined in \eqref{defn of D} \\
shared component & \\
\\
Additive part of the & $R_A^i$ -- defined by \eqref{remainder R1} + \eqref{remainder R2} \\
residual & \\
\\
Dominance part of the & $R_D^i$ -- defined by \eqref{remainder S1} + \eqref{remainder S2} \\
residual & \\
\\
Genetic component of trait value & $Z^i=\bar{z}_0 + {\cal A}^i + {\cal D}^i + R_A^i + R_D^i$ \\
\\
Observed trait value & $\widetilde{Z}^i=Z^i + E^i$, $\quad E^i\sim \mathcal{N}(0,\sigma_E^2)$\\
\\
\hline
\end{tabular}
\caption{Coefficients of classical quantitative genetics (top) and elements of individual trait decomposition (bottom). We use $\widehat{\chi}_l$ to denote an allelic state sampled from the distribution $\widehat{\nu}_l$ of possible allelic states at locus $l$ in the ancestral population; $\widehat{\chi}_l^1$, $\widehat{\chi}_l^2$ are independent draws from the same distribution.}
\label{QG coefficients}
\end{table}
Under the infinitesimal model, conditional on the pedigree, the components
$({\cal A}^i+{\cal D}^i)$ and $(R_A^i+R_D^i)$ of
the trait values of individuals in a family
follow independent multivariate normal distributions.
In Appendix~\ref{QG derivations}
the expressions presented in this section will be justified by taking
the trait values
determined by~(\ref{expression for genetic component of trait})
under a model of Mendelian inheritance. In writing down the
infinitesimal model, we shall assume
that as the number of loci tends to infinity, the quantities defined in the top part of
Table~\ref{QG coefficients}
converge to well defined limits.

To simplify notation, we shall use $1$ and $2$ in place of $i[1]$ and
$i[2]$ in our expressions for identity; thus, for example,
$F_{12}\equiv F_{i[1],i[2]}$, and $F_{11}$ will be the probability of
identity by descent of the two genes in parent $i[1]$.
The mean and variance of $({\cal A}^i+{\cal D}^i)$ are then
\begin{equation}
\label{mean A+D}
\IE[{\cal A}^i+{\cal D}^i]=\iota F_{12},
\end{equation}
and
\begin{align}
\label{variance A+D unconditioned}
\mathtt{Var}({\cal A}^i+{\cal D}^i)= &\ \frac{\sigma_A^2}{2}
\left(1+\frac{F_{11}+F_{22}}{2}+2F_{12}\right)
+\sigma_{ADI}\left(F_{12}+\frac{F_{112}+F_{122}}{2}\right) \nonumber
\\ & +\frac{(\sigma_{DI}^2+\iota^*)}{4}\left(F_{12}+F_{112}+F_{122}+F_{1122}\right) +\frac{\iota^*}{4}\widetilde{F}_{1212}-\iota^* F_{12}^2 \nonumber
\\
& +\frac{\sigma_D^2}{4}\left(1-F_{12}+F_{22}-F_{122}+
F_{11}-F_{112}+\widetilde{F}_{1122}+\frac{1}{2}\widetilde{F}_{1212}\right).
\end{align}
In this expression, the term proportional to $\sigma_A^2$ is the variance of ${\cal A}^i$, the term proportional to $\sigma_{ADI}$ is twice the covariance of ${\cal A}^i$ and ${\cal D}^i$ and the remaining sum gives the variance of ${\cal D}^i$. Recall that we are assuming here that the ancestral population is
in linkage equilibrium. With linkage disequilibrium there is an additional term, \emph{c.f.}~the remark below Eq.~(\ref{variance Z}). The components $({\cal A}+{\cal D})$ are also correlated across families. For individuals
labelled $i$ and $j$ respectively,
\begin{align}
\label{covariance A plus D}
\mathtt{Cov}(({\cal A}^i+{\cal D}^i),({\cal A}^j+{\cal D}^j))=&\ 2F_{ij}\sigma_A^2
+(F_{ijj}+F_{iij})\sigma_{ADI}\nonumber\\
& +\widetilde{F}_{ijij}\sigma_D^2
+F_{iijj}(\sigma_{DI}^2+\iota^*)-\iota^2F_{ii}F_{jj}
+\iota^*\widetilde{F}_{iijj}.
\end{align}
Note that, in contrast to our expression for the variance of $Z^i$, in
this expression, the subscripts $i$ and $j$ in the identities
refer to the individuals themselves, not their parents; for example
the expression $F_{ij}$ is the probability of
identity of two genes, one sampled at random from individual $i$ and one
sampled at random from individual $j$.
We reserve letters for individuals in the current generation, and
numbers for their parents.

If we combine the components $R_A^i$ and $R_D^i$ that segregate within families, we have that the sums $(R_A^i+R_D^i)$ are independent of each other (due to the independence of the variables encoding Mendelian inheritance), mean zero, normally distributed random variables with variance
\begin{align}
\mathtt{Var}(R_A^i+R_D^i) = &
\left(1-\frac{F_{11}+F_{22}}{2}\right)\frac{\sigma_A^2}{2}
+
\frac{1}{4}\left(3F_{12}-F_{1122}-F_{112}-F_{122}\right)
\left(\sigma_{DI}^2+\iota^*\right) \nonumber\\
& +\frac{1}{4}\left(3(1-F_{12})-(F_{11}-F_{112})-(F_{22}-F_{122})
-\widetilde{F}_{1122}-\frac{1}{2}\widetilde{F}_{1212}\right)\sigma_D^2
\nonumber\\
&+\left(F_{12}-\frac{F_{112}+F_{122}}{2}\right)\sigma_{ADI}-\frac{\iota^*}{4}\widetilde{F}_{1212}. \label{var residual}
\end{align}
Here again, the term proportional to $\sigma_A^2$ is the variance of $R_A^i$, the term proportional to $\sigma_{ADI}$ is twice the covariance of $R_A^i$ and $R_D^i$, and the remaining sum equals the variance of $R_D^i$.
We calculate the mean, variance and covariance of these different components
in Appendix~\ref{QG derivations}. In order to recover the mean and variance of the trait values, we add
the contributions of $({\cal A}^i+{\cal D}^i)$ and $(R_A^i+R_D^i)$ and observe that the identity
$F_{12}$ in our expressions for the variances of these quantities (which
we recall was
the probability of identity of one gene sampled at random from each of
the parents $i[1]$, $i[2]$ of our individual)
corresponds to $F_{ii}$. This yields that,
conditional on the pedigree,
\begin{equation}
\label{mean Z}
\IE[Z^i]=\bar{z}_0+\iota F_{ii},
\end{equation}
\begin{equation}
\label{covariance Z}
\mathtt{Cov}(Z^i,Z^j)=2F_{ij}\sigma_A^2
+(F_{ijj}+F_{iij})\sigma_{ADI}+\widetilde{F}_{ijij}\sigma_D^2
+F_{iijj}(\sigma_{DI}^2+\iota^*)-\iota^2F_{ii}F_{jj} +\iota^*\widetilde{F}_{iijj},
\end{equation}
and
\begin{equation}
\label{variance Z}
\mathtt{Var}(Z^i)=\sigma_A^2(1+F_{ii})+\sigma_D^2(1-F_{ii}) +(\sigma_{DI}^2+\iota^*)F_{ii} +2\sigma_{ADI}F_{ii}-\iota^*F_{ii}^2.
\end{equation}
For a single individual, its trait value can only depend on the two alleles that it carries at each locus, so it is no surprise that this expression
depends only on pairwise identities between those two genes.
We remark that~(\ref{variance Z}) differs from the corresponding expression
(Eq.~11.6c) in Walsh \& Lynch~(2018).
\nocite{walsh/lynch:2018}
To recover exactly their expression, one must
add $(\tilde{f}-F_{ii}^2)(\iota^2-\iota^*)$ to the right hand side, where
$\tilde{f}$ is the probability of identity at two distinct loci
in individual $i$.
We see how to recover this term in
Remark~\ref{source of discrepancy from WL},
but because we have assumed
linkage equilibrium in our base population, for the
period over which the infinitesimal model remains a good approximation,
under our assumptions we have
$\tilde{f}\approx F_{ii}^2$. This is not to say that there is not a
significant contribution to the trait value from linkage disequilibrium;
it is just that for any specific pair of loci it is negligible. We shall
see a toy example that reinforces this point at the
beginning of Section~\ref{mendelian inheritance}.

We emphasize again that our partition of the trait values into a
contribution that is shared by all individuals in a family and residuals
differs from the conventional split into an additive part and a
dominance deviation. The additive part of the trait is
$A^i={\cal A}^i+R_A^i$ and the dominance component is $D^i={\cal D}^i+R_D^i$.
From our calculations in Appendix~\ref{QG derivations}, we can read off
\begin{equation}\label{prediction AD}
\IE[A^i]=0,\qquad \IE[D^i]=\iota F_{ii},
\end{equation}
\begin{equation}\label{prediction Var AD}
\mathtt{Var}(A^i)=\sigma_A^2\big(1+F_{ii}\big),
\qquad\mathtt{Cov}(A^i,D^i)=\sigma_{ADI}F_{ii},
\end{equation}
and
\begin{equation}
\label{variance dominance deviation}
\mathtt{Var}(D^i)=\sigma_D^2\big(1-F_{ii}\big)+\sigma_{DI}^2F_{ii}+
\iota^*\big(F_{ii}-F_{ii}^2\big).
\end{equation}

\begin{remark}
Notice that the purely additive case can be simply recovered by taking $\phi_l\equiv 0$, so that ${\cal D}^i=0=R_D^i$, and $\sigma_A^2$ is the only nonzero variance coefficient. This yields
\begin{align*}
& \IE[{\cal A}^i+{\cal D}^i]= 0, \quad \mathtt{Var}({\cal A}^i+{\cal D}^i)= \frac{\sigma_A^2}{2} \bigg(1+\frac{F_{11}+F_{22}}{2}+2F_{12}\bigg), \\
& \mathtt{Cov}(({\cal A}^i+{\cal D}^i),({\cal A}^j+{\cal D}^j))=2F_{ij}\sigma_A^2, \quad \mathtt{Var}(R_A^i+R_D^i)= \bigg(1-\frac{F_{11}+F_{22}}{2}\bigg)\, \frac{\sigma_A^2}{2},
\end{align*}
and finally
$$\IE[Z^i] =\bar{z}_0,\quad \mathtt{Var}(Z^i)=\sigma_A^2(1+F_{ii}), \quad \mathtt{Cov}(Z^i,Z^j)= 2F_{ij}\sigma_A^2.
$$
\end{remark}

\subsection*{Conditioning on trait values of parents}

Under the infinitesimal model,
the trait values of individuals across the pedigree are given
by a multivariate normal. Therefore standard results on conditioning multivariate
normal random vectors on their marginal values, which for ease of reference
we record in Appendix~\ref{conditioning normals}, allow us to
read off the effect on the distribution of $Z^i$ of conditioning on
$Z^{i[1]}$ and $Z^{i[2]}$.
However, a little care is needed; we shall be justifying the normal
distribution within families
as an approximation as the number of loci tends to infinity,
and we must be sure that asymptotic normality is
preserved under this conditioning.
We shall see that if, for example, parental trait values are
too extreme, then the conditioning pushes us to a part of the
probability space where the normal approximation breaks down. This is
particularly evident in the toy example that we present in
Section~\ref{mendelian inheritance}.
A justification for asymptotic normality even after conditioning is
outlined in Section~\ref{mendelian inheritance}, and details are
presented in
the appendices.

Just as in the classical infinitesimal model, the mean and variance of the residuals $R_A^i+R_D^i$ are unchanged by conditioning on the trait
values of the parents (recall that these residuals encode the stochasticity due to Mendelian inheritance at each locus; expressions for $R_A^i$ and $R_D^i$ are given in Eq.~(\ref{remainder R1})-(\ref{remainder S2})). For the shared components, the mean and
variance will be distorted by quantities determined by the covariances
between $({\cal A}^i+{\cal D}^i)$ and $Z^{i[1]}$, $Z^{i[2]}$.
Let us write
\begin{equation}
\label{covariance A+D and parental trait}
C(i, i[1]):=\mathtt{Cov}(({\cal A}^i+{\cal D}^i), Z^{i[1]}),
\end{equation}
with a corresponding definition for $C(i, i[2])$.
Then, once again using $1$ and $2$ in place of $i[1]$ and $i[2]$ in our
expressions for identities,
\begin{align}
\label{expression for C([1])}
C(i, i[1])=&\ \frac{\sigma_A^2}{2}\left(1+F_{11}+2F_{12}\right)
+\frac{\sigma_{ADI}}{2}\left(F_{11}+F_{12}+2F_{112}\right)\nonumber \\
& +\sigma_D^2(F_{12}-F_{112})
+(\sigma_{DI}^2+\iota^*)F_{112}-\iota^2F_{11}F_{12},
\end{align}
with $C(i,i[2])$ given by the corresponding expression with the roles
of the subscripts $1$ and $2$ interchanged.
(A derivation of this expression is provided in Appendix~\ref{QG derivations}.)
With this notation,
\begin{multline}
\label{mean A+D given parental traits}
\IE[({\cal A}^i+{\cal D}^i)|Z^{i[1]},Z^{i[2]}]=\IE[({\cal A}^i+{\cal D}^i)]
+\frac{1}
{\mathtt{Var}(Z^{i[1]}) \mathtt{Var}(Z^{i[2]})
-\mathtt{Cov}(Z^{i[1]},Z^{i[2]})^2}\\
\times\Bigg\{
\left(C(i, i[1])\mathtt{Var}(Z^{i[2]})
-C(i, i[2])\mathtt{Cov}(Z^{i[1]},Z^{i[2]})\right)(Z^{i[1]}-\IE[Z^{i[1]}])
\\
+\left(C(i, i[2])\mathtt{Var}(Z^{i[1]})
-C(i, i[1])\mathtt{Cov}(Z^{i[1]},Z^{i[2]})\right)(Z^{i[2]}-\IE[Z^{i[2]}])
\Bigg\},
\end{multline}
and
\begin{multline}
\label{variance A+D given parental traits}
\mathtt{Var}(({\cal A}^i+{\cal D}^i)|Z^{i[1]}, Z^{i[2]})
=\mathtt{Var}({\cal A}^i+{\cal D}^i)
\\-
\frac{\mathtt{Var}(Z^{i[1]})
C(i, i[2])^2+\mathtt{Var}(Z^{i[2]})
C(i,i[1])^2-2\mathtt{Cov}(Z^{i[1]},Z^{i[2]})C(i,i[1])C(i,i[2])}
{\mathtt{Var}(Z^{i[1]})\mathtt{Var}(Z^{i[2]})-
\mathtt{Cov}(Z^{i[1]},Z^{i[2]})^2}.
\end{multline}
(We have implicitly assumed that $i[1]\neq i[2]$; in the case
$i[1]=i[2]$ the expression is simpler as we are then conditioning a bivariate
normal on one of its marginals.)

\begin{remark}
In the purely
additive case, things simplify greatly. From the expressions above,
before conditioning, the mean of
${\cal A}^i+{\cal D}^i$ is zero (since $\iota=0$), and the variance is
$$\frac{\sigma_A^2}{2}\left(1+\frac{(F_{11}+F_{22})}{2}+2F_{12}\right).$$
Moreover,
$$\mathtt{Var}(Z^{i[1]})=\sigma_A^2(1+F_{11}),\quad
\mathtt{Var}(Z^{i[2]})=\sigma_A^2(1+F_{22}),\quad
\mathtt{Cov}(Z^{i[1]}, Z^{i[2]})=2\sigma_A^2 F_{12},$$
and
$$C(i, i[1])=\frac{1}{2}\sigma_A^2\left(1+F_{11}+2F_{12}\right),
\quad
C(i, i[2])=\frac{1}{2}\sigma_A^2\left(1+F_{22}+2F_{12}\right).$$
Substituting
into~(\ref{mean A+D given parental traits})
and~(\ref{variance A+D given parental traits}),
and observing that
\begin{align*}
& (1+F_{11})(1+F_{22}+2F_{12})^2+(1+F_{22})(1+F_{11}+2F_{12})^2
-4F_{12}(1+F_{11}+2F_{12})(1+F_{22}+2F_{12})\\
& =2\left((1+F_{11})(1+F_{22})-4F_{12}^2\right)\left(1+\frac{F_{11}+F_{22}}{2}
+2F_{12}\right),
\end{align*}
we find that
conditional on the trait values of the parents, the mean
and variance of ${\cal A}^i+{\cal D}^i$
reduce to $(Z^{i[1]}+Z^{i[2]})/2$ and zero, respectively, and
we recover the classical infinitesimal model.
\end{remark}

Although in the presence of dominance
the expressions~(\ref{mean A+D given parental traits})
and~(\ref{variance A+D given parental traits})
are rather complicated, we emphasize that they
are derived from knowledge of just the ancestral population and the pedigree,
and are expressed in terms of familiar quantities from classical
quantitative genetics.

\section{Numerical examples}\label{s:numerics}

In this section, we present numerical examples to illustrate the
accuracy of the predictions of the infinitesimal model, again disregarding the environmental component of the trait.

We first generated a pedigree for a population of constant size of $N=30$
diploid individuals over $50$ discrete generations. Mating is
random, but with no selfing.
In order to facilitate comparison of different scenarios, the same
pedigree was used for all subsequent simulations. In this way, the
identity coefficients are held constant. As expected, the mean probability
of identity between pairs of genes sampled from different individuals
in generation $t$ is close to $1-(1-1/2N)^t$.

We define a trait, $Z$, which depends on $M=1000$ biallelic loci.
There is no epistasis, so that the trait value is a
sum across loci. In the examples here, we assume
complete dominance, so that the effects of the three genotypes at each
locus are either $-\alpha:-\alpha:+\alpha$ or $-\alpha:+\alpha:+\alpha$.
In order to ensure that the inbreeding depression $\iota$ is bounded, we
need to have some `balance' and so we choose the effects at each locus
according to an independent Bernoulli random variable with parameter $H$; that
is, the probability that the effects across the three genotypes at locus $l$ is
$-\alpha:-\alpha:+\alpha$ is $1-H$, independently for each locus.
The effect size $\alpha$ is taken to be $1/\sqrt{M}$
for all loci and $H=\frac{1}{2}+\frac{2}{\sqrt{M}}$. With these choices
the additive and dominance variances will be $\mathcal{O}(1)$.

In the ancestral population, the allele frequencies were
generated to mimic neutral allele frequencies with very low mutation rates, but
conditioned to segregate at each locus. Thus, allele frequencies at every locus were sampled independently and according to a distribution with density proportional to $(p(1-p))^{1-\epsilon}$,
with $\epsilon=0.001$, but with those in $[0, 1/60]$ and $[1-1/60, 1]$
discarded (and the distribution renormalised). Then for each population replicate, these frequencies were used to endow each individual in the base population with an allelic type at every locus.

Variance components are defined with respect to
this reference set of allele frequencies. For the population generated for the
examples presented here, these values were $\sigma_A^2 =0.269$,
$\sigma_D^2 =0.073$, and the inbreeding depression
$\iota =-0.531$. The additive and dominance components are
uncorrelated in the base population
($\mathtt{Cov}(A, D)=0$).
In the numerical experiments that follow, each replicate population is
started at time zero from a different collection of genotypes, sampled
from this base distribution.

\begin{figure}
\centerline{\includegraphics[width=5in]{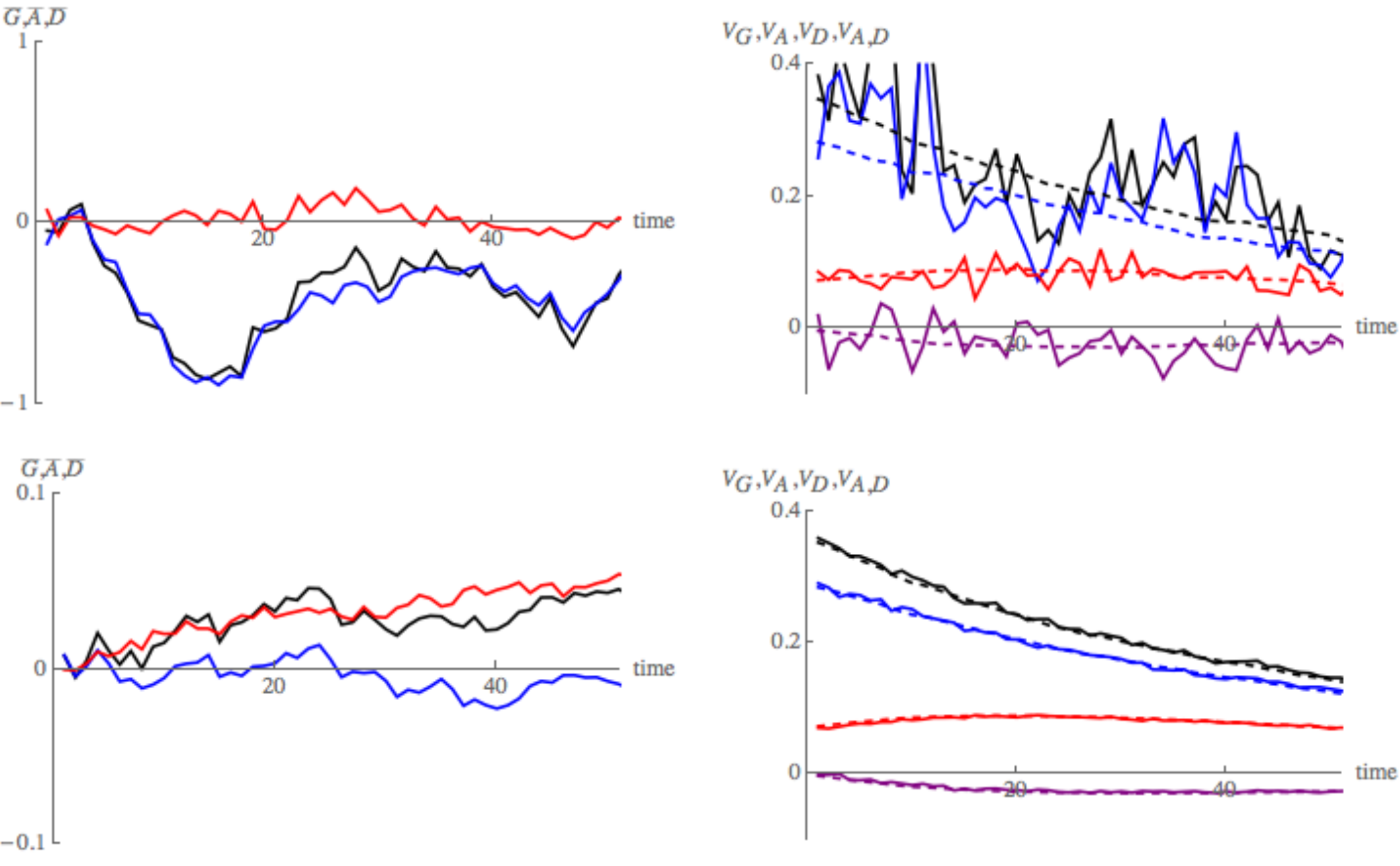}}
\caption{
Changes of the mean and variance of the additive part of the trait, the dominance part, and their sum over $50$ generations of neutral evolution. The top row
shows a single replicate, whilst the bottom row shows the average over
$300$ replicates using the same sequence of individuals spanning the 50 generations. The left
column shows the means ($\bar{G} = \bar{A} + \bar{D}$, $\bar{A}$, $\bar{D}$; black, blue, red),
whilst the right column shows the
variance components ($V_G=\mathtt{Var}(G)$, $V_A=\mathtt{Var}(A)$, $V_D =\mathtt{Var}(D)$, $V_{A,D}=\mathtt{Cov}(A,D)$; black, blue, red, purple).
On the right, solid lines show the
total variances and covariance, whilst the dashed lines show the genic
component. These differ
through the contribution of linkage disequilibrium, which generates
substantial variation. The
genic component changes smoothly, as expected with a large number
($M = 1000$) of loci. With $M=1000$ loci, we expect the infinitesimal model to be accurate for about $\sqrt{M}\sim 30$ generations. Simulations are made on a single pedigree with $30$ individuals; variance components are measured relative
to the ancestral population. The predicted values for these means and variances under the infinitesimal model are given in Eq.~\eqref{prediction AD}-\eqref{variance dominance deviation} (note that the identity coefficients $F_{ii}$ increase through time due to genetic drift.)}
\label{change in mean and variance}
\end{figure}

We first simulated a neutral model.
Figure~\ref{change in mean and variance} illustrates how the different
components of the trait values change over fifty generations of neutral
evolution. Recall that we always use the same realisation of the pedigree.
For each replicate, we take an independent sample of allelic types
at time zero. For each individual in the pedigree we evaluate the additive
and dominance components $A$ and $D$ and then in each generation we
calculate the mean and
variance of these quantities across the $30$ individuals in the population.
This is only intended to give some feeling for
the ways in which the components fluctuate through time.
Of course the infinitesimal model is only providing a prediction for the \emph{distribution} of
trait values within families; a single realisation will see substantial contributions to trait
values from linkage disequilibrium (\emph{c.f.} the toy example in Section~\ref{mendelian inheritance}
and Theorem~\ref{shared parent theorem}).
In the following
figures we compare these quantities to the detailed predictions of the
infinitesimal model.
The top row in Figure~\ref{change in mean and variance}
is a single replicate, while the bottom is the
average over three hundred replicates. On the left we have the mean of the
additive and dominance components and their sum; on
the right we have plotted the variance components.
For a single replicate,
there is indeed a substantial contribution from linkage disequilibrium.
When we
plot just the genic components (that is the sum over variances at each locus,
ignoring the contribution from linkage disequilibrium),
as expected, the picture is much smoother
and we see that the predictions of the infinitesimal model are close to the values obtained by averaging over $300$ replicates. Since
linkage disequilibrium will dissipate rapidly, halving in each generation, it is the genic
component that determines the long term evolution.

All components are measured relative to the base population.
In practice, in natural populations, one does not have access to
the ancestral population and so one measures
components relative to the current population.
This amounts to a change of reference (Hill et al.~2006).
\nocite{hill/barton/turelli:2006}
We do not do this in our setting as it would result in different
variance components for every replicate.

\begin{figure}
\centerline{
\includegraphics[width=3in]{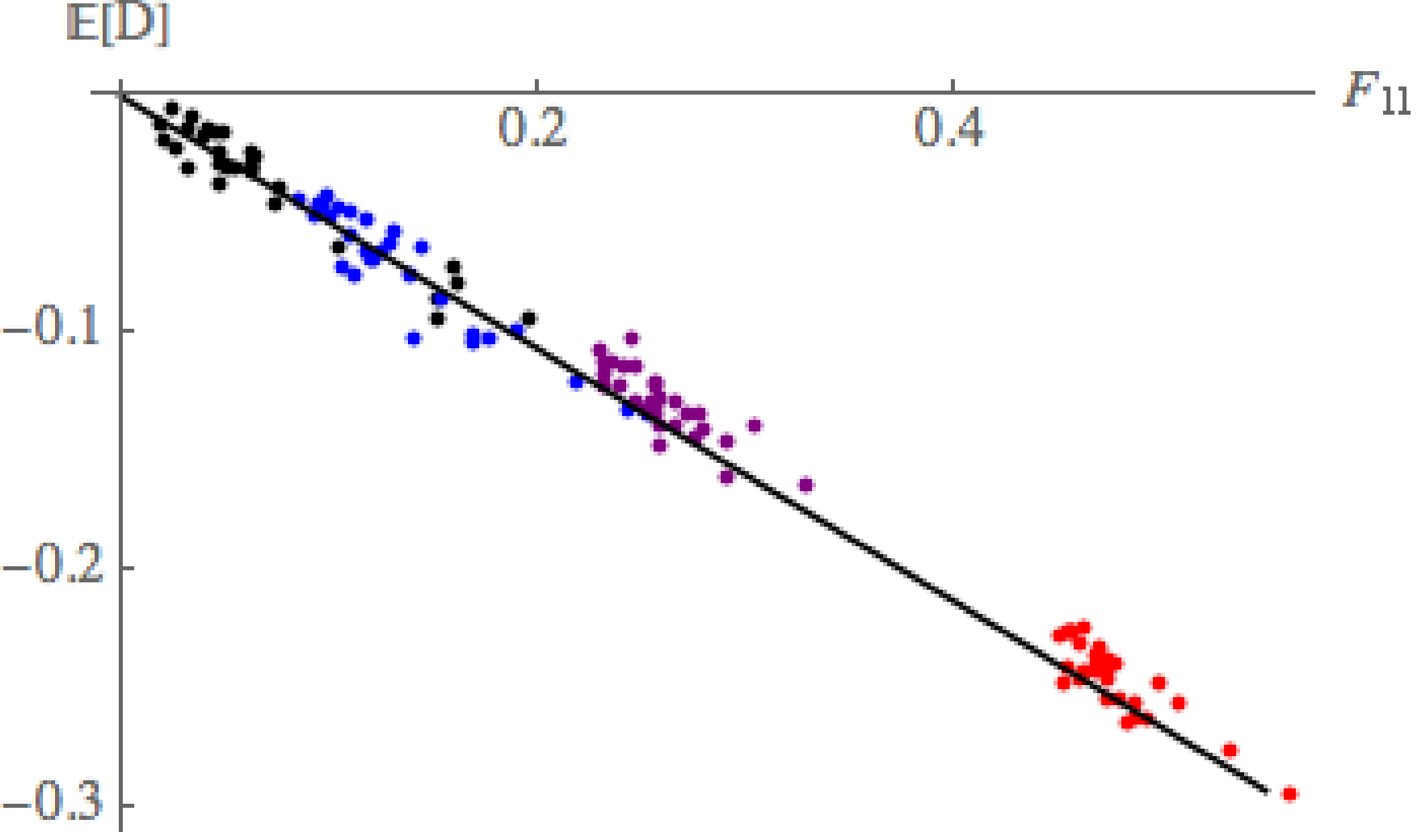}}
\caption{The relation between the dominance deviation and
the probability of identity of the two genes within an individual.
There is one point for the average over $1000$ replicates for
each of the thirty individuals in generations
$5$, $10$, $20$, $40$ (black, blue, purple, red). (Recall that the
pedigree is fixed, so identities are the same for each replicate.)
The mean of $D$ decreases as $\iota F_{ii} = -0.53 F_{ii}$
(black line), in accordance with Eq.~\eqref{prediction AD}.
}
\label{D vs F}
\end{figure}

In Figure~\ref{D vs F}
we explore the relationship between the dominance deviation and inbreeding.
Since we use the same pedigree for all our experiments, each individual
is characterised by a single $F_{ii}$ (the probability of identity of
the two genes at a given locus). For each of 1000 replicates (that is
independent samples of allelic types for the
individuals in generation zero), we calculated the dominance deviation for
each individual in the pedigree.
The plot in Figure~\ref{D vs F}
shows the dominance deviation
averaged over those $1000$ replicates for each individual in
the pedigree. Thus there are $30$ points in each generation, one for
each individual in the population.
As expected, the mean of the
dominance component decreases in proportion to $F_{ii}$,
$\IE[D]=-0.53 F_{ii}$ (recall that $\iota=-0.53$ for our base
population).

\begin{figure}
\centerline{\includegraphics[width=3in]{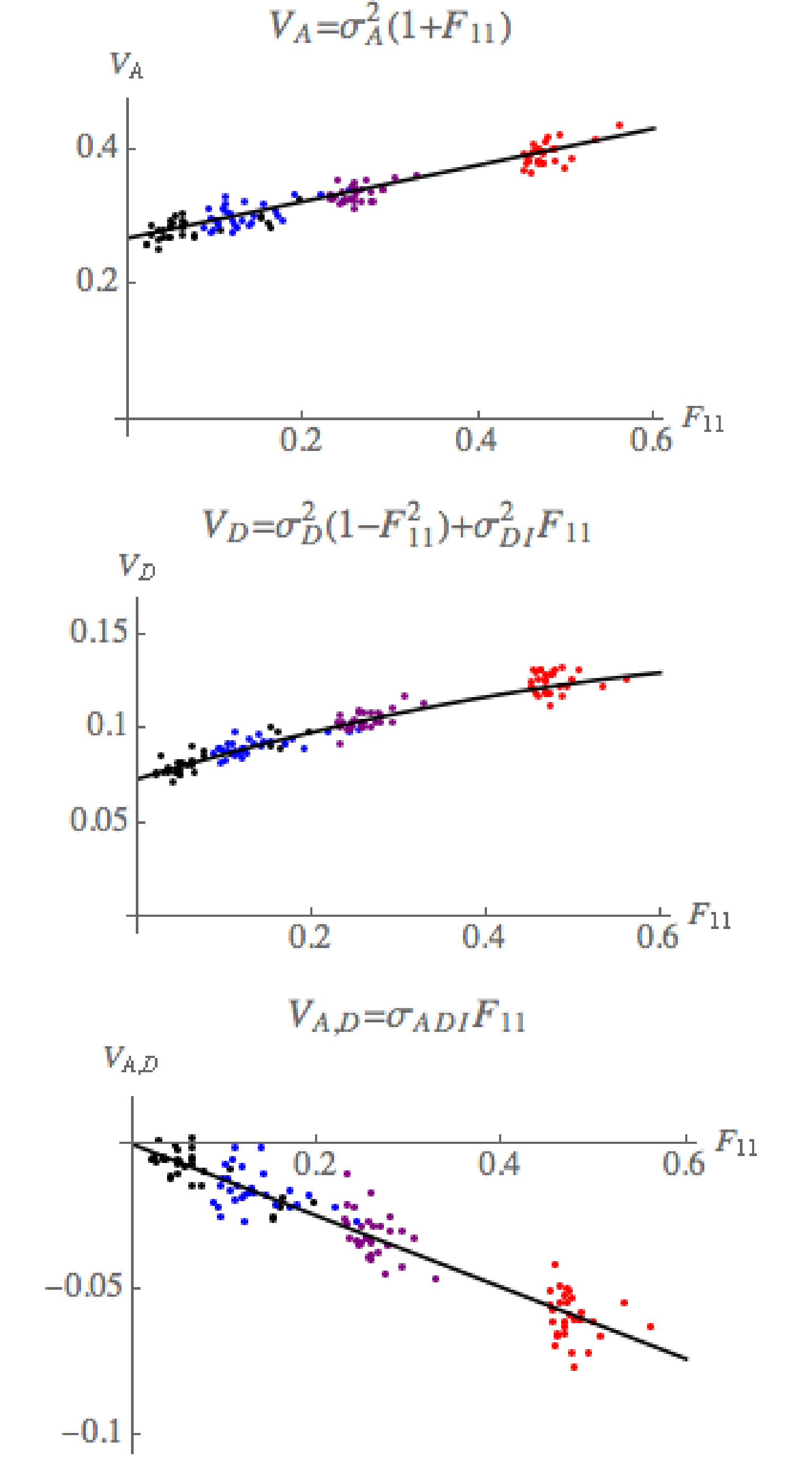}}
\caption{The variance and covariance of $A$ and $D$ versus identity
$F_{ii}$ for
individuals in the pedigree. As in Figure~\ref{D vs F},
there are $30$ points in each generation,
one corresponding to each of the thirty individuals in the population.
Generations $5$, $10$, $20$, $40$ (black, blue, purple, red). Here again we use the shorter notation $V_A=\mathtt{Var}(A)$, $V_D =\mathtt{Var}(D)$, $V_{A,D}=\mathtt{Cov}(A,D)$ and the theoretical predictions were derived in Eq.~\eqref{prediction Var AD} and \eqref{variance dominance deviation}.}
\label{covariance A D}
\end{figure}

Figure~\ref{covariance A D} shows how the (co)variance of $A$ and
$D$ depends on identity $F_{ii}$ for pairs of individuals in
the pedigree. As in Figure~\ref{D vs F},
for each individual in the pedigree, $A$ and $D$ are calculated for
each of the 1000 replicates; Figure~\ref{covariance A D} shows the
variances and covariances of the resulting values for each of the
thirty individuals in generations $5$, $10$, $20$ and $40$ and these
are compared to the theoretical predictions.
Note that since in the biallelic case $\sigma_D^2=\iota^*$, the
expression~(\ref{variance dominance deviation})
for the variance of the dominance component reduces to
$$\sigma_D^2\big(1-F_{ii}^2\big)+\sigma_{DI}^2F_{ii}.$$

\begin{figure}
\centerline{\includegraphics[width=6.5in]{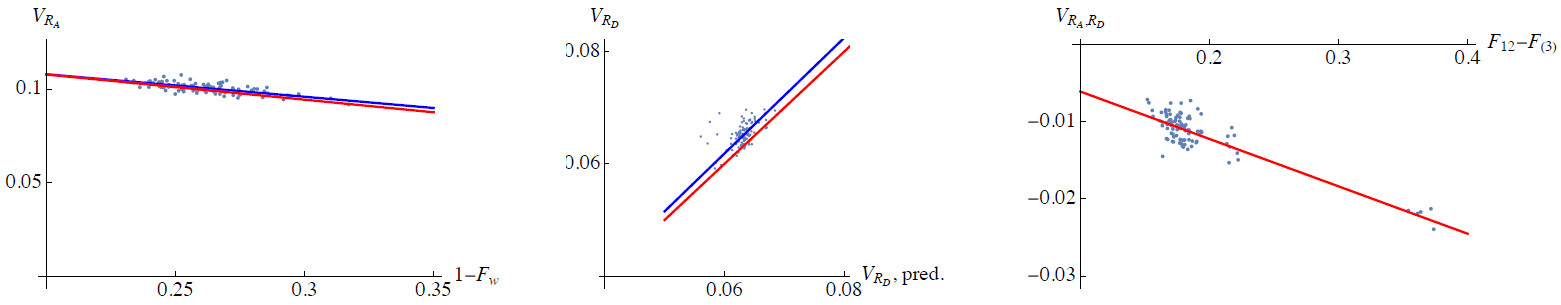}}
\caption{The variance and covariance within families between the
residual additive and dominance deviations $R_A$ and $R_D$ ($V_{R_A}=\mathtt{Var}(R_A)$, $V_{R_D} =\mathtt{Var}(R_D)$, $V_{R_A,R_D}=\mathtt{Cov}(R_A,R_D)$).
One hundred pairs of parents were chosen at random from the ancestral population
and from each one
thousand offspring were generated. The within family variances obtained
in this way were averaged over ten replicates (with the same pedigree
and parents). Each of the 100 points in each plot corresponds to one
pair of parents. The five outliers are families produced by selfing.
The blue lines show a least-squares regression; the red lines are the
theoretical predictions (see Eq.~\eqref{var residual}). The two lines exactly coincide in the plot on the right.
}
\label{R and S}
\end{figure}

Next we consider the variances of the residuals $R_A$ and $R_D$ within families.
One hundred pairs of parents were chosen at random from the population, and
from each 1000 offspring were generated. This was repeated for ten replicates
made with the same pedigree and the same set of parents; within family
variances were then averaged over replicates.
In Figure~\ref{R and S}, in each plot there are 100 points, one for each
pair of parents.
The two lines correspond to least square regression (blue) and
theoretical predictions (red) which can be read off from
Eq.~\eqref{var residual}. For readability, in the figure we use the notation $V_{R_A}$, $V_{R_D}$ and $V_{R_A,R_D}$ to denote the variance of $R_A$, the variance of $R_D$ and the covariance between $R_A$ and $R_D$ respectively. Using Eq.~(\ref{var residual}) and the explanation below, together with the fact that $\sigma_D^2=\iota^*$ in our bi-allelic case, we have $V_{R_A}=\sigma_A^2(1-F_{W})/2$, where
$F_W=(F_{i[1]i[1]}+F_{i[2]i[2]})/2$ is the within-individual identity
averaged over parents $1$ and $2$;
$$V_{R_D}=\frac{\sigma_{DI}^2}{4}\big(3F_{12}-F_{1122}-F_{112}-F_{122}\big)+
\frac{\sigma_D^2}{4}\bigg(3-F_{11}-F_{22}-F_{1122}-\tilde{F}_{1122}
-\frac{3}{2}\tilde{F}_{1212}\bigg);$$
and
$$V_{R_A,R_D}=\frac{\sigma_{ADI}}{2}\big(F_{12}-F_{(3)}\big),$$
where $F_{(3)}$ is defined as:
$$F_{(3)}=\frac{F_{112}+F_{122}}{2}.$$

\begin{figure}
\centerline{
\includegraphics[width=3in]{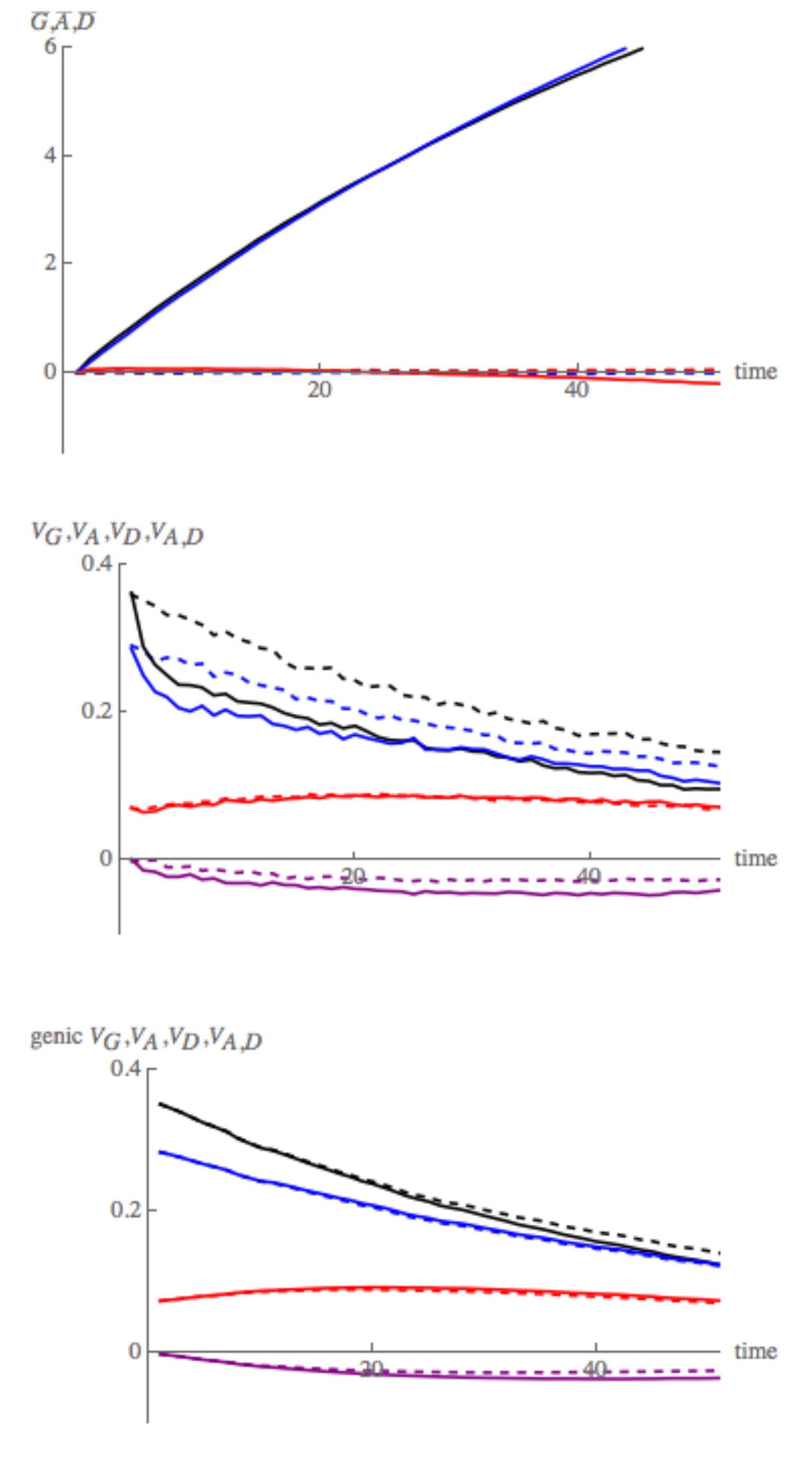}}
\caption{Comparison between a neutral population (dashed lines) and one subject to
truncation selection (solid lines). Top row: change in means relative to the
initial value ($\overline{G}=\overline{A+D}$, $\overline{A}$,$\overline{D}$;
black, blue, red); middle: variances, including linkage disequilibria
($V_G=\mathtt{Var}(A+D)$,
$V_A=\mathtt{Var}(A)$,
$V_D=\mathtt{Var}(D)$,
$V_{A,D}=\mathtt{Cov}(A,D)$; black, blue, red, purple). The bottom row is the changes to genic variances with time against
predictions of the infinitesimal model.
The values are averages over 300 replicates for the neutral case, 1000 for the selected case, made with the same pedigree. There are $M=1000$ loci, and thus we expect the infinitesimal model to be accurate for about $\sqrt{M}\sim 30$ generations. Selection is made within families; for each offspring, two individuals are generated from the corresponding parents,
and the one with the larger trait value retained.}
\label{selected vs neutral}
\end{figure}

The full force of our theoretical results is that even if we condition on
the trait values of parents, the within family distribution of their
offspring will consist of two normally distributed components and, in
particular, the variance components will be independent of the trait
values of the parents.
We test this by imposing strong truncation selection on the population.
We retain the same pedigree relatedness, but working down the pedigree,
each individual's genotype is determined by generating two possible
offspring from its parents and retaining the one with the larger trait
value.
In Figure~\ref{selected vs neutral} we compare the results with simulations
of the neutral population. Dashed lines are for the neutral simulations, solid
ones for the simulation with selection.
For the population under selection, we see an
immediate drop in the total genetic variation, caused by the strong
selection; there is significant negative linkage disequilibrium between individual
loci, as predicted by Bulmer~(1971).
\nocite{bulmer:1971}
The blue is the additive component. We see that about
one third of the variance is dominance variance.
The bottom row shows that the genic components are hardly affected by selection,
as predicted by the infinitesimal model. With or without selection, the variance
components change as a result of inbreeding.

\begin{figure}
\centerline{
\includegraphics[width=4.5in]{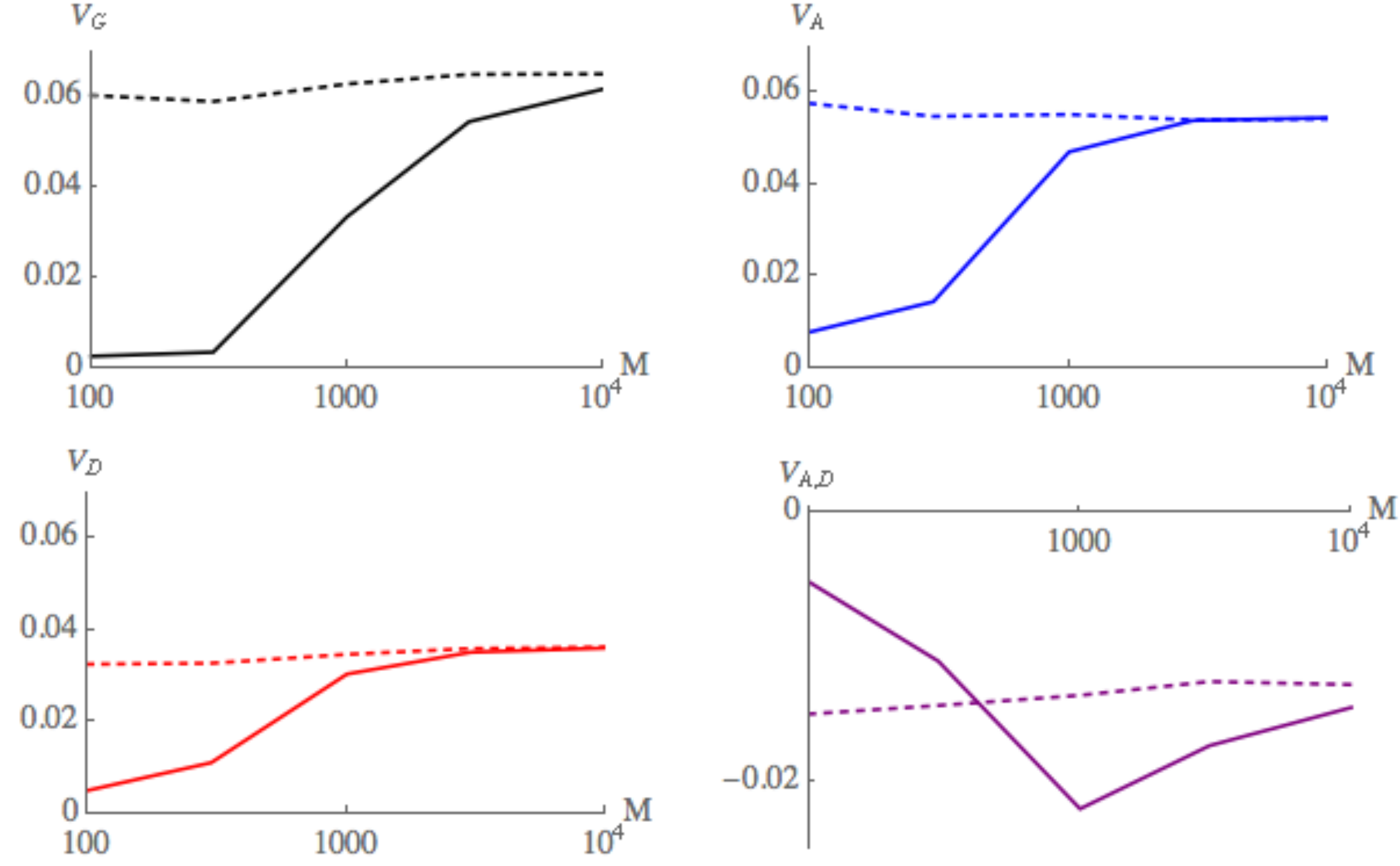}}
\caption{Convergence of the variance components at 50 generations,
as the number of loci increases from
$M=100$ to $M=10^4$ (same notation as in Figure~\ref{selected vs neutral}). Simulations with 50\% truncation selection are compared
with neutral simulations (solid, dashed lines).
The replicate simulations were generated as
in Figure~\ref{selected vs neutral} (see main text). Regressions of the log absolute difference between selected and neutral variance components against $\ln(M)$ have slopes $-0.62$, $-0.72$, $-0.70$, $-0.66$ for $V_G$, $V_A$, $V_D$, $V_{A,D}$ respectively (see Supplementary Material for details). Thus, convergence is somewhat faster than $\sqrt{M}$.}
\label{rate of convergence figure}
\end{figure}

Finally, Figure~\ref{rate of convergence figure}
compares the variance components at 50 generations for neutral simulations
with those with truncation selection
as the number of loci increases from $M=100$ to
$M=10^4$. Replicate simulations were generated as
in Figure~\ref{selected vs neutral}. Under the infinitesimal model,
these components should take the same values with and without selection.
This is reflected in the simulations, with the covariance between the
additive and dominance effects being the slowest to settle down to
the infinitesimal limit.

\section{The infinitesimal model with dominance as a limit of Mendelian
inheritance}
\label{mendelian inheritance}

In this section, we turn to the justification of our model as a limit of a model of Mendelian inheritance as the number $M$ of loci tends to infinity. Although we shall focus on the distribution of the genetic components of the trait values in the pedigree, in this section we consider the general situation where the \emph{observed} trait of an individual, $\widetilde{Z}^i$, is the sum of a genetic component $Z^i$ and an environmental component $E^i$. Our mathematical assumptions on $E^i$ are detailed in the paragraph `Main results' below.

Our work is an extension of that of Abney et al.~(2000),
\nocite{abney/mcpeek/ober:2000}
which in turn builds on Lange~(1978).
\nocite{lange:1978}
The distinctions here are that we explicitly
model the component of the trait value that is shared by all
individuals in a family
separately from the part that segregates within that family; we identify
the effect on each of these components of conditioning on knowing the trait
values of the parents of the family; and we estimate the error that we are
making in taking the normal approximation, thus providing information on when
the infinitesimal approximation breaks down.

The fact that the genetic component of trait values within families is
normally distributed is a consequence of the Central Limit Theorem.
That this remains valid even when we condition on the trait values of
the parents stems from the fact that knowing the trait value of an
individual actually provides very little information about the
allelic state at any particular locus. This in turn is because, typically,
there
are a large number of different genotypes that are consistent with a
given phenotype.
In Barton et al.~(2017), this was illustrated
through a simple example which can be found
on p.402 of Fisher~(1918), which concerned an additive trait in
a haploid population.
\nocite{fisher:1918}
\nocite{barton/etheridge/veber:2017}
Here we adapt that example to the model for which we performed our
numerical experiments.

Suppose then that we have $M$ biallelic loci. We denote the alleles at
locus $l$ by $a_l$ and $A_l$. The contributions to the trait of the
three genotypes $a_la_l$, $a_lA_l$ and $A_lA_l$ are
$-\alpha$, $-\alpha$, $\alpha$ respectively with probability
$\frac{1}{2}-\frac{2}{\sqrt{M}}$ and they are
$-\alpha$, $\alpha$, $\alpha$ with probability
$\frac{1}{2}+\frac{2}{\sqrt{M}}$.
The effect size $\alpha=1/\sqrt{M}$.
For simplicity, in contrast to our numerical experiments, we suppose that
the probabilities of genotypes $a_la_l$, $a_lA_l$, $A_lA_l$ are
$1/4$, $1/2$, $1/4$ respectively.

Now suppose that
we observe the trait value to be $k/\sqrt{M}$.
What is the conditional probability that the allelic
types at locus $l$, which we denote $\chi^1_l\chi^2_l$ are $A_lA_l$?
For definiteness, we take $M$ and $k$ both to
be even and $l=1$.

First consider the probability that the contribution to the trait
value from locus $1$ is $+1/\sqrt{M}$. Let us write $p_{+}$ for the
(unconditional) probability that the contribution from locus $1$
is $1/\sqrt{M}$, that is
$$p_{+}=\frac{1}{4}+\frac{1}{2}\bigg(\frac{1}{2}+\frac{2}{\sqrt{M}}\bigg)
=\frac{1}{2}\bigg(1+\frac{1}{\sqrt{M}}\bigg),$$
and $p_{-}=1-p_{+}$.
Let us write $\Psi_l/\sqrt{M}$ for the contribution to the trait
from locus $l$. We have
\begin{eqnarray*}
\frac{ \IP\left[\left. \sum_{l=1}^M\Psi_l=k\right|  \Psi_1=1\right]
}{\IP\left[\sum_{l=1}^M\Psi_l=k\right]}
&=& \frac{\IP\left[\sum_{l=2}^M\Psi_l=k-1\right]
}{\IP\left[\sum_{l=1}^M\Psi_l=k\right]}
\\
&=&\frac{p_{+}^{(M+k-2)/2}p_{-}^{(M-k)/2}}{p_+^{(M+k)/2}p_{-}^{(M-k)/2}}\frac{\binom{M-1}{(M+k-2)/2}}
{\binom{M}{(M+k)/2}}
\\&=& \left(1+\frac{k}{M}\right)\frac{1}{2p_+}
\\&=& \left(1+\frac{k}{M}\right)\frac{1}{(1+1/\sqrt{M})}.
\end{eqnarray*}
An application of Bayes' rule then gives
\begin{eqnarray*}
\IP\left[\chi^1_1=A_1, \chi^2_{1}=A_1\left|
\sum_{l=1}^M\frac{\Psi_l}{\sqrt{M}}=\frac{k}{\sqrt{M}}\right.\right]
&=&\frac{ \IP\left[\left. \sum_{l=1}^M\Psi_l=k\right|  \Psi_1=1\right]
}{\IP\left[\sum_{l=1}^M\Psi_l=k\right]}\IP\left[\chi^1_1=A_1, \chi^2_1=A_1\right]
\\
&=&
\left(1+\frac{k}{M}\right)\frac{1}{(1+1/\sqrt{M})}
\IP\left[\chi^1_1=A_1, \chi^2_1=A_1\right].
\end{eqnarray*}
Similarly,
\begin{equation*}
\IP\left[\chi^1_1=a_1, \chi^2_{1}=a_1\left|
\sum_{l=1}^M\frac{\Psi_l}{\sqrt{M}}=\frac{k}{\sqrt{M}}\right.\right]
=
\left(1-\frac{k}{M}\right)\frac{1}{(1-1/\sqrt{M})}
\IP\left[\chi^1_1=a_1, \chi^2_1=a_1\right],
\end{equation*}
and
\begin{multline*}
\IP\left[\chi^1_1=a_1, \chi^2_{1}=A_1\left|
\sum_{l=1}^M\frac{\Psi_l}{\sqrt{M}}=\frac{k}{\sqrt{M}}\right.\right]
\\=
\left\{\left(1+\frac{k}{M}\right)\frac{(1/2+2/\sqrt{M})}{(1+1/\sqrt{M})}
+\left(1-\frac{k}{M}\right)\frac{(1/2-2/\sqrt{M})}{(1-1/\sqrt{M})}\right\}
\IP\left[\chi^1_1=a_1, \chi^2_1=A_1\right].
\end{multline*}

In view of the Central Limit Theorem,
we would expect a `typical' value of $k$ to be on the
order of $\sqrt{M}$; conditioning has only perturbed the
probability that $\Psi_1=1$ by a factor $k/M+\mathcal{O}(1/\sqrt{M})$,
which we expect
to be of order $1/\sqrt{M}$.
In the purely additive case, which corresponds to taking
$p_+=p_{-}=1/2$, at the extremes of what is possible ($k=\pm M$),
we recover complete
information about the values of $\chi^1_1$, $\chi_2^1$; however,
with dominance that is no longer true.

Notice that
for the difference between the trait value of an individual
and the mean over the population to be order one requires order
$\sqrt{M}$ of the loci to be `non-random', but observing the trait does
not tell us which of the possible $M$ loci these are.
Similarly, performing the entirely analogous calculation for pairs of loci,
and observing that
$$\frac{\binom{M-2}{(M+k-4)/2}}{\binom{M}{(M+k)/2}}
=\frac{1}{4}\left(1+\frac{k}{M}\right)\left(1+\frac{k-1}{M-1}\right),$$
we deduce that,
\begin{align}
& \IP\left[\chi^1_1=A_1, \chi^2_1=A_1; \chi^1_2=A_2, \chi^2_2=A_2\bigg|
\sum_{l=1}^M\frac{\Psi_l}{\sqrt{M}}=\frac{k}{\sqrt{M}}\right] \nonumber\\
& =\left(1+\frac{k}{M}\right)\left(1+\frac{k-1}{M-1}\right)
\frac{1}{(1+1/\sqrt{M})^2}
\IP[\chi^1_1=A_1, \chi^2_1=A_1; \chi^1_2=A_2, \chi^2_2=A_2] \nonumber
\\
& = \IP\left[\chi^1_1=A_1, \chi^2_1=A_1\bigg|
\sum_{l=1}^M\frac{\Psi_l}{\sqrt{M}}=\frac{k}{\sqrt{M}}\right]
\times \IP\left[\chi^1_2=A_2, \chi^2_2=A_2\bigg|
\sum_{l=1}^M\frac{\Psi_l}{\sqrt{M}}=\frac{k}{\sqrt{M}}\right] \nonumber
\\ & \qquad+
\IP[\chi^1_1=A_1, \chi^2_1=A_1; \chi^1_2=A_2, \chi^2_2=A_2]
\left(1+\frac{k}{M}\right)
\frac{1}{(1+1/\sqrt{M})^2}
\left(\frac{k-1}{M-1}-\frac{k}{M}\right).
\label{toy model calculation}
\end{align}
For a `typical' trait value the last term
in~(\ref{toy model calculation})
is order $1/M$.
When we sum over loci, this is enough to give a nontrivial contribution to the trait
value coming from the linkage disequilibrium. However,
although observing the trait of a typical
individual tells us
something about linkage disequilibria, it does not tell us enough to identify
which of the order $M^2$ pairs of loci are in linkage disequilibrium.

Essentially the same argument will apply to the much more
general models that we develop below.
In particular, for the infinitesimal model to be a
good approximation, the observed parental trait
values must not contain too much information about the allelic
effect at any given locus, which requires that
the parental traits must not be too extreme (corresponding to $k$ in
our toy model being ${\mathcal O}(\sqrt{M})$).

In the additive case, it was enough to control the additional
information that we gained about any particular locus from knowledge
of the trait value in the parents. This is
because, in that case, the variance of the shared contribution
within a family
is zero and independent Mendelian inheritance at each locus ensures that
linkage disequilibria do not distort the variance of the
residual component that
segregates within families. With dominance, we must estimate the
(non-trivial) variance
of the shared component, and for this
we shall see that we need to control the build up of linkage disequilibrium
between pairs of loci.
It will turn out that since all pairs of loci are in linkage equilibrium
in the ancestral population, any \emph{given} pair of loci will be approximately
in linkage equilibrium for the order $\sqrt{M}$ generations for which the
infinitesimal approximation is valid.

This does not mean that the linkage disequilibria do not affect the
trait values, but because of the very many different
combinations of alleles in an individual that are consistent with a given
trait, observing the trait tells us very little about the allelic
state at a particular locus. The allele at that locus can only ever
contribute $\mathcal{O}(1/\sqrt{M})$ to the overall trait value.

As the population evolves, and we are able to observe more and more
traits on the pedigree, we gain more and more information about the
allele that an individual carries at a particular locus.
In Barton et al.~(2017),
\nocite{barton/etheridge/veber:2017}
we considered an additive
trait in a population of haploid individuals. In that setting we showed
that for a given individual, one does not gain any more information
about the state at a given locus from looking at the trait values on
the whole of the rest of the pedigree than one does from observing just
the parents of that individual. In our model for diploid individuals
with dominance, this is no longer the case; observing the trait values of any relatives, no matter how distant,
provides some additional information about the allelic state at a locus. The difference arises from the fact that the contribution that a gene makes to the trait value of an individual depends not only on its own allelic state, but also on that of the other copy of the gene at that locus. As a result, we gain information about the allelic state in a focal individual by observing trait values in any other individuals in the pedigree with which it may be identical by descent at that locus.
However, the amount of information gleaned about the allelic state
of an individual from observing new individuals in the pedigree
will decrease in proportion to
the probability of identity, and so for distant relatives in the pedigree is very small;
provided our pedigree is not too inbred,
and trait values are not too extreme, we
can still expect the infinitesimal model to be a good approximation
for order $\sqrt{M}$ generations.

\subsection*{Environmental noise}

Our derivations will depend on two approaches to proving asymptotic
normality. The first, which we apply to the portion $R_A^i+R_D^i$ of the
trait values, uses a generalised Central Limit theorem (which allows
for the summands to have different distributions),
which provides control over the rate of convergence
as $M\rightarrow\infty$.
(It is this control that tells us for how many generations
we can expect the infinitesimal model to be valid.)
However, the Central Limit Theorem guarantees only the rate of
convergence of the cumulative distribution function of the normalised
sum of effects at different loci. Our proofs exploit convergence to
the corresponding probability density function, which may
not even be defined.
To get around this, we
can follow the approach of Barton et al.~(2017)
\nocite{barton/etheridge/veber:2017}
and
make the (realistic) assumption that rather than observing the genetic
component of a trait directly, the observed trait has an environmental
component with a smooth density. This results in the trait distribution having a
smooth density which is enough to guarantee the faster rate of
convergence. In addition to the benefit in terms of regularity of the trait distribution, an environmental noise with a smooth distribution also reinforces the property that observing the trait value gives us very little information on the allelic state at a given locus: a continuum of combinations of genetic and environmental components may have led to the observed trait, in which each given locus contributes an infinitesimal amount. (To ensure sufficient regularity of the trait density, we could instead make the assumption that the distribution of allelic effects at every locus has a smooth
probability density function.)
The approach to proving asymptotic normality of the shared component uses
an extension of Stein's method of exchangeable pairs. Once
again in the presence of environmental noise (to ensure that the trait distribution
has a smooth density)
we recover convergence with an error of order
$1/\sqrt{M}$.

If the environmental component is taken to be normally
distributed, then exactly as in Barton et al.~(2017),
\nocite{barton/etheridge/veber:2017}
we can adapt our application of
Theorem~\ref{conditioning multivariate normals} in Appendix~\ref{conditioning normals}
to write down the conditional distribution of the genetic
components given \emph{observed} traits; \emph{i.e.}, traits distorted by a small
environmental noise, \emph{c.f.}~Remark~\ref{remark on breeder's equation}.

\subsection*{Assumptions and notation}

Recall that we assume that in generation zero,
the individuals that found the pedigree are unrelated and
sampled from an ancestral population in which all loci are assumed to be
in linkage equilibrium. The allelic states at locus $l$ on the
two chromosomes drawn from the ancestral population will be denoted
$\widehat{\chi}^1_l, \widehat{\chi}^2_l$. They are independent draws from
a distribution on possible allelic states that we denote by
$\widehat{\nu}_l(dx)$.
Without loss of generality,
by replacing $\phi_l(\widehat{\chi}_l^1, \widehat{\chi}_l^2)$ by
$$
\phi_l(\widehat{\chi}_l^1, \widehat{\chi}_l^2) -
\IE[\phi_l(\widehat{\chi}_l^1, \widehat{\chi}_l^2)|\widehat{\chi}_l^1]
-\IE[\phi_l(\widehat{\chi}_l^1, \widehat{\chi}_l^2)|\widehat{\chi}_l^2]
+\IE[\phi_l(\widehat{\chi}_l^1, \widehat{\chi}_l^2)],$$
and observing that the second and third terms on the right hand side are
functions of $\widehat{\chi}_l^1$ and $\widehat{\chi}_l^2$ respectively,
which we may therefore
subsume into $\eta_l(\widehat{\chi}_l)$, we may assume that for any
value $x'$ of the allelic state at locus $l$, the
conditional expectation
\begin{equation}
\label{vanishing conditional distribution phi}
\IE[\phi_l(\widehat{\chi}_l,x')]=\int\phi_l(x,x')\widehat{\nu}_l(dx)=0
=\IE[\phi_l(x',\widehat{\chi}_l)].
\end{equation}
As a consequence, partitioning over the possible values of
$\widehat{\chi}_l^2$,
we have that the cross variation term
\begin{equation}
\label{vanishing cross variation}
\IE[\eta_l(\widehat{\chi}_l^1)\phi_l(\widehat{\chi}_l^1, \widehat{\chi}_l^2)]
=\int \IE[\eta_l(x')\phi_l(x',\widehat{\chi}_l^2)]\widehat{\nu}_l(dx')
=\int \eta_l(x')\IE[\phi_l(x',\widehat{\chi}_l^2)]\widehat{\nu}_l(dx')
=0.
\end{equation}
With this modification of $\phi_l(x,x')$,
\begin{equation}
\label{mean phi zero}
\IE[\phi_l(\widehat{\chi}_l^1,\widehat{\chi}_l^2)]=0.
\end{equation}
Moreover, still without loss of generality, by absorbing the mean into
$\bar{z}_0$, we may assume that
\begin{equation}
\label{mean eta zero}
\IE[\eta_l(\widehat{\chi}_l)]=\int\eta_l(x)\widehat{\nu}_l(dx)=0.
\end{equation}
In this notation, the genetic component of the trait of an individual in
the ancestral population (which we denote by $\hat{Z}$ to make it clear that the following property is specific to individuals in generation $0$) is
$$\hat{Z}=\bar{z}_0+\frac{1}{\sqrt{M}} \sum_{l=1}^M
\left(\eta_l(\widehat{\chi}_l^1)+
\eta_l(\widehat{\chi}_l^2) +\phi_l(\widehat{\chi}_l^1,
\widehat{\chi}_l^2)\right),$$
and by \eqref{mean phi zero} and \eqref{mean eta zero}, we have  $\IE[\hat{Z}]=\bar{z}_0$.

We assume that the scaled allelic effects $\eta_l$, $\phi_l$ are bounded;
$|\eta_l|$, $|\phi_l|\leq B$, for all $l$.
We also assume that all the quantities in the top part of Table~\ref{QG coefficients}
exist in the limit as $M\to\infty$.

\subsection*{Inheritance}

We now need some notation for Mendelian inheritance.
Recall that $i[1]$ and $i[2]$ are the labels of the parents of individual
$i$ in our pedigree, each of which contributes exactly one gene at each locus in a given offspring. Mendelian inheritance translates into the property that the gene passed on by parent~$i[1]$ was the one inherited from its own `first' parent $(i[1])[1]$ with probability $1/2$, or from its `second' parent $(i[1])[2]$ with probability $1/2$. Even though we do not distinguish between males and females, it is convenient to think of
the chromosomes in individual $i$ as being labelled $1$ and $2$, according to whether they are inherited from $i[1]$ or $i[2]$. In particular, $\chi_l^{i[1],1}$ and $\chi_l^{i[1],2}$ will denote the allelic states of the two genes at locus $l$ in parent $i[1]$, respectively inherited from its own `first' and `second' parent. Again following the conventions of Barton et al.~(2017), \nocite{barton/etheridge/veber:2017}
extended to account for the fact that we are now considering diploid individuals, we use independent Bernoulli$(1/2)$ random variables, $X_l^i$, $Y_l^i$ to determine the inheritance of genes $1$ and $2$, respectively, at locus $l$ in individual $i$. Thus,
$X_l^i=1$ if the allelic state of gene~$1$ at locus $l$ in individual $i$ is inherited from gene $1$ in $i[1]$, and $X_l^i=0$ if it is inherited from gene $2$ in $i[1]$. Likewise, $Y_l^i=1$ if the allelic state of gene~$2$ at locus $l$ in individual $i$ is inherited from gene $1$ in $i[2]$, and $Y_l^i=0$ if it is inherited from gene $2$ in $i[2]$.

In this notation, the trait of individual $i$ in generation $t$ is given by
\begin{eqnarray}
\nonumber
Z^i&=&\bar{z}_0+ {\cal A}^i+{\cal D}^i
\\
\label{remainder R1}
&&+\frac{1}{\sqrt{M}}\sum_{l=1}^M\Bigg\{
\bigg(X_l^i-\frac{1}{2}\bigg)\eta_l(\chi_l^{i[1],1})
+\bigg(\frac{1}{2}-X_l^i\bigg)\eta_l(\chi_l^{i[1],2})
\\
\label{remainder R2}
&& \qquad \qquad+\bigg(Y_i-\frac{1}{2}\bigg)\eta_l(\chi_l^{i[2],1})
+\bigg(\frac{1}{2}-Y_i\bigg)\eta_l(\chi_l^{i[2],2})\Bigg\}
\\
\label{remainder S1}
&&+\frac{1}{\sqrt{M}}\sum_{l=1}^M\Bigg\{
\bigg(X_l^iY_l^i-\frac{1}{4}\bigg)\phi_l(\chi_l^{i[1],1}, \chi_l^{i[2],1})
+\bigg(X_l^i(1-Y_l^i)-\frac{1}{4}\bigg)\phi_l(\chi_l^{i[1],1}, \chi_l^{i[2],2})
\\
\label{remainder S2}
&&+\bigg((1-X_l^i)Y_l^i-\frac{1}{4}\bigg)\phi_l(\chi_l^{i[1],2}, \chi_l^{i[2],1})
+\bigg((1-X_l^i)(1-Y_l^i)-\frac{1}{4}\bigg)\phi_l(\chi_l^{i[1],2}, \chi_l^{i[2],2})
\Bigg\},
\end{eqnarray}
where
\begin{equation}
\label{defn of A}
{\cal A}^i=\frac{1}{2\sqrt{M}}\sum_{l=1}^M
\left(\eta_l(\chi_l^{i[1],1})+\eta_l(\chi_l^{i[1],2})+
\eta_l(\chi_l^{i[2],1})+\eta_l(\chi_l^{i[2],2})\right)
\end{equation}
and
\begin{equation}
\label{defn of D}
{\cal D}^i=\frac{1}{4\sqrt{M}}\sum_{l=1}^M\left\{
\phi_l(\chi_l^{i[1],1}, \chi_l^{i[2],1})
+\phi_l(\chi_l^{i[1],1}, \chi_l^{i[2],2})
+\phi_l(\chi_l^{i[1],2}, \chi_l^{i[2],1})
+\phi_l(\chi_l^{i[1],2}, \chi_l^{i[2],2})
\right\}.
\end{equation}

The terms ${\cal A}^i$ and ${\cal D}^i$
are shared by all descendants of the
parents $i[1]$ and $i[2]$.
In Section~\ref{setting out the model}, we presented
the mean and variance of their sum, conditional on the pedigree ${\cal P}(t)$.
The sums~(\ref{remainder R1})+(\ref{remainder R2})
and~(\ref{remainder S1})+(\ref{remainder S2}),
comprise what we previously called $R_A^i$ and $R_D^i$ respectively;
each has mean zero. They capture the randomness of Mendelian inheritance.
They are uncorrelated with ${\cal A}^i+{\cal D}^i$. Again, in Section~\ref{setting out the model} we gave expressions for the variances and covariance
of $R_A^i$ and $R_D^i$
in terms of the ancestral population and
identities generated by the pedigree.
These calculations allowed us to identify the mean and variance of
the parts ${\cal A}^i+{\cal D}^i$ and
$R_A^i+R_D^i$ in terms of the classical quantities of quantitative genetics
in Table~\ref{QG coefficients}.
Since we are assuming unlinked loci,
the asymptotic normality of these quantities when we condition on the
pedigree, but not on the trait values within that pedigree, is an
elementary application of Theorem~\ref{rinott clt} in Appendix~\ref{generalised CLTS}, a generalised
Central Limit Theorem which allows for non-identically distributed summands.

In Barton et al.~(2017),
\nocite{barton/etheridge/veber:2017}
we showed that in the purely additive case, the vector
$(R_A^i)_{i=1}^{N_t}$ which determines the joint distribution of
the trait values within families in generation $t$ (recalling that
in the additive case $R_D^i=0$),
is asymptotically a multivariate normal, even when we condition not just
on the pedigree relatedness of the individuals in generation $t$, but
also on knowing the observed
trait values of all individuals in the pedigree up to
generation $t-1$, which we denote by $\widetilde{Z}(t-1)$ (notice the difference between this notation and the notation $\widetilde{Z}_t$ for the observed trait of an individual living in generation $t$).
Our main result extends this to include dominance, at least under the assumption that the ancestral population was
in linkage equilibrium.

With dominance, the expression for the distribution of the mean and
variance-covariance matrix of the multivariate normal
$Z^1,\ldots ,Z^{N_t}$ conditioned on the pedigree up to generation $t$
and some
collection of the observed trait values of individuals in that pedigree
up to generation $t-1$ is a sum of the quantities of
classical quantitative genetics in Table~\ref{QG coefficients}, weighted by
four-way identities and deviations of trait values from the mean.
In principle, they can be read off from
Theorem~\ref{conditioning multivariate normals} in Appendix~\ref{conditioning normals}.

We will focus on proving that
conditional on knowing just
the trait values of the parents of individual $i$
and the pedigree, the components $({\cal A}^i+{\cal D}^i)$ and $(R_A^i+R_D^i)$
are both
asymptotically normal, but we explain why our proof allows us to extend to the
case in which we also know trait values of other individuals.
The importance (and surprise) is that given the pedigree relationships
between the parents and classical coefficients of quantitative genetics
for a base population (assumed to be in linkage equilibrium), knowing the
traits of the parents distorts the distribution of their offspring in an
entirely predictable way.
In particular, this is what we mean when we say
that the infinitesimal model continues to hold even with
dominance.

The extra challenge compared to the additive case is that, in contrast to the
part $R_A^i+R_D^i$, where Mendelian inheritance ensures independence of the
summands corresponding to different loci even after conditioning on trait
values, when we condition on trait values the terms in ${\cal A}^i+{\cal D}^i$ will be
(weakly) dependent and proving a Central Limit Theorem becomes more involved.

\subsection*{Main results}

Recall that the trait values that we
observe, and therefore on which we condition, are the sum of a genetic
component and an independent environmental component; that is,
the observed trait value is
$$\widetilde{Z}^i:=Z^i+E^i,$$
where, for convenience, the $\{E^i\}$ are independent $N(0,\sigma_E^2)$-valued
random variables. 
We suppose that the
environmental noise is shared by individuals in a family (so we can think
of it as part of the component ${\cal A}^i+{\cal D}^i$ of the trait
value, whose distribution therefore also has a smooth density).

We write $N_t$ for the number of individuals in the population in generation
$t$, $\big(Z_t^1,\ldots , Z_t^{N_t}\big)$ for the corresponding vector
of trait values,
and ${\cal P}(t)$ for the pedigree up to and including generation~$t$.
A simple application of the Central Limit Theorem gives that
$$\left.\big(Z_t^1,\ldots ,Z_t^{N_t}\big)\right| {\cal P}(t)$$
is asymptotically distributed as a multivariate normal random variable
as $M\to\infty$. More precisely, let $(\beta_1,\beta_2,\ldots, \beta_{N_t})
\in \IR^{N_t}$, and write
$Z_\beta=\sum_{i=1}^{N_t}\beta_i Z_t^i$, then
using Theorem~\ref{rinott clt},
$$\left|\IP\left[\frac{Z_{\beta}-\IE[Z_{\beta}]}{\sqrt{\mathtt{Var}(Z_\beta)}}
\leq z\right]-{\cal N}(z)\right|
\leq\frac{C}{\sqrt{M}\sqrt{\mathtt{Var}(Z_\beta)}}
\left(1+\frac{\widetilde{C}}{\mathtt{Var}(Z_\beta)}\right),$$
for suitable constants $C,\,\widetilde{C}$ (which can be made explicit),
where ${\cal N}(z)$ is the cumulative distribution function for
a standard normal random variable. The
mean and variance of $Z_\beta$ can be read off from
Eq.~(\ref{mean Z}), (\ref{covariance Z}), and~(\ref{variance Z}).

Our main results concern the components of the trait values of
offspring when
we condition on the observed trait values of their parents.
The following result follows in essentially the same way as the
additive case of Barton et al.~(2017).
\nocite{barton/etheridge/veber:2017}
\begin{thm}
\label{convergence of residuals}
The conditioned residuals
$(R_A^i+R_D^i)|{\cal P}(t), \widetilde{Z}^{i[1]}, \widetilde{Z}^{i[2]}$
are asymptotically normally distributed, with an error of
order $1/\sqrt{M}$. More precisely, for all $z\in\IR$,
\begin{align}
 \bigg|\IP\bigg[&\frac{R_A^i+R_D^i}{\sqrt{\mathtt{Var}(R_A^i+R_D^i)}}
\leq z\Big| {\cal P}(t), \widetilde{Z}^{i[1]},
\widetilde{Z}^{i[2]}\bigg]-{\cal N}(z)\bigg|
\nonumber\\ & \leq \frac{1}{\sqrt{M}}
\frac{C'}{\sqrt{\mathtt{Var}(R_A^i+R_D^i)}}\left(1+\frac{\widetilde{C'}}{\mathtt{Var}(R_A^i+R_D^i)}\right)\big(1+ \mathfrak{C}\big(i[1],i[2]\big)\big)
\end{align}
where
\begin{align}
\mathfrak{C}\big(i[1],i[2]\big)& =
C''\frac{|\widetilde{Z}^{i[1]}-\IE[\widetilde{Z}^{i[1]}|{\cal P}(t-1)]|}
{\sqrt{\mathtt{Var}(\widetilde{Z}^{i[1]})}}+
C''\frac{|\widetilde{Z}^{i[2]}-\IE[\widetilde{Z}^{i[2]}|{\cal P}(t-1)]|}
{\sqrt{\mathtt{Var}(\widetilde{Z}^{i[2]})}}
\nonumber \\
& \quad +
C'''\frac{1}{\sqrt{\mathtt{Var}(\widetilde{Z}^{i[1]})}
\, p\big(\mathtt{Var}(\widetilde{Z}^{i[1]}), |Z^{i[1]}-\IE[Z^{i[1]}|{\cal P}(t-1)]|\big)}
\left(1+\frac{1}{\mathtt{Var}(\widetilde{Z}^{i[1]})}\right)
\nonumber\\ & \quad +
C'''\frac{1}{\sqrt{\mathtt{Var}(\widetilde{Z}^{i[2]})}
\,p\big(\mathtt{Var}(\widetilde{Z}^{i[2]}), |Z^{i[2]}-\IE[Z^{i[2]}|{\cal P}(t-1)]|\big)}
\left(1+\frac{1}{\mathtt{Var}(\widetilde{Z}^{i[2]})}\right),
\end{align}
and we have used $p(\sigma^2,x)$ to denote the density at $x$ of a mean zero normal random variable with variance $\sigma^2$. The constants $C'$, $\widetilde{C'}$, $C''$, $C'''$ depend only on the bound $B$ on the scaled allelic
effects. The variances in the expressions above are all calculated conditional on ${\cal P}(t-1)$, but not on
observed parental trait values.
\end{thm}
Put simply, the normal approximation is good to an error of order $1/\sqrt{M}$; the constant in the error term will be large, meaning that the approximation will be poor, if the within family variance somewhere in the pedigree is small or if the observed trait values are very different from their expected values. Just as in the additive case, we could prove an entirely analogous result when we condition on any number of observed trait values in the pedigree, except that with dominance this is at the expense of picking up an extra term in the error for each observed trait value on which we condition. The justification required for this is provided by Appendix~\ref{appendix: accumulation of information}.

What is at first sight more surprising is that the shared component of the trait value within a family, \emph{i.e.}, the random variable ${\cal A}+{\cal D}+E$, is also asymptotically normally distributed, even when we condition on observed parental trait values. Note that the randomness of the shared component comes from the fact that the allelic states underlying the parental traits are still random (they are unobserved). In the case of a purely additive trait, it turns out that the shared component can be simply expressed as the average of the two parental traits and therefore conditioning on these traits renders the shared contribution totally deterministic, but such a simplification no longer occurs when we add dominance, due to the nonlinearity of the allelic contributions in $\mathcal{D}$ (see~\eqref{defn of D}). Our proof of normality uses the fact that we consider the environmental noise to be shared by individuals within the family; in this way we can guarantee that the shared component of the observed trait value also has a smooth density.

We are only going to prove the result for the shared component of a family in generation one that was produced by selfing ($i[1]=i[2]$). In what follows, for a given function $h$ we write $\|h\|$ for the supremum norm of $h$, and ${\cal N}_{\mu,\sigma^2}(h)$ for the integral of $h$ with respect to the distribution of an $\mathcal{N}(\mu,\sigma^2)$ random variable (whenever this quantity makes sense):
$$
{\cal N}_{\mu,\sigma^2}(h) = \frac{1}{\sqrt{2\pi \sigma^2}}\int_{-\infty}^{+\infty}h(z) e^{-(z-\mu)^2/(2\sigma^2)}dz.
$$
\begin{thm}
\label{shared parent theorem}
Let $W={\cal A}+{\cal D}+E$ denote the shared component of the trait
value in a family
in generation one. Let $h$ be an
absolutely continuous function with $\|h'\|<\infty$, then
\begin{equation}
\label{result with h}
\left|\IE\big[h(W)|i[1]=i[2], \widetilde{Z}^{i[1]}\big]-{\cal N}_{\mu_W,\sigma_W^2}(h)\right|
\leq \frac{C\|h'\|}{\sqrt{M}},
\end{equation}
where $\mu_W$
is given
by~(\ref{mean gen one identical parents}),
and $\sigma_W^2$ is the sum of the variance of the environmental noise and
the expression
in~(\ref{cond variance gen one same parent}).
\end{thm}
\begin{remark}
\begin{enumerate}
\item
Although we only prove that ${\cal A}^i+{\cal D}^i+E^i$ is asymptotically
normal in this special case of an individual in generation one that is
produced by selfing, the same arguments will apply in general. However, the
expressions involved become extremely cumbersome. By considering selfing, we
capture all the complications that arise in later generations (when distinct
parents may nonetheless be related).
\item
We do not record the exact bound on the constant $C$. It takes the same form
as the error function $\mathfrak{C}$ in Theorem~\ref{convergence of residuals},
except that the constants $C'$, $\widetilde{C'}$, $C''$, $C'''$ depend on the inbreeding
depression $\iota$, as well as the bound $B$ on the scaled allelic effects.
In particular, just as there, the asymptotic normality will break down
if the trait value of the parent is too extreme, or if the variance of the
trait values among offspring is too small.
\item
Since we are assuming that the environmental noise has a smooth density,
convergence in
the sense of~(\ref{result with h})
is sufficient to deduce
that the
cumulative distribution of ${\cal A}^i+{\cal D}^i+E^i$ converges.
\end{enumerate}
\end{remark}

\begin{figure}
\centerline{\includegraphics[width=6in]{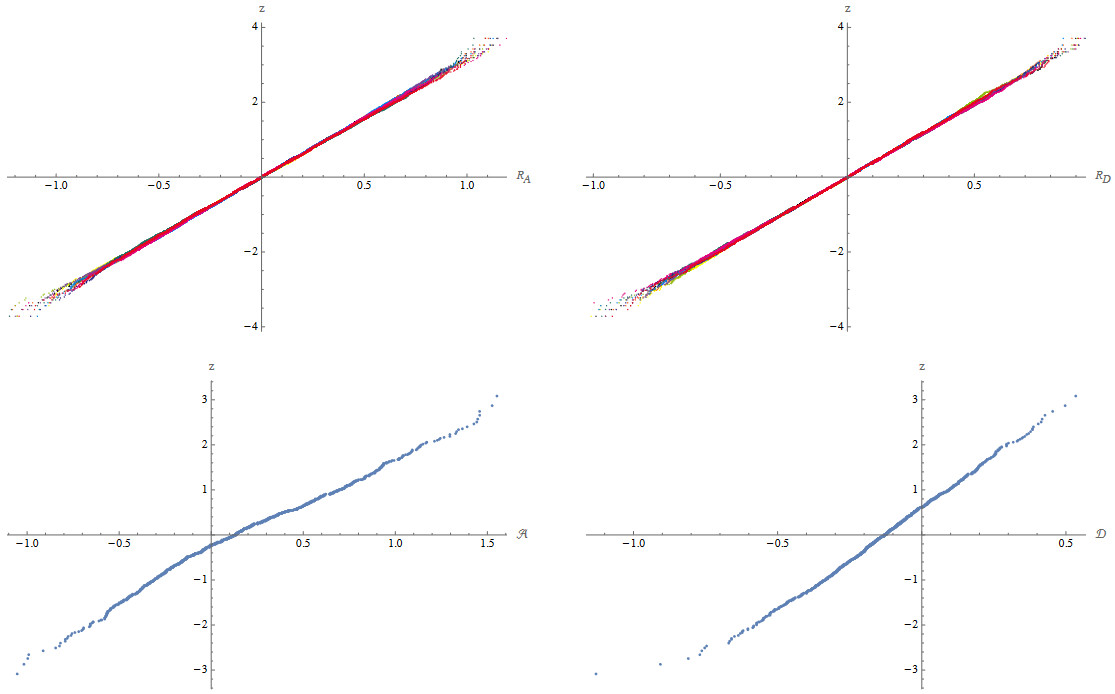}}
\caption{The distributions of the residual (top row: $R_A$ , $R_D$) and shared (bottom row: $\mathcal{A}$, $\mathcal{D}$) components of phenotype ($M=1000$ loci); for each, the CDF is plotted as standard deviations of a Gaussian, $z$, so that a normal distribution appears as a straight line. These are calculated from families of 1000 offspring, from multiple pairs of parents, each replicated 10 times, drawn after 20 generations without selection. The residuals are calculated by subtracting values from the family mean, and pooling across the 10 replicates. Thus, for each family there are 10000 values; the CDF is shown for 10 pairs of
parents, in 10 colours. The shared component is calculated by taking the mean of each family, and pooling across 100 pairs of parents and across the 10 replicates. Thus, for each plot there are 1000 points. There is now some deviation from a Gaussian.}
\label{fig:cdf}
\end{figure}

In Figure~\ref{fig:cdf}, we show the cumulative distribution functions of the additive and dominance parts of the shared and residual components of trait within 10 families after 20 generations of neutral evolution, with $M=1000$ loci. All 10 within-family distributions of $R_A$, $R_D$ are close to Gaussian; they vary somewhat in slope, since families vary in identity coefficients (see Figure~\ref{R and S}), but this is not apparent in
these plots. The normal approximation is better for the residual components than for the shared component. This may be due to the fact that the random variables encoding Mendelian inheritance at different loci are independent and identically distributed, which makes the summands in the expressions for $R_A$ and $R_D$ more weakly dependent than the summands in $\mathcal{A}$ and $\mathcal{D}$, leading to faster convergence to a Gaussian distribution. This also explains why we need a more elaborate approach to show convergence of the shared parts to Gaussians.

\subsection*{Strategy of the derivation}

Our first task will be to show that conditional on the pedigree,
the distribution of the trait values
in generation $t$ is approximately multivariate normal (with an
appropriate error bound).
Since Mendelian inheritance ensures that (before we condition on
knowing any of the previous trait values in the pedigree) the allelic
states at different loci are independent, this is a straightforward
application of a generalised Central Limit Theorem (generalised
because the summands are not required to all have the same distribution).
Just as in Barton et al.~(2017), \nocite{barton/etheridge/veber:2017}
we can keep track of the error that we are making
in assuming a normal approximation at each generation.
In this way we see that, under our assumptions, the
infinitesimal model can be expected to be a good approximation
for order $\sqrt{M}$ generations.

The same Central Limit Theorem guarantees that the joint
distribution of $(Z^{i[1]}, Z^{i[2]}, {\cal A}^i+{\cal D}^i)$ is asymptotically
normally distributed as the number of loci tends to infinity.
This certainly suggests that the conditional distribution of ${\cal A}^i+{\cal D}^i$ given
$Z^{i[1]}$, $Z^{i[2]}$ should be (approximately) normal with
mean and variance predicted by standard results on conditioning
a multivariate normal distribution on some of its marginals
(which we recall in Theorem~\ref{conditioning multivariate normals}).
However, this is not immediate. It is possible that the conditioning forces
the distribution on to the part of our probability space where the normal
approximation breaks down.

To verify that the conditional distribution is asymptotically normal,
we shall show that observing the
trait value of
an individual provides very little information
about their allelic state at any particular locus, or
any particular pair of loci,
and consequently conditioning on parental trait values provides very
little information about allelic states in their offspring.
This is (essentially)
achieved through an application of Bayes' rule, although some care is needed
to control the cumulative error across loci.
We use this to calculate the first and second moments of
${\cal A}^i+{\cal D}^i$ conditional on $\widetilde{Z}^{i[1]}$,
$\widetilde{Z}^{i[2]}$. The fact that they agree
with the predictions of Theorem~\ref{conditioning multivariate normals}
depends crucially on the assumption that dominance is `balanced', in the
sense that the inbreeding depression $\iota$ is well-defined. This
quantity enters not just in the expression for the expected
trait value of inbred individuals, but also in our error bounds,
\emph{c.f.}~Remark~\ref{where we use inbreeding}.

Of course checking that the first two moments of the
conditional distribution of ${\cal A}^i+{\cal D}^i$ are (approximately)
consistent with asymptotic normality is not enough to prove
that the conditioned random variable is indeed (approximately) normal.
Moreover, we cannot apply our generalised Central Limit Theorem
to this term.
Instead we use a generalisation of
Stein's method of `exchangeable pairs' (outlined in
Appendix~\ref{generalised CLTS}), which relies on our ability
to control the (weak) dependence between the contributions to
${\cal A}^i+{\cal D}^i$
from different loci that is induced by the conditioning.
We present the details in the case of identical parents (which is
the case in which normality is most surprising) in
Appendix~\ref{convergence to normal}.

We only present our results in the case in which we condition on the
parental traits of a single individual in generation $t$.
Just as in the additive case, this
can be extended to conditioning on any combination of traits in the pedigree
up to generation $t-1$, but the expressions involved become
unpleasantly complex. Instead of writing them out, we content ourselves
with explaining the only step that requires a new argument.
We must show that
knowing the traits of all individuals up to generation $t-1$
does not provide enough information
about the allelic states at any particular locus in an individual
in generation $t$ to destroy the
asymptotic normality of its trait value. This is
justified
in Appendix~\ref{appendix: accumulation of information} using the
fact that, because of Mendelian inheritance,
the amount of information gleaned about an allele carried
by individual $i$ from looking at the trait
value of one its relatives, is proportional to the
probability of identity with that individual as dictated by the
pedigree.

\subsection*{Asymptotic normality conditional on the pedigree}

We first illustrate the application of the generalised
Central Limit Theorem by
showing that in the ancestral population, the distribution
of $(Z_0^1,\ldots ,Z_0^{N_0})$ is multivariate normal with mean
vector $(\bar{z}_0,\ldots ,\bar{z}_0)$ and variance-covariance matrix
$(\sigma_A^2+\sigma_D^2)\,\mathrm{Id}$,
where $\mathrm{Id}$ is the
identity matrix and $\sigma_A^2$ and $\sigma_D^2$ were defined in
Table~\ref{QG coefficients}.

To prove this, it is enough to show that for any choice of
$\beta=(\beta_1,\ldots,\beta_{N_0})\in\IR^{N_0}$,
\begin{equation*}
\sum_{j=1}^{N_0}\beta_jZ^j\rightarrow Z_{\beta},
\end{equation*}
where $Z_{\beta}$ is normally distributed with mean
$\bar{z}_0\sum_{j=1}^{N_0}\beta_j$ and
variance $(\sigma_A^2+\sigma_D^2)\sum_{j=1}^{N_0}\beta_j^2$.
We apply Theorem~\ref{rinott clt}, due to Rinott~(1994), which
provides control of the rate of convergence as $M\rightarrow\infty$.
\nocite{rinott:1994}
It is convenient to write
$\|\beta\|_1=\sum_{j=1}^{N_0}|\beta_j|$ and $ \|\beta\|_2^2=\sum_{j=1}^{N_0}\beta_j^2$.
Let us write
$$\Psi_l= \left(\eta_l(\widehat{\chi}_l^1)+
\eta_l(\widehat{\chi}_l^2) +\phi_l(\widehat{\chi}_l^1,
\widehat{\chi}_l^2)\right),$$
and we abuse notation by writing $\Psi_l^j$ for this quantity in the
$j$th individual in generation zero.
Set
$E_l=\sum_{j=1}^{N_0}\beta_j\Psi_l^j$. Recalling our assumption that all $\eta_l$ and $\phi_l$ are bounded by some constant $B$, so that the sum of the scaled effects at each locus is bounded
by $3B$, we have that $|E_l|$ is bounded
by $3B\|\beta\|_1$ for all $l$.
Moreover, since the individuals that found the pedigree are assumed to be
unrelated and sampled from an ancestral population in which all loci
are in linkage equilibrium, using~(\ref{mean phi zero})
and~(\ref{mean eta zero}), we find that
$$\IE\Bigg[\sum_{l=1}^ME_l\Bigg]=0,\qquad \mathtt{Var}\Bigg(\sum_{l=1}^ME_l\Bigg)
=M\|\beta\|_2^2\left(\sigma_A^2+\sigma_D^2\right).$$
Theorem~\ref{rinott clt} then yields
\begin{multline*}
	\left|\IP\left[
	\frac{\sum_{i=1}^{N_0}\beta_i(Z^i-\bar{z}_0)}{
	\|\beta\|_2 
\sqrt{\sigma_A^2+\sigma_D^2}}
\leq z\right]
	-{\cal N}(z)\right|\leq
\frac{1}{\sqrt{M}\|\beta\|_2\sqrt{\sigma_A^2+\sigma_D^2}}
\Bigg\{\sqrt{\frac{1}{2\pi}}3B\|\beta\|_1
\\
+\frac{16}{\|\beta\|_2\sqrt{\sigma_A^2+\sigma_D^2}}(3B)^2\|\beta\|_1^2
	+10\left(\frac{1}{\|\beta\|_2^2(\sigma_A^2+\sigma_D^2)}\right)(3B\|\beta\|_1)^3\Bigg\}.
\end{multline*}
Here ${\cal N}$ is the cumulative distribution function of a standard normal
random variable.
The right hand side can be bounded above by
\begin{equation}
\label{rinott clt bound generation zero}
	\frac{C(\|\beta\|_1)}{\|\beta\|_2
\sqrt{M}\sqrt{\sigma_A^2+\sigma_D^2}}
\left(1+\frac{1}{\|\beta\|_2^2{(\sigma_A^2+\sigma_D^2)}}\right),
\end{equation}
for a suitable constant $C$.
In particular, taking $\beta_k=0$ for $k\neq j$ and $\beta_j=1$, we read
off that the
rate of convergence to the normal distribution of $Z_0^j$ as the number
of loci tends to
infinity is order $1/\sqrt{M}$. Note that the normal approximation is
poor if the variance $\sigma_A^2+\sigma_D^2$ is small.

Exactly the same argument shows that the
distribution of
$(Z^1,\ldots ,Z^{N_t})$ of the individuals in generation $t$ converges
to that of a multivariate normal, with
mean vector $(\bar{z}_0+\iota F_{11},\ldots ,\bar{z}_0+\iota F_{N_tN_t})$ and
variance-covariance matrix
determined by Eq.~(\ref{covariance Z})
and~(\ref{variance Z}).

Our proof of asymptotic normality of ${\cal A}^i+{\cal D}^i$ conditional
on the observed trait
values of parents will exploit that the joint distribution of
$({\cal A}^i+{\cal D}^i, Z^{i[1]}, Z^{i[2]})$ is asymptotically normal, also with an
error of order $1/\sqrt{M}$. This time we show that
$\beta_1Z^{i[1]}+\beta_2 Z^{i[2]}+\beta_3({\cal A}^i+{\cal D}^i)$ is asymptotically
normal for every choice of the vector $(\beta_1,\beta_2,\beta_3)\in\IR^3$.
We apply Theorem~\ref{rinott clt}
with
$$\widetilde{E}_l=\beta_1\Psi_l(i[1])
+\beta_2\Psi_l(i[2])+\beta_3 \Phi_l^i$$
where
$$\Psi_l(i[1])=\eta_l(\chi_l^{i[1],1})+\eta_l(\chi_l^{i[1],2})
+\phi_l(\chi_l^{i[1],1},\chi_l^{i[1],2}),$$
with a symmetric expression for $\Psi_l(i[2])$, and
\begin{multline*}
\Phi^i_l=\frac{1}{2}\left(\eta_l(\chi_l^{i[1],1})+
\eta_l(\chi_l^{i[1],2})+\eta_l(\chi_l^{i[2],1})
+\eta_l(\chi_l^{i[2],2})\right)
\\
+\frac{1}{4}\left(\phi_l(\chi_l^{i[1],1},\chi_l^{i[2],1})+
\phi_l(\chi_l^{i[1],1},\chi_l^{i[2],2})+
\phi_l(\chi_l^{i[1],2},\chi_l^{i[2],1})+
\phi_l(\chi_l^{i[1],2},\chi_l^{i[2],2})\right).
\end{multline*}
Theorem~\ref{rinott clt} then shows that the difference between the
cumulative distribution function of $\beta_1Z^{i[1]}+\beta_2 Z^{i[2]}
+\beta_3\big({\cal A}^i+{\cal D}^i\big)$ and that of a normal
random variable with the corresponding mean and variance can be
bounded by~(\ref{rinott clt bound generation zero}) with
$\|\beta\|_2^2\big(\sigma_A^2+\sigma_D^2\big)$ replaced by
$\mathtt{Var}\Big(\beta_1Z^{i[1]}+\beta_2Z^{i[2]}+\beta_3\big({\cal A}^i
+{\cal D}^i\big)\Big)$, which
can be deduced from the expressions for the variance and
covariance of $\Psi_l^{i[1]}$, $\Psi_l^{i[2]}$ and $\Phi_l^i$
that are calculated in Appendix~\ref{QG derivations} and
recorded in~(\ref{covariance Z}), (\ref{variance Z}),
and~(\ref{expression for C([1])}).

\subsubsection*{Conditioning on trait values of the parents}

We suppose that for each $i$, we know the parents of the individual $i$ and
their trait values $Z^{i[1]}$ and $Z^{i[2]}$.
We shall treat the shared components $({\cal A}^i+{\cal D}^i)$ and the residuals
$(R_A^i+R_D^i)$ separately. Both will converge to multivariate normal
distributions which are independent of one another.

Mendelian inheritance ensures that the contributions to $R_A^i+R_D^i$ from
different loci are independent and so normality becomes an easy
consequence of Theorem~\ref{rinott clt}
once we have shown that the information gleaned from knowing the
trait values only perturbs the distribution by order
$1/\sqrt{M}$. This is checked in~(\ref{diagonal terms behave well})
and the proof then
closely resembles the proof in the additive setting of
Barton et al.~(2017) and so we omit the details.
\nocite{barton/etheridge/veber:2017}

The proof that $({\cal A}^i+{\cal D}^i)$ is normal is more involved as once we condition
on the trait values in the parents,
the contributions
$\Phi_l^i$ for $l=1,\ldots , M$ will all be (weakly) correlated.
Our approach uses an extension of
Stein's method of exchangeable pairs which we recall
in Appendix~\ref{generalised CLTS} and
apply to our setting in Appendix~\ref{convergence to normal}.
This calculation is more delicate, but the key is that our
conditioning induces very weak dependence between loci. The
deviation from normality is controlled by
$$\frac{1}{\IP\Big[\widetilde{Z}^{i[1]}=z_1, \widetilde{Z}^{i[2]}=z_2,
{\cal A}^i+{\cal D}^i+E^i=w\Big]}\frac{\partial}{\partial z_1}
\IP\Big[\widetilde{Z}^{i[1]}=z_1, \widetilde{Z}^{i[2]}=z_2,
{\cal A}^i+{\cal D}^i+E^i=w\Big],$$
and the corresponding quantity for the partial derivative with respect
to $z_2$ (both to be interpreted as ratios of densities)
evaluated at $\widetilde{Z}^{i[1]}$, $\widetilde{Z}^{i[2]}$ respectively.
(We recall that $\widetilde{Z}$ denotes observed trait value.)
The normal
approximation will break down if the trait values are too extreme
or if the pedigree is too inbred.

\section{Discussion}
\label{discussion}

The essence of the infinitesimal model is that the distribution of a polygenic trait across a pedigree is multivariate normal. Necessarily, if some individuals are selected (that is, if we condition on their trait values), there can be an arbitrary distortion away from Gaussian \emph{across the population}. However, conditional on parental values and on the pedigree, offspring within each family still follow a Gaussian distribution. This was shown in Barton et al. (2017) in the purely additive case, and is extended here to the case with dominance; the only difference being that with dominance, the part of the trait shared by all siblings, ${\cal A} + {\cal D}$, is now still random even when conditioning on the parental traits (observing the parental traits does not give us full information on the contribution of the parental alleles to the average offspring trait as it did in the purely additive case), and the most difficult part of our analysis consists in showing that this shared contribution is also Gaussian. Our results strongly rely on our assumption that inbreeding depression, $\iota$, is finite (it is zero in the purely additive case). Armed with these results, the classic theory for neutral evolution of quantitative traits can be used to predict evolution, even under selection. Theorems~\ref{convergence of residuals} and~\ref{shared parent theorem} show that this infinitesimal limit holds with dominance, at least over timescales of order square root of the number of loci. Indeed, they show that conditional on the parental traits, the distance between the distributions of the components of the offspring trait and a normal distribution is of the order of $1/\sqrt{M}$. Hence, the distance between the trait distribution of an individual and the infinitesimal approximation increases in every generation by a factor of order $1/\sqrt{M}$, and the error bound becomes macroscopic (\emph{i.e.}, order $1$) after of the order of $\sqrt{M}$ of generations.

Our work provides some mathematical justification for the ubiquity of the
Gaussian, and the empirical success of quantitative genetics - a success which is
remarkable, given the complex interactions that underlie most traits.
The limit is not universal: a non-linear transformation of a Gaussian trait
leads to a non-Gaussian distribution, and failure of the infinitesimal model.
This is because epistatic and dominance interactions then have a systematic
direction, which violates the terms of the Central Limit Theorem.
(Recall that in our toy example in Section~\ref{mendelian inheritance},
we needed a `balance' in the dominance component, which we see
reflected in our main results in the requirement that $\iota$ be
well-defined.)
Nevertheless, if the population is restricted to a range that is narrow
relative to the extremes that are genetically possible,  then the
infinitesimal model may be accurate, even if the genotype-phenotype map is
not linear.  This links to another way to understand our results: if very
many genotypes can generate the same phenotype, then knowing the trait
value gives us negligible information about individual allele frequencies.
To put this another way, the infinitesimal limit implies that selection on
individual alleles is weak relative to random drift ($N_e s \sim 1$), so
that neutral evolution at the genetic level is barely perturbed by selection
on the trait (Robertson, 1960).
\nocite{robertson:1960}

If traits truly evolve in this infinitesimal regime, then it will be
impossible to find any genomic trace of their response to selection. This
extreme view is contradicted by finding an excess of `signatures' of
selection in candidate genes, though it might nevertheless be that these
signals are generated by alleles with modest $N_es$, such that the
infinitesimal model remains accurate for the trait.  Indeed,
Boyle et al.~(2017)
\nocite{boyle/li/pritchard:2017}
argue that the very large numbers of SNPs that are typically implicated in
GWAS for complex traits implies an `omnigenic' view, in which trait variance
is largely due to genes with no obvious functional relation to the trait.
Frequencies of non-synonymous and synonymous mutations suggest that selection
on deleterious alleles is typically much stronger than drift
($N_e s\gg 1$; Charlesworth, 2015).
\nocite{charlesworth:2015}
However, it might still be that selection on the focal trait is comparable
with drift, even if the total selection on alleles is much stronger.
Whether the infinitesimal model accurately describes trait evolution under
such a pleiotropic model is an interesting open question.

In principle, we can simulate the infinitesimal model exactly, by generating
offspring from the appropriate Gaussian distributions. For the additive case,
this is straightforward, since we only need follow the breeding value of each
individual, and the matrix of relationships amongst individuals
(e.g.~Barton \& Etheridge 2011, 2018).
\nocite{barton/etheridge:2011, barton/etheridge:2018}
However, to simulate the infinitesimal model with dominance, we need to track
four-way identities, which is only feasible for small populations
($<30$, say).

We have not set out the extension of the infinitesimal model to structured
populations in detail. In principle, this just requires that we track the
identities within and between the various classes of individual.  One
motivation for the present theoretical work was to extend our infinitesimal
model of `evolutionary rescue' (Barton \& Etheridge, 2018)
\nocite{barton/etheridge:2018}
to include inbreeding depression and partial selfing.  This should be
feasible, provided that we do not need to track identities between specific
individuals, but instead, group individuals according to the time since
their most recent outcrossed ancestor - an approach applied successfully by
Sachdeva (2019).
\nocite{sachdeva:2019}
\nocite{lande/porcher:2015}
Already, Lande and Porcher~(2015) applied the infinitesimal model to a
deterministic model of partial selfing, whilst  Roze~(2016)
\nocite{roze:2016}
analysed an explicit multi-locus model of partial selfing, allowing for
dominance and drift, assuming that all loci are equivalent, and that linkage
disequilibria are weak.

One of the most obviously unreasonable assumptions of the classical
infinitesimal model, and the extension described here, is that there are
an infinite number of unlinked loci.  Santiago~(1998)
\nocite{santiago:1998}
showed how loose linkage could be approximated by averaging over pairwise
linkage disequilibria.  In the additive case, the infinitesimal model can
be defined precisely for a linear genome, by assuming that very many genes
are spread uniformly over the genome (Sachdeva \& Barton,~2018).
\nocite{sachdeva/barton:2018}
The techniques used in our approach are not robust to (even moderately) high levels of linkage, as groups of genes passed on together will decrease the number of `independent' units of heritable contributions to the trait value, leading to an effective number of loci $M_{\mathrm{eff}}$ too low for the Gaussian approximation to be valid (or more precisely, for the bound between the trait distribution and the appropriate Gaussian distribution in Theorems~\ref{convergence of residuals} and \ref{shared parent theorem} to be small). In this case, one needs to consider explicit models of recombination that are out of the scope of this work.

The main value of the infinitesimal model may be to show that trait evolution
depends on only a few macroscopic parameters; even if we still make explicit
multi-locus simulations, this focuses attention on those key parameters,
and gives confidence in the generality of our results. Quantitative genetics
has developed quite separately from population genetics. Although the
theoretical synthesis half a century ago (e.g.~Robertson, 1960; Bulmer,~1971;
\nocite{robertson:1960, bulmer:1971, lande:1975}
Lande,~1975) stimulated much subsequent work (empirical as well as
theoretical), the failure to find a practicable approximation for the
evolution of the genetic variance (e.g.~Turelli \& Barton,~1994)
\nocite{turelli/barton:1994}
was an obstacle to further progress. The infinitesimal model provides a
justification for neglecting the intractable effects of selection on the
variance components, and treating them as evolving solely due to drift and
migration.  This approach may be helpful for understanding evolution in
the short and even medium term.

\paragraph{Data availability.} The code and data produced for this work and used in this article can be found in the public repository \cite{barton:2023}.

\paragraph{Acknowledgements.} We thank the two Reviewers and the Associate Editor for their very useful detailed comments, which helped us to improve the presentation of the results.

\paragraph{Funding.} NHB was supported in part by ERC Grants 250152 and 101055327. AV was partly supported by the chaire Mod\'elisation Math\'ematique
et Biodiversit\'e of Veolia Environment - Ecole Polytechnique - Museum National d'Histoire Naturelle - Fondation X.

\paragraph{Conflicts of interest.} The authors declare no conflict of interest.

\begin{appendix}
\section*{Appendices}
The appendices are organised as follows. Appendix~\ref{appendix:id coef} discusses a simple algorithm to compute identity coefficients. In Appendix~\ref{QG derivations}, we derive the mean and covariances of the shared and residual parts of the offspring trait knowing the pedigree (but not the parental traits). In Appendix~\ref{conditioning normals}, we recall a standard result for conditioning multivariate normal random vectors on their marginal values, while in Appendix~\ref{generalised CLTS} we recall the generalised Central Limit Theorems that will be needed to obtain the normal distribution of the offspring trait components conditional on the parental traits. In Appendix~\ref{key lemmas}, we prove some key lemmas on conditional allelic distributions that we use in Appendix~\ref{cond mean and var} to compute the mean and variance of trait values conditional on the pedigree and on parental traits. The convergence of the shared component of the trait to a Gaussian random variable, as the number of loci tends to infinity, is obtained in Appendix~\ref{convergence to normal}. Finally, in Appendix~\ref{appendix: accumulation of information} we investigate how information accumulates when we condition on knowing more ancestral traits than those of the parents.

\section{Calculating identity coefficients}\label{appendix:id coef}

\subsection*{Recursions for pairwise identity by descent}
Two-way identities are readily expressed as solutions to a recurrence. The recursion for $F$ can be written in terms of a \emph{pedigree matrix}, $P_{i,k}(t)$, which gives the probability that a gene in individual $i$ in generation $t$ came from parent $k$ in generation $(t-1)$; each row has two non-zero entries each with
value 1/2 (the entries corresponding to the indices of the two parents, since the gene may have been inherited from either parent with the same probability), unless the individual is produced by selfing, in which case there is a single entry with value 1 (that corresponding to the index of the single parent). Observe that the matrices $P(t)$ are totally determined by knowledge of the pedigree. In contrast to Barton et al.~(2017), \nocite{barton/etheridge/veber:2017} where we focused on haploids, here we necessarily have to deal with diploids. For diploids, the recursion for $F$ is
\begin{equation}
\label{diploid recursion for identity}
F_{ij}(t)=\sum _{k,l}P_{i,k}(t)P_{j,l}(t)F_{kl}^*(t-1),
\end{equation}
where
\begin{equation*}
F_{kl}^*=F_{kl} \quad\mbox{if }k\neq l,
\qquad F_{kk}^*=\frac{1}{2}\left(1+F_{kk}\right).
\end{equation*}
The quantity $F_{kl}^*$ is the probability of identity of two genes drawn \emph{independently} from individuals $k$ and $l$ (this independent drawing corresponds to Mendelian inheritance); if $k=l$, then we may either pick the same gene twice, which happens with probability $1/2$ (and since the two genes are identical, they are also identical by descent), or pick the two genes of individual $k$, again with probability $1/2$, and their probability of identity by descent is then $F_{kk}$ by definition. Restating \eqref{diploid recursion for identity} in words, the probability that a gene taken in individual $i$ and a gene taken in individual $j$, both in generation $t$, are identical by descent is equal to the sum over all potential pairs $(k,l)$ of parents in the previous generation ($t-1$) of the probability that the gene in $i$ descends from $k$, the gene in $j$ descends from $l$ and that the `parental' genes in $k$ and $l$ are themselves identical by descent.

\subsection*{Calculating two-, three- and four-way identities}
Several papers have developed algorithms for calculating identity coefficients, given a pedigree (Karigl, 1981; Abney, 2009; Garcia-Cortes, 2015; Kirkpatrick et al., 2018). \nocite{karigl:1981, abney:2009, garcia-cortes:2015, kirkpatrick/ge/wang:2019} These assume a single genetic locus, and primarily consider the nine condensed identity coefficients of Figure~\ref{picture for all identities} that describe the relationship between two diploid individuals. This body of work has developed algorithms that can efficiently calculate identity coefficients involving two individuals, across large pedigrees. Karigl (1982) \nocite{karigl:1982} considers (but does not implement) calculation of identities amongst more than two individuals.

Here, we define and implement a (fairly) simple algorithm that deals with multiple sets of genes across multiple individuals. The corresponding code in \emph{Mathematica} can be found in Supplementary Material \cite{barton:2023}. This is unlikely to be as efficient as existing algorithms for identities amongst one set of genes across two individuals; it is limited by the need to calculate and store identities amongst very many sets of ancestral genes, corresponding to the very many routes by which genes may descend through the pedigree.

First we establish our notation. The two genes in each individual each receive a separate label. Thus a gene in
individual $i$ will have label \(\mathbf{i}=\{i,1\}\) or \(\mathbf{i}=\{i,2\}\). Sets of genes will be generically denoted by $S=\{\mathbf{i}_1,\ldots,\mathbf{i}_k\}$. We define \(F\left[S_1,S_2,\ldots , S_n\right]\) to be the probability that the genes contained in each set $S_1$, $S_2$, $\ldots$ , $S_n$ are identical by descent, tracing back to $n$ distinct founders in the ancestral population. For example, $F[\{\mathbf{i}_1\},\{\mathbf{i}_2,\mathbf{i}_3\},\{\mathbf{i}_4,\mathbf{i}_5\}]$ is the probability that these 3 sets of genes, $S_1=\{\mathbf{i}_1\}$, $S_2=\{\mathbf{i}_2,\mathbf{i}_3\}$ and $S_3=\{\mathbf{i}_4,\mathbf{i}_5\}$, each trace back to 3 distinct founders: one ancestral to $\mathbf{i}_1$, another one ancestral to $\mathbf{i}_2$ and $\mathbf{i}_3$, and a last one ancestral to $\mathbf{i}_4$ and $\mathbf{i}_5$. Necessarily, $F[\{\mathbf{i}\}]=1$ (a single gene traces back to a unique founder), and the probability of identity of genes $\mathbf{i}_1$ and $\mathbf{i}_2$ satisfies $F[\{\mathbf{i}_1,\mathbf{i}_2\}]= 1-F[\{\mathbf{i}_1\},\{\mathbf{i}_2\}]$. Identities in generation $t$ are denoted $F_t$.

Given the pedigree, the identities are defined recursively; $F_t$ is a linear combination of identities $F_{t-1}$ in the previous generation. 
Here we simply outline the algorithm. A detailed explanation in terms of the \emph{Mathematica} code is in the
Supplementary Material \cite{barton:2023}.

In generation $t=0$ all individuals are assumed unrelated and so $F_0[S_1,\ldots,S_n]$ is set to be $1$ if each $S_k$ comprises a single gene and these $n$ genes are all distinct.
Otherwise it is set to zero.

The algorithm proceeds in two steps, first identifying the possible parents from which each gene is descended and then the possible genes within that parent. In this way, a list of all possible scenarios is generated, with
each scenario having equal probability. A slight twist here is that if a set contains a single gene in a given individual, that gene traces back to one or other parent of the individual, with equal probability; two genes in the same individual must trace back to the two parents, although those may be the same individual if there is selfing. This list contains many permutations that are equivalent, differing only by order; these are tallied to reduce the number of configurations that need to be stored, resulting in a weighted list.
This gives a recursion back to the founder generation.
The number of generations and size of pedigree is limited by the amount of memory needed to store the intermediate lists. 

\section{Conditioning on the pedigree}
\label{QG derivations}

In this section, we illustrate how to recover the expressions for the mean and
variance of the two parts $({\cal A}^i+{\cal D}^i)$ and $(R_A^i+R_D^i)$ of the trait
of individual $i$ from identity coefficients
of its parents $i[1]$ and
$i[2]$ and the classical coefficients of Table~\ref{QG coefficients}.
Covariances between families are calculated in the same way.
We also calculate the covariance between $({\cal A}^i+{\cal D}^i)$ and $Z^{i[1]}$
and $Z^{i[2]}$ (given the pedigree) which will be important for
establishing the effect of conditioning on the trait values of the
parents.
Although these expressions are
well known, it seems to be hard to find an explicit
derivation such as that presented here.
Note that at this stage we are only conditioning on the pedigree, not
on the observed trait values and the results in this section do not require us
to assume the presence of an environmental noise term.

\subsection*{Notation}

Throughout this section we are going to be calculating quantities conditional
on the pedigree. We shall suppress that in our notation.

\subsection*{Mean and variance of ${\cal A}^i+{\cal D}^i$}

\begin{figure}
\centerline{\includegraphics[width=4in]{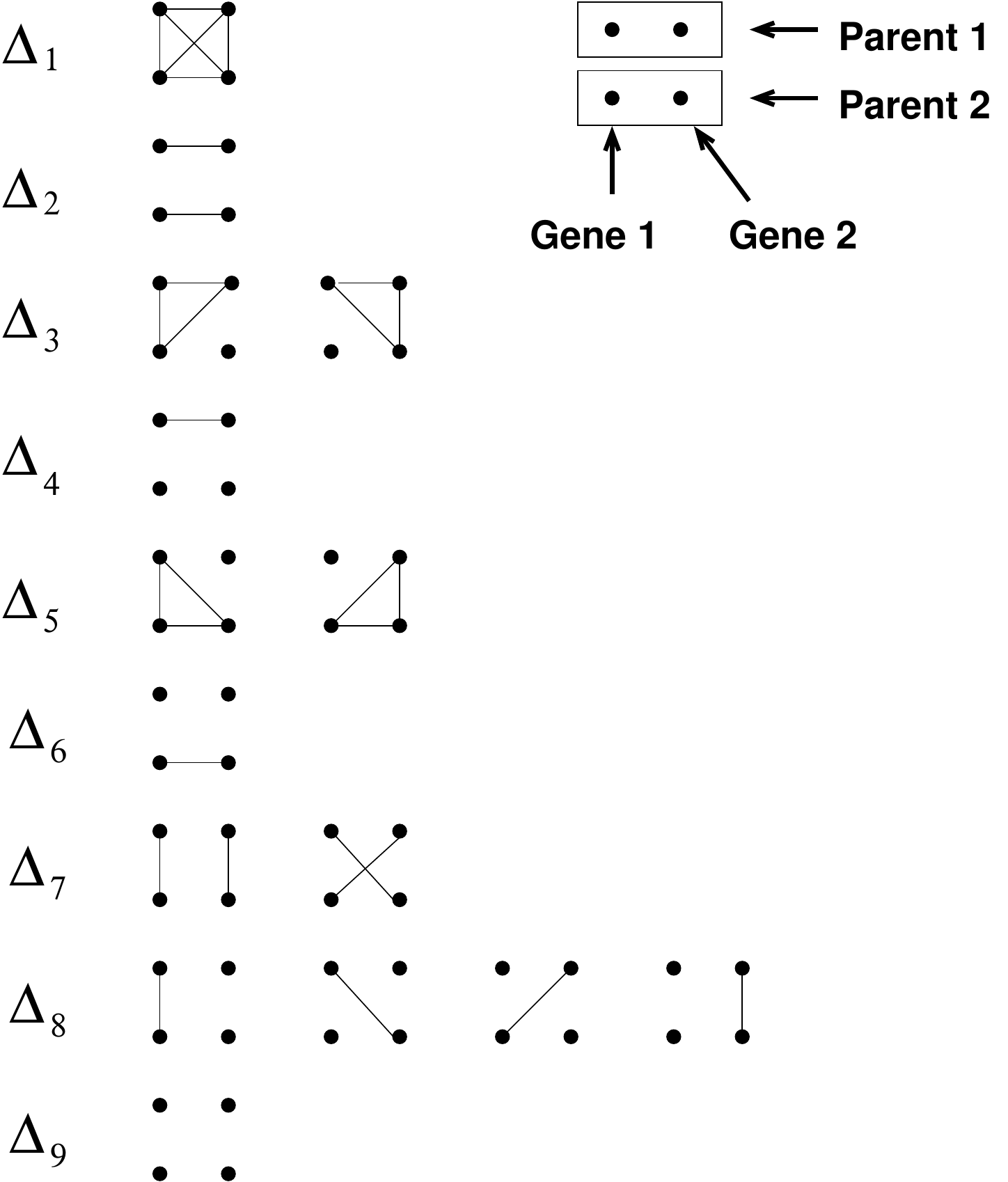}}
\caption{All possible four way identities. The dots represent the
four genes across the two parents (each parent corresponding to
a row) and lines indicate identity (\emph{c.f.}~Abney et al., 2000).
}
\label{picture for all identities}
\end{figure}

The contribution to the trait $Z^{i}$ from the $l$th locus
is determined by the four alleles $\chi_l^{i[1],1}$,
$\chi_l^{i[1],2}$,$\chi_l^{i[2],1}$ and $\chi_l^{i[2],2}$ and
the independent Bernoulli random variables $X_l^i$ and $Y_l^i$.
The mean and variance of $({\cal A}^i+{\cal D}^i)$ and $(R_A^i+R_D^i)$ will
depend on which combinations of these alleles are identical.
First we introduce some notation for the nine possible identity
classes. In Figure~\ref{picture for all identities}, the two copies
of each gene in each individual are represented by two (horizontally
adjacent) dots. Lines between dots represent identity by descent.
It is convenient to think of the genes within an individual
as being ordered.

\nocite{abney/mcpeek/ober:2000}

Let us define
\begin{multline}
\label{defn of Phil}
\Phi(l)=\frac{1}{2}\left(\eta_l(\chi_l^{i[1],1})+
\eta_l(\chi_l^{i[1],2})+\eta_l(\chi_l^{i[2],1})
+\eta_l(\chi_l^{i[2],2})\right)
\\
+\frac{1}{4}\left(\phi_l(\chi_l^{i[1],1},\chi_l^{i[2],1})+
\phi_l(\chi_l^{i[1],1},\chi_l^{i[2],2})+
\phi_l(\chi_l^{i[1],2},\chi_l^{i[2],1})+
\phi_l(\chi_l^{i[1],2},\chi_l^{i[2],2})\right),
\end{multline}
\begin{equation}
\Psi_l(i[1])=\eta_l(\chi_l^{i[1],1})+
\eta_l(\chi_l^{i[1],2})+
\phi_l(\chi_l^{i[1],1},\chi_l^{i[1],2}),
\end{equation}
and
\begin{equation}
\Psi_l(i[2])=\eta_l(\chi_l^{i[2],1})+
\eta_l(\chi_l^{i[2],2})+
\phi_l(\chi_l^{i[2],1},\chi_l^{i[2],2}).
\end{equation}
For each of the nine possible identity classes between $i[1]$ and
$i[2]$, we calculate two quantities from which the mean and
variance of $({\cal A}^i+{\cal D}^i)$ will readily follow.

$$
\begin{array}{ccc}
\mbox{identity state}&\IE\big[\frac{1}{M}\sum_{l=1}^M\Phi(l)^2|\Delta_{\cdot}\big]&
\IE\big[\frac{1}{\sqrt{M}}\sum_{l=1}^M\Phi(l)|\Delta_{\cdot}\big]
\\
\\
\hline\hline
\\

\Delta_1&2\sigma_A^2+2\sigma_{ADI}+\sigma_{DI}^2+\iota^* & \iota
\\
\\
\Delta_2 & \sigma_A^2+\sigma_D^2 & 0
\\
\\
\Delta_3 & \frac{5}{4}\sigma_A^2+\frac{3}{4}\sigma_{ADI}+
\frac{\sigma_{DI}^2+\iota^*}{4}+\frac{\sigma_D^2}{4} &
\frac{\iota}{2}
\\
\\
\Delta_4 & \frac{3}{4}\sigma_A^2+\frac{1}{2}\sigma_D^2 &
0
\\
\\
\Delta_5 & \frac{5}{4}\sigma_A^2+\frac{3}{4}\sigma_{ADI}+
\frac{\sigma_{DI}^2+\iota^*}{4}+\frac{\sigma_D^2}{4} & \frac{\iota}{2}
\\
\\
\Delta_6 & \frac{3}{4}\sigma_A^2+\frac{1}{2}\sigma_D^2 & 0\\
\\
\Delta_7 & \sigma_A^2+\frac{\sigma_{DI}^2+\iota^*}{8}+
\frac{\sigma_{ADI}}{2}+\frac{\sigma_D^2}{4} & \frac{1}{4}\iota^* \\
\\
\Delta_8 & \frac{3}{4}\sigma_A^2+\frac{\sigma_{ADI}}{4}+
\frac{\sigma_{DI}^2+\iota^*}{16}+\frac{3}{16}\sigma_D^2 &
\frac{1}{4}\iota\\
\\
\Delta_9 & \frac{\sigma_A^2}{2}+\frac{\sigma_D^2}{4} & 0
\end{array}
$$

To see where these expressions come from, consider for example
identity state $\Delta_3$, with, say,
$\chi_l^1:=\chi_l^{i[1],1}= \chi_l^{i[1],2}= \chi_l^{i[2],1}\neq
\chi_l^{i[2],2}=:\chi_l^2$, where `$=$' here means identical
by descent. Then,
using~(\ref{vanishing cross variation})--(\ref{mean eta zero}),
\begin{eqnarray*}
\IE\left[\left.\frac{1}{M}\sum_{l=1}^M\Phi(l)^2\right|\Delta_3\right]&=&
\frac{1}{M}\sum_{l=1}^M\IE\left[\left(\frac{3\eta(\chi_l^1)+\eta(\chi_l^2)}{2}+
\frac{2\phi(\chi_l^1, \chi_l^1)+2\phi(\chi_l^1,\chi_l^2)}{4}\right)^2\right]
\\
&=&
\frac{5}{4}\sigma_A^2+\frac{3}{4}\sigma_{ADI}+\frac{1}{4}(\sigma_{DI}^2+\iota^*)
+\frac{1}{4}\sigma_D^2.
\end{eqnarray*}
The following quantities can be calculated in the same way. They
are important for calculating the
covariance between the trait values of parent and offspring
(in particular the covariance between $({\cal A}^i+{\cal D}^i)$ and $Z^{i[1]}$
and $Z^{i[2]}$) which
will dictate the change in distribution of the trait values within
families arising from conditioning on knowing the traits of the
parents. We record them here for later reference.
$$
\begin{array}{ccc}
\mbox{identity state}&
\IE\big[\frac{1}{M}\sum_{l=1}^M\Phi(l)\Psi_l(i[1])|\Delta_{\cdot}\big]&
\IE\big[\frac{1}{M}\sum_{l=1}^M\Phi(l)\Psi_l(i[2])|\Delta_{\cdot}\big]\\
\\
\hline\hline\\
\Delta_1&
2\sigma_A^2+2\sigma_{ADI}+\sigma_{DI}^2+\iota^*&
2\sigma_A^2+2\sigma_{ADI}+\sigma_{DI}^2+\iota^*\\
\\
\Delta_2 & 
\sigma_A^2+\frac{\sigma_{ADI}}{2}
&
\sigma_A^2+\frac{\sigma_{ADI}}{2}\\
\\
\Delta_3 & 
\frac{3}{2}\sigma_A^2+\frac{\sigma_{DI}^2+\iota^*}{2}+
\frac{5}{4}\sigma_{ADI} &
\sigma_A^2+\frac{\sigma_{ADI}}{4}+\frac{\sigma_D^2}{2}
\\
\\
\Delta_4 & \sigma_A^2+\frac{1}{2}\sigma_{ADI} & \frac{\sigma_A^2}{2}\\
\\
\Delta_5 & \sigma_A^2+\frac{\sigma_{ADI}}{4}+\frac{\sigma_D^2}{2} &
\frac{3}{2}\sigma_A^2+\frac{\sigma_{DI}^2+\iota^*}{2}+
\frac{5}{4}\sigma_{ADI}\\
\\
\Delta_6 & \frac{\sigma_A^2}{2} & \sigma_A^2 +\frac{1}{2}\sigma_{ADI}\\
\\
\Delta_7 & \sigma_A^2+\frac{\sigma_{ADI}}{4}+\frac{\sigma_D^2}{2}
& \sigma_A^2+\frac{\sigma_{ADI}}{4}+\frac{\sigma_D^2}{2}\\
\\
\Delta_8 &
\frac{3}{4}\sigma_A^2+\frac{1}{8}\sigma_{ADI}+\frac{1}{4}\sigma_D^2&
\frac{3}{4}\sigma_A^2+\frac{1}{8}\sigma_{ADI}+\frac{1}{4}\sigma_D^2\\
\\
\Delta_9 & \frac{1}{2}\sigma_A^2&
\frac{1}{2}\sigma_A^2
\end{array}
$$

We can express two and three way identities between the parents in
terms of the four way identities $\Delta_1,\ldots ,\Delta_9$.
Recall
that we write, for example, $F_{11}$ for the probability of identity of
the two genes in $i[1]$ and $F_{12}$ for the probability of
identity of two genes, one selected at random from $i[1]$ and
one from $i[2]$. In terms of the nine identity states we have
\begin{eqnarray*}
F_{11}&=& \IP[\Delta_1]+\IP[\Delta_2]+\IP[\Delta_3]+\IP[\Delta_4]\\
F_{22}&=&\IP[\Delta_1]+\IP[\Delta_2]+\IP[\Delta_5]+\IP[\Delta_6]\\
F_{12}&=&\IP[\Delta_1]+\frac{1}{2}\left(\IP[\Delta_3]+\IP[\Delta_5]
+\IP[\Delta_7]\right)+
\frac{1}{4}\IP[\Delta_8]\\
F_{112}&=&\IP[\Delta_1]+\frac{1}{2}\IP[\Delta_3]\\
F_{122}&=&\IP[\Delta_1]+\frac{1}{2}\IP[\Delta_5]\\
F_{1122}&=&\IP[\Delta_1]\\
\widetilde{F}_{1122}&=&\IP[\Delta_2]\\
\widetilde{F}_{1212}&=&\IP[\Delta_7].
\end{eqnarray*}
Combining the above, we find
\begin{align*}
\IE\left[\frac{1}{M}\sum_{l=1}^M \Phi(l)\Psi_l(i[1])\right]
= &\ \frac{\sigma_A^2}{2}\left(1+F_{11}+2F_{12}\right)
+\frac{\sigma_{ADI}}{2}\left(F_{11}+F_{12}+2F_{122}\right) \nonumber\\
&+\sigma_D^2\left(F_{12}-F_{112}\right)+
\left(\sigma_{DI}^2+\iota^*\right)F_{112},
\end{align*}
with a symmetric expression for
$\frac{1}{M}\sum_{l=1}^M\IE\left[\Phi(l)\Psi_l(i[2])\right]$.
Similarly,
$$\IE[({\cal A}^i+{\cal D}^i)]=\frac{1}{\sqrt{M}}\sum_{l=1}^M\IE[\Phi(l)]=\iota F_{12}$$
and
\begin{align*}
\frac{1}{M}\sum_{l=1}^M\IE[\Phi(l)^2]= &\
\frac{\sigma_A^2}{2}\left(1+\frac{F_{11}+F_{22}}{2}+2F_{12}\right)
+\sigma_{ADI}\left(F_{12}+\frac{F_{112}+F_{122}}{2}\right)\\
& +\frac{\sigma_{DI}^2+\iota^*}{4}\left(F_{12}+F_{112}+F_{122}+F_{1122}\right)\\
& +\frac{\sigma_D^2}{4}\left(1-F_{12}+F_{11}-F_{112}+
F_{22}-F_{122}+\tilde{F}_{1122}+\frac{1}{2}\tilde{F}_{1212}\right)
+\frac{1}{4}\iota^*\widetilde{F}_{1212},
\end{align*}
from which, since for $l\neq m$ we are assuming
$\IE[\Phi(l)\Phi(m)]=\IE[\Phi(l)]\IE[\Phi(m)]$,
\begin{equation}
\nonumber
\frac{1}{M}\IE\Bigg[\mathop{\sum_{l=1}^M\sum_{m=1}^M}_{l\neq m}
\Phi(l)\Phi(m)\Bigg]=(\iota F_{12})^2-\iota^*F_{12}^2,
\end{equation}
and the expression~(\ref{variance A+D unconditioned})
for the variance of $({\cal A}^i+{\cal D}^i)$ follows.

\begin{remark}
\label{source of discrepancy from WL}
Walsh \& Lynch~(2018) give an expression for the variance when there is
linkage disequilibrium. In their notation,
$\tilde{f}$ is the probability of identity at two distinct loci.
Then for $l\neq m$,
$$\IE[\Phi(l)\Phi(m)]=\tilde{f}\IE[\Phi_l(\hat{\chi}_l,\hat{\chi}_l)]\IE[\Phi_m(\hat{\chi}_m,\hat{\chi}_m)],$$
so that our expression for
\begin{equation}
\nonumber
\frac{1}{M}\IE\Bigg[\mathop{\sum_{l=1}^M\sum_{m=1}^M}_{l\neq m}
\Phi(l)\Phi(m)\Bigg]
\end{equation}
will be multiplied by
$(\tilde{f}/F_{12}^2)$, resulting (when we subtract
$\IE[{\cal A}^i+{\cal D}^i]^2$)
in an overall expression
of $(\tilde{f}-F_{12}^2)\iota^2-\tilde{f}\iota^*$
in place of $-\iota^*F_{12}^2$.
Correcting for this by adding
$(\tilde{f}-F_{12}^2)(\iota^2-\iota^*)$ to our
expression~(\ref{variance Z}) for the
variance of $Z^i$ (for which we recall that $F_{12}$ becomes $F_{ii}$),
we recover the expression of Walsh \& Lynch~(2018).
\end{remark}

\subsection*{The covariance between ${\cal A}^i+{\cal D}^i$ and
${\cal A}^j+{\cal D}^j$.}

To understand the expression~(\ref{covariance A plus D})
for the covariance between ${\cal A}^i+{\cal D}^i$and
${\cal A}^j+{\cal D}^j$ for $i\neq j$,
consider
\begin{multline*}
\IE\Bigg[
\Bigg\{
\frac{
\left(\eta_l(\chi_l^{i[1],1})+\eta_l(\chi_l^{i[1],2})+
\eta_l(\chi_l^{i[2],1})+\eta_l(\chi_l^{i[2],2})\right)}{2}
\\+
\frac{
\phi_l(\chi_l^{i[1],1}, \chi_l^{i[2],1})
+\phi_l(\chi_l^{i[1],1}, \chi_l^{i[2],2})
+\phi_l(\chi_l^{i[1],2}, \chi_l^{i[2],1})
+\phi_l(\chi_l^{i[1],2}, \chi_l^{i[2],2})
}{4}\Bigg\}
\\
\Bigg\{
\frac{
\left(\eta_l(\chi_l^{j[1],1})+\eta_l(\chi_l^{j[1],2})+
\eta_l(\chi_l^{j[2],1})+\eta_l(\chi_l^{j[2],2})\right)}{2}
\\+
\frac{
\phi_l(\chi_l^{j[1],1}, \chi_l^{j[2],1})
+\phi_l(\chi_l^{j[1],1}, \chi_l^{j[2],2})
+\phi_l(\chi_l^{j[1],2}, \chi_l^{j[2],1})
+\phi_l(\chi_l^{j[1],2}, \chi_l^{j[2],2})
}{4}\Bigg\}
\Bigg].
\end{multline*}
The sixteen terms corresponding to products of additive effects
correspond to the sixteen different possibilities for the allelic
types at locus $l$ if we choose one allele at random
from individual
$i$ and one from individual $j$, and the contribution to the
expectation will be nonzero precisely if the chosen alleles are identical,
in which case they contribute $\IE[\eta_l(\widehat{\chi}_l)^2]$.
Summing over $l$, the overall contribution of such terms to the
covariance will therefore be $2\sigma_A^2 F_{ij}$.

Similarly, terms involving one factor of $\eta_l$ and one $\phi_l$
will only be non-zero if all evaluated on the same allelic type, hence
the terms multiplied by $F_{iij}$ and $F_{ijj}$ in
Eq.~(\ref{covariance A plus D}).

Continuing in this way and using that $\IE[{\cal A}^i+{\cal D}^i]=\iota F_{ii}$, we
recover Eq.~(\ref{covariance A plus D}).

\subsection*{The residuals $R_A^i+R_D^i$}

The corresponding calculations for the mean and variance of the residuals,
$R_A^i+R_D^i$ follow exactly the same pattern. It is convenient to consider
$R_A^i$ and $R_D^i$ separately, and then calculate the covariance. The first of
these, corresponding to the additive part
is very straightforward since it is only going to depend on pairwise
identities.

Recall first that
$$
R_A^i\!=\!\frac{1}{\sqrt{M}}\sum_{l=1}^M\left\{
\big(X_i-\frac{1}{2}\big)\eta_l(\chi_l^{i[1],1})
+\big(\frac{1}{2}-X_l^i\big)\eta_l(\chi_l^{i[1],2})
+\big(Y_i-\frac{1}{2}\big)\eta_l(\chi_l^{i[2],1})
+\big(\frac{1}{2}-Y_i\big)\eta_l(\chi_l^{i[2],2})\right\}.
$$
Since the Mendelian inheritance is independent of the
allelic states, $R_A^i$ has mean zero; to establish the
variance, we must calculate its square. Since
inheritance is independent at distinct loci, only the diagonal
terms contribute and we find
\begin{eqnarray}
\nonumber
\IE[(R_A^i)^2]
&=&
\frac{1}{M}\sum_{l=1}^M\IE\Bigg[\Bigg\{
\big(X_i-\frac{1}{2}\big)\eta_l(\chi_l^{i[1],1})
+\big(\frac{1}{2}-X_l^i\big)\eta_l(\chi_l^{i[1],2})
\\
\nonumber
&&
\qquad\qquad +\big(Y_i-\frac{1}{2}\big)\eta_l(\chi_l^{i[2],1})
+\big(\frac{1}{2}-Y_i\big)\eta_l(\chi_l^{i[2],2})\Bigg\}^2
\Bigg]\\
\nonumber
&=&
\frac{1}{4M}\sum_{l=1}^M\IE\left[
(\eta_l(\chi_l^{i[1],1}))^2
+(\eta_l(\chi_l^{i[1],2}))^2
+(\eta_l(\chi_l^{i[2],1}))^2
+(\eta_l(\chi_l^{i[2],2}))^2
\right] \\
\nonumber
&&\qquad\qquad
-
\frac{1}{2M}\sum_{l=1}^M\IE\left[
\eta_l(\chi_l^{i[1],1})
\eta_l(\chi_l^{i[1],2})
+\eta_l(\chi_l^{i[2],1})
\eta_l(\chi_l^{i[2],2})
\right]\\
\nonumber
&=&
\frac{1}{M}\sum_{l=1}^M\mathtt{Var}(\eta_l(\widehat{\chi}_l))
-\frac{1}{2M}\sum_{l=1}^M\left(F_{11}+F_{22}\right)
\mathtt{Var}(\eta(\widehat{\chi}_l))
\\
&=&\left(1-\frac{F_{11}+F_{22}}{2}\right)\frac{\sigma_A^2}{2}.
\label{star one}
\end{eqnarray}
This is, of course, exactly the expression we would obtain in the
purely additive case.

The second residual, $R_D^i$, also has mean zero, but its variance will now
involve higher order identities.
Recall that
\begin{align*}
R_D^i=& \frac{1}{\sqrt{M}}\sum_{l=1}^M\Bigg\{
\big(X_l^iY_l^i-\frac{1}{4}\big)\phi_l(\chi_l^{i[1],1}, \chi_l^{i[2],1})
+\big(X_l^i(1-Y_l^i)-\frac{1}{4}\big)\phi_l(\chi_l^{i[1],1}, \chi_l^{i[2],2})
\\
& \qquad \qquad +\big((1-X_l^i)Y_l^i-\frac{1}{4}\big)\phi_l(\chi_l^{i[1],2}, \chi_l^{i[2],1})
+\big((1-X_l^i)(1-Y_l^i)-\frac{1}{4}\big)\phi_l(\chi_l^{i[1],2}, \chi_l^{i[2],2})
\Bigg\}.
\end{align*}
Once again, since Mendelian inheritance is independent at different loci,
$\IE[(R_D^i)^2]$ will be entirely determined by the diagonal terms.
Note that for independent Bernoulli (parameter $1/2$)
random variables $X$ and $Y$,
$$\IE\left[\bigg(XY-\frac{1}{4}\bigg)^2\right]
=\IE\left[\frac{1}{2}XY+\frac{1}{16}\right]
=\frac{3}{16},$$
and
$$
\IE\left[\bigg(XY-\frac{1}{4}\bigg)\bigg(X(1-Y)-\frac{1}{4}\bigg)\right]
=-\frac{1}{16}.$$
So, taking expectations over the variables $X_l^i$ and $Y_l^i$, we find
\begin{align}
\label{raw equation for Si squared}
& \IE\big[(R_D^i)^2\big] \nonumber\\
& =\frac{3}{16M}\sum_{l=1}^M\IE\left[
\phi_l(\chi_l^{i[1],1},\chi_l^{i[2],1})^2+
\phi_l(\chi_l^{i[1],1},\chi_l^{i[2],2})^2+
\phi_l(\chi_l^{i[1],2},\chi_l^{i[2],1})^2
+\phi_l(\chi_l^{i[1],2},\chi_l^{i[2],2})^2
\right] \nonumber\\
&\quad  -\frac{2}{16M}\sum_{l=1}^M\IE\Bigg[
\phi_l(\chi_l^{i[1],1},\chi_l^{i[2],1})
\phi_l(\chi_l^{i[1],1},\chi_l^{i[2],2})
+\phi_l(\chi_l^{i[1],1},\chi_l^{i[2],1})
\phi_l(\chi_l^{i[1],2},\chi_l^{i[2],1})
\nonumber\\
& \qquad \qquad\qquad +\phi_l(\chi_l^{i[1],1},\chi_l^{i[2],1})
\phi_l(\chi_l^{i[1],2},\chi_l^{i[2],2})
+\phi_l(\chi_l^{i[1],1},\chi_l^{i[2],2})
\phi_l(\chi_l^{i[1],2},\chi_l^{i[2],1})
\nonumber\\
& \qquad \qquad\qquad+\phi_l(\chi_l^{i[1],1},\chi_l^{i[2],2})
\phi_l(\chi_l^{i[1],2},\chi_l^{i[2],2})
+\phi_l(\chi_l^{i[1],2},\chi_l^{i[2],1})
\phi_l(\chi_l^{i[1],2},\chi_l^{i[2],2})
\Bigg].
\end{align}
The first term depends only on pairwise identities and we see
immediately that it is
\begin{equation*}
\frac{3}{4M}F_{12}\sum_{l=1}^M
\IE[\phi_l(\widehat{\chi}_l,\widehat{\chi}_l)^2]
+\frac{3}{4M}(1-F_{12})\sum_{l=1}^M
\IE[\phi_l(\widehat{\chi}^1_l,\widehat{\chi}^2_l)^2]
=
\frac{3}{4}F_{12}(\sigma_{DI}^2+\iota^*)+\frac{3}{4}(1-F_{12})\sigma_D^2.
\end{equation*}

The second term in~(\ref{raw equation for Si squared}) is most easily
calculated conditional on identity class.
Let us write $\Xi(l)$ 
for the summand corresponding to locus $l$.

$$
\begin{array}{cc}
\mbox{identity state}&\IE\big[\frac{1}{M}\sum_{l=1}^M\Xi(l)|\Delta_{\cdot}\big]
\\
\\
\hline\hline
\\

\Delta_1& \frac{3}{4}\left(\sigma_{DI}^2+\iota^*\right)\\
\\
\Delta_2 & \frac{3}{4}\sigma_D^2\\
\\
\Delta_3 & \frac{1}{8}\left(\sigma_D^2+\sigma_{DI}^2+\iota^*\right)
\\
\\
\Delta_4 & \frac{1}{4}\sigma_D^2
\\
\\
\Delta_5 & \frac{1}{8}\left(\sigma_D^2+\sigma_{DI}^2+\iota^*\right)
\\
\\
\Delta_6 & \frac{1}{4}\sigma_D^2
\\
\\
\Delta_7 & \frac{1}{8}\left(\sigma_D^2+\iota^*\right)
\\
\\
\Delta_8 & 0
\\
\\
\Delta_9 & 0
\end{array}
$$

Using our notation for identities, this becomes
\begin{align*}
&-\frac{1}{4}\left(F_{1122}+F_{122}+F_{112}\right)\left(\sigma_{DI}^2
+\iota^*\right) -\frac{1}{4}\left(F_{11}-F_{112}+F_{22}-F_{122}+\widetilde{F}_{1122}
+\frac{1}{2}\widetilde{F}_{1212}\right)\sigma_D^2 \nonumber \\
& \qquad  -\frac{1}{4}\iota^*\widetilde{F}_{1212}.
\end{align*}
Thus
\begin{align}
\label{star two}
\IE\left[(R_D^i)^2\right] = &\ \frac{1}{4}\left(3F_{12}-F_{1122}-F_{122}-F_{112}\right)
\left(\sigma_{DI}^2+\iota^*\right) -\frac{1}{4}\iota^*\widetilde{F}_{1212} \nonumber
\\
& +\frac{1}{4}\left(3(1-F_{12})-(F_{22}-F_{122})-(F_{11}-F_{112})
-\widetilde{F}_{1122}-\frac{1}{2}\widetilde{F}_{1212}\right)\sigma_D^2.
\end{align}

\subsection*{The covariance of $R_A^i$ and $R_D^i$.}

Since $R_A^i$ has mean zero, it suffices to calculate
$\IE[R_A^i R_D^i]$. 
We need to establish the mean of
\begin{multline}
\label{summand for covariances}
\left\{ \big(X-\frac{1}{2}\big)\eta_l(\chi_l^{i[1],1})
+\big(\frac{1}{2}-X\big)\eta_l(\chi_l^{i[1],2})
+\big(Y-\frac{1}{2}\big)\eta_l(\chi^{i[2],1}) +
\big(\frac{1}{2}-Y\big)\eta_l(\chi_l^{i[2],2})\right\}
\\
\times
\Bigg\{XY\phi_l(\chi_l^{i[1],1},\chi_l^{i[2],1})+X(1-Y)
\phi_l(\chi_l^{i[1],1},\chi_l^{i[2],2})+(1-X)Y
\phi_l(\chi_l^{i[1],2},\chi_l^{i[2],1})
\\
+(1-X)(1-Y)\phi_l(\chi_l^{i[1],2},\chi_l^{i[2],2})\Bigg\}.
\end{multline}
We have been able to drop the `$-1/4$' terms in the second bracket
since $\IE[R_A^i]=0$.

Now
\begin{eqnarray*}
\IE\left[\bigg(X-\frac{1}{2}\bigg)XY\right] &=& \frac{1}{2}\IE[XY]=\frac{1}{8},\\
\IE\left[\bigg(X-\frac{1}{2}\bigg)(1-X)Y\right] &=& -\frac{1}{2}\IE[(1-X)Y]
=-\frac{1}{8},
\end{eqnarray*}
and so the mean of~(\ref{summand for covariances}) is that of
\begin{multline}
\frac{1}{8}\Big\{\big(\eta_l(\chi_l^{i[1],1})
-\eta_l(\chi^{i[1],2})\big)\phi_l(\chi_l^{i[1],1},\chi_l^{i[2],1})
+\big(\eta_l(\chi_l^{i[2],1})-\eta_l(\chi_l^{i[2],2}))
\phi_l(\chi_l^{i[1],1},\chi_l^{i[2],1})
\\
+\big(\eta_l(\chi_l^{i[1],1}-\eta_l(\chi_l^{i[1],2})\big)
\phi_l(\chi_l^{i[1],1},\chi_l^{i[2],2})
+\big(\eta_l(\chi_l^{i[2],2})-\eta_l(\chi_l^{i[2],1})\big)
\phi_l(\chi_l^{i[1],1},\chi_l^{i[2],2})
\\
+\big(\eta_l^{i[1],2}-\eta_l(\chi_l^{i[1],1}\big)
\phi_l(\chi_l^{i[1],2},\chi_l^{i[2],1})
+\big(\eta_l(\chi_l^{i[2],1})-\eta_l(\chi_l^{i[2],2})\big)
\phi_l(\chi_l^{i[1],2},\chi_l^{i[2],1})
\\
+\big(\eta_l(\chi_l^{i[1],2})-\eta_l(\chi^{i[1],1})\big)
\phi_l(\chi_l^{i[1],2},\chi_l^{i[2],2})
+\big(\eta_l(\chi_l^{i[2],2})-\eta_l(\chi_l^{i[2],1})\big)
\phi_l(\chi_l^{i[1],2},\chi_l^{i[2],2})\Big\}.
\end{multline}
Taking expectations (conditional on the pedigree)
and summing over loci, we find
\begin{equation}
\label{star three}
\IE\left[R_A^i R_D^i\big|{\cal P}(t)\right]
=\left(F_{12}-\frac{F_{112}+F_{122}}{2}\right)\frac{\sigma_{ADI}}{2}.
\end{equation}

Finally, for two distinct parents, we have found that in generation $t$,
conditional on the pedigree up to time $t$,
\begin{align*}
\mathtt{Var}(R_A^i+R_D^i)= & \
\left(1-\frac{F_{11}+F_{22}}{2}\right)\frac{\sigma_A^2}{2}
+\frac{1}{4}\left(3F_{12}-F_{112}-F_{122}-F_{1122}\right)
\left(\sigma_{DI}^2+\iota^*\right)
\\ & +\frac{1}{4}\left(3(1-F_{12})-(F_{11}-F_{112})-(F_{22}-F_{122})
-\widetilde{F}_{1122}-\frac{1}{2}\widetilde{F}_{1212}\right)\sigma_D^2 \\
& +\left(F_{12}-\frac{F_{112}+F_{122}}{2}\right)\sigma_{ADI} -\frac{1}{4}\iota^*\widetilde{F}_{1212}.
\end{align*}

We can also read off the result for when the two parents are the same
from this formula.
In that case
$$F_{1122}=F_{11}=F_{22}=F_{112}=F_{122},\quad
F_{12}=\frac{1}{2}(1+F_{11}),\mbox{ and }\widetilde{F}_{1212}=1-F_{11}.$$
Thus $\mathtt{Var}(R_A^i+R_D^i)$ reduces to
$$(1-F_{11})\left(\frac{\sigma_A^2}{2}+\frac{3}{8}\big(\sigma_{DI}^2+
\iota^*\big)
+\frac{1}{4}\sigma_D^2+\frac{1}{2}\sigma_{ADI}\right)
-\frac{1}{4}\iota^* .$$

\section{Conditioning multivariate Gaussian vectors}
\label{conditioning normals}

For ease of reference, we record here
a standard result for conditioning multivariate normal
random vectors on their marginal values.
\begin{thm}
\label{conditioning multivariate normals}
Suppose that
$$\begin{bmatrix}
x_A\\ x_B
\end{bmatrix}
\sim {\cal N}\left(\begin{bmatrix} \mu_A \\ \mu_B \end{bmatrix},
		\begin{bmatrix} \Sigma_{AA} & \Sigma_{AB}\\
\Sigma_{BA} & \Sigma_{BB}\end{bmatrix}\right).$$
Then
$$x_A | x_B \sim {\cal N} \left(\mu_A+\Sigma_{AB}\Sigma_{BB}^{-1}
	(x_B-\mu_B), \Sigma_{AA}-\Sigma_{AB}\Sigma_{BB}^{-1}\Sigma_{BA}
\right).$$
\end{thm}
The proof can be found, for example, in Brockwell \& Davis~(1996)
(Proposition~1.3.1
in Appendix~A).\nocite{brockwell/davis:1996}

\section{Generalised Central Limit Theorems}
\label{generalised CLTS}

We shall exploit known techniques for proving both convergence to
a normal distribution, and for establishing the rate of
convergence, in situations which go beyond the classical setting of
independent identically distributed random variables. For convenience
we recall the key results that we need here.

We begin with a result of Rinott~(1994)
\nocite{rinott:1994}
on the rate of convergence in a generalised Central Limit Theorem;
generalised because the summands are not identically distributed and
it allows some dependence between elements in the sum.
We do not use this second feature here, but it would be needed to
extend our results to include effects that depend on
more than one locus, and so for completeness
we include it in the statement of the result.
It also gives an idea of how quickly the rate of convergence deteriorates
if one includes epistasis or higher order dominance effects.
This result can be
used both to prove asymptotic normality when we condition only on the
pedigree (and not on any observed trait values), and to prove
asymptotic normality of the residuals (that is
the part of the trait distribution within families that is not shared
among offspring) conditional on the observed traits of ancestors in
the pedigree.

The dependence is captured by a \emph{dependency graph}.
\begin{defn}
\label{dependency graph}
Let $\{X_l; l\in {\cal V}\}$ be a collection of random variables. The graph
${\cal G}=({\cal V}, {\cal E})$, where ${\cal V}$ and ${\cal E}$ denote the vertex set
and edge set respectively, is said to be a \emph{dependency graph} for the
collection if for any pair of disjoint subsets $A_1$ and $A_2$ of ${\cal V}$ such that no
edge in ${\cal E}$ has one endpoint in $A_1$ and the other in $A_2$, the sets of
random variables $\{X_l; l\in A_1\}$ and $\{X_l; l\in A_2\}$ are independent.
\end{defn}
The degree of a vertex in the graph is the number of edges connected to it and the maximal
degree of the graph is just the maximum of the degrees of the vertices in it.
\begin{thm}[Theorem~2.2, Rinott~(1994)]
\label{rinott clt}
Let $E_1,\ldots ,E_M$ be random variables having a dependency graph whose maximal degree is strictly
less than $D$, satisfying $|E_l-\IE[E_l]|\leq B$ a.s.,~$l=1,
\ldots ,M$, $\IE[\sum_{l=1}^ME_l]=\lambda$
and $\mathtt{Var}\left(\sum_{l=1}^ME_l\right)=\sigma^2>0$. Then, for
every $w\in\IR$,
\begin{equation}
\label{CLT bound}
\left|\IP\left[\frac{\sum_{l=1}^ME_l-\lambda}{\sigma}\leq w\right]-
{\cal N}(w)\right|
\leq\frac{1}{\sigma}\left\{\sqrt{\frac{1}{2\pi}}DB+16\left(\frac{M}{\sigma^2}\right)^{1/2}D^{3/2}B^2
	+10\left(\frac{M}{\sigma^2}\right)D^2B^3\right\},
\end{equation}
where ${\cal N}$ is the distribution function of a standard normal random variable.
\end{thm}
In particular, when $D$ and $B$ are order one
and $\sigma^2$ is of order $M$,
the bound is of order $1/\sqrt{M}$.

Since we are only allowing for dominance effects that depend on allelic
states at a single locus, and we have no epistasis, our dependency
graphs will have no edges and so the maximal degree of any vertex will
be zero and we may take $D=1$.
Epistasis or higher order dominance effects, will increase the degree.
This bound on the accuracy of the normal approximation will
decrease rapidly as the number
of combinations through which the
allelic state at a single locus
can influence the trait grows.

\subsection*{Exchangeable pairs}

In order to prove the asymptotic normality of the part of the trait
value that is shared by all the offspring in a family conditional on parental traits, we require a
different approach. Because we are conditioning on the trait values
of the parents, there will be weak dependence between all the
pairs of loci within the sums defining ${\cal A}^i+{\cal D}^i$ (and so the
dependency graph for the summands would be the complete graph).
To check that nonetheless
the limit is Gaussian we shall use a variant of Stein's method of
exchangeable pairs, originally introduced in Stein~(1986).
\nocite{stein:1986}

Recall that the pair of random variables $(W, W')$ is called
an exchangeable
pair if their joint distribution is symmetric.
Suppose that $\IE[W]=0$, $\IE[W^2]=1$, $(W,W')$ is an exchangeable pair
and
\begin{equation}
\label{approximate regression}
\IE[W-W'|W]=\lambda (W-R),
\end{equation}
for some $0<\lambda <1$, where $R$ is a random variable of small order.

Let us write $\Delta =W-W'$ and define
$$\widehat{K}(t)=\frac{\Delta}{2\lambda}\Big(\mathbf{1}_{\{-\Delta\leq t\leq 0\}}
-\mathbf{1}_{\{0\leq t\leq -\Delta\}}\Big).$$
Note that $\int_{-\infty}^\infty \widehat{K}(t)dt=\Delta^2/(2\lambda)$.
\nocite{chen/goldstein/shao:2011}
In this case, one can show (see Chen et al.~2011, \S2.3) that
\begin{equation}
\label{modified Stein equation}
\IE[Wf(W)]=\IE\left[\int_{-\infty}^\infty f'(W+t)\widehat{K}(t)dt\right]
+\IE[Rf(W)].
\end{equation}

\begin{propn}[Chen et al.~2011, Proposition~2.4i]
\label{chen propn}
Let $h$ be an absolutely continuous function with $\|h'\|<\infty$,
and ${\cal F}$ any $\sigma$-algebra containing $\sigma(W)$.
If~(\ref{modified Stein equation}) holds, then
\begin{equation}
\label{bound in exchangeable pair proposition}
\left|\IE[h(W)]-{\cal N}(h)\right| \leq \|h'\|\left(\sqrt{\frac{2}{\pi}}
\IE\left[|1-\widehat{K}_1|\right]+2\IE[\widehat{K}_2]+2\IE[|R|]\right),
\end{equation}
where
\begin{equation}
\label{K1 and K2}
\widehat{K}_1=\IE\left[\left.\int_{-\infty}^\infty\widehat{K}(t)dt\right| {\cal F}\right]
=
\IE\left[\left.
\frac{\Delta^2}{2\lambda}\right|{\cal F}\right]\quad\mbox{and}\quad
\widehat{K}_2=\int_{-\infty}^\infty\left|t\widehat{K}(t)\right|dt=
\frac{|\Delta |^3}{4\lambda}
\end{equation}
\end{propn}

\begin{corollary}
\label{chen corollary}
Suppose that $(W,W')$ is an exchangeable pair with $\IE[W]=\mu_W$ and
$\mathtt{Var}(W)=\sigma_W^2$ with
\begin{equation}
\label{exchangeable pair requirement}
\IE[W'|W]=(1-\lambda)W+\lambda \IE[W]-\lambda R
\end{equation}
where $R$ is a random variable of small order. Then defining $\widehat{K}_1$,
$\widehat{K}_2$, $h$ and ${\mathcal F}$
as in Proposoition~\ref{chen propn},
\begin{equation}
\label{unnormalised difference}
\left|\IE[h(W)]-{\cal N}_{\mu_W,\sigma_W^2} (h)\right|
\leq \|h'\|\left(\sqrt{\frac{2}{\pi}}\frac{1}{\sigma_W}
\IE\left[|\sigma_W^2-\widehat{K}_1|\right]+
\frac{2}{\sigma_W^2}\IE[\widehat{K}_2]+2\IE[|R|]\right),
\end{equation}
where ${\cal N}_{\mu_W,\sigma_W^2}$ denotes the distribution of a normal random
variable with mean $\mu_W$ and variance $\sigma_W^2$.
\end{corollary}

\begin{remark}
Although this result is enough to guarantee that $W$ is asymptotically
normal, because we require $\|h'\|<\infty$,
it is not enough to bound even the
distance between the cumulative distribution function
of $W$ and that of a standard normal random variable with
an error of order $1/\sqrt{M}$.
To propagate our argument from one generation to the next
requires convergence of the density function of the observed trait value, and
once again it is our
assumption that there is some environmental noise (with
a smooth density) that allows us to guarantee this convergence
based on the result proved here.
\end{remark}

\section{Key Lemmas}
\label{key lemmas}

\begin{notn}
\label{noise in notation}
Throughout the rest of the appendices, to ease the notation we shall assume that the
(Gaussian) environmental noise is subsumed into the trait value $Z$, so that
its distribution can be assumed to have a smooth density. That is, what we call $Z$ below is the observed trait $\widetilde{Z}$ discussed in the main text. Moreover, when we write $\IP[Z=z]$, we actually mean the density function of the distribution of $\widetilde{Z}$ evaluated at the value $z$ (in formula, $\IP[Z=z]:=\varphi_{\widetilde{Z}}(z)$ with $\varphi_{\widetilde{Z}}$ the density of $\widetilde{Z}$). This notation allows us to cover both the case when the allelic distributions are general (potentially concentrated on a finite number of values) and the environmental component is smooth enough that the distribution of their sum is also smooth, and the case when there is no environmental noise but the scaled allelic distributions have a smooth density over [-B,B] (in which case the distribution of the genetic component $Z$ is itself smooth enough for the method below to be employed).
\end{notn}
In this section we prove two key lemmas which will underpin our proof.
They will allow us to estimate the effect on the distribution of the
allelic types at a particular locus, or particular
pair of loci, of knowing the
trait value. We shall be using Bayes' rule. With a slight abuse of notation
$$\IP[(\chi_l^1,\chi_l^2)=(x,x')|Z=z]=
\frac{\IP[Z=z|(\chi_l^1,\chi_l^2)=(x,x')]}{\IP[Z=z]}\IP[(\chi_l^1,\chi_l^2)=
(x,x')].$$
Let us write $\Psi_l(x,x')=\eta_l(x)+\eta_l(x')+\phi_l(x,x')$
and $Z_{-l}$ for the trait value of an individual with the effect of
locus $l$ removed, then the ratio in this expression becomes
$$\frac{\IP[Z_{-l}=z-\Psi_l(x,x')]}{\IP[Z=z]}.$$
Of course this ratio of probabilities should be interpreted as a
ratio of density functions. Moreover, bearing in mind
our remarks on environmental noise, we are going to suppose that these
density functions are sufficiently smooth that we can justify an
application of Taylor's Theorem.
Of course, we know that $Z_{-l}$ is approximately normally distributed,
using exactly the same argument as for $Z$, and it is no surprise that the
ratio differs from one by something of order $1/\sqrt{M}$. The
importance of the next lemma will become evident when we sum conditional
expectations over loci; \emph{c.f.}~Remark~\ref{remark on troublesome}.

\begin{lemma}
\label{replacement for troublesome equation}
In the notation above,
\begin{multline*}
\IP[Z_{-l}=z]=\IP[Z=z]+\frac{1}{\sqrt{M}}\IE[\Psi_l(\chi_l^1,\chi_l^2)]
\frac{d}{dz}\IP[Z=z]\\
+\frac{1}{M}\IE[\Psi_l(\chi_l^1,\chi_l^2)]^2\frac{d^2}{dz^2}\IP[Z=z]
-\frac{1}{2M}\IE[\Psi_l(\chi_l^1,\chi_l^2)^2]\frac{d^2}{dz^2}\IP[Z=z]
+C_l(z)\frac{1}{M^{3/2}},
\end{multline*}
where the function $C_l(z)$ in the error term can be bounded independent of
$l$ and $z$.
\end{lemma}
\begin{remark}[Conditioning on the pedigree]
Although we have suppressed it in the notation, this lemma holds
in any generation, but the expressions
$\IE[\Psi(\chi_l^1,\chi_l^2)]^2$ and
$\IE[\Psi(\chi_l^1,\chi_l^2)^2]$ should be interpreted as being
calculated conditional on the pedigree (which will determine the
probability of identity of $\chi_l^1$, $\chi_l^2$).
\end{remark}
\noindent
{\bf Proof of Lemma~\ref{replacement for troublesome equation}}

We are going to abuse notation (still further) and imagine that
$\IP[\chi_l^1=x, \chi_l^2=x', Z=z]$ has a density with respect to $x$, $x'$.
Of course we do not expect that to be true (even with environmental noise),
but it makes our expressions
easier to parse than using a more mathematically accurate notation.
We begin with an application of Taylor's Theorem (with respect to $z$):
\begin{eqnarray}
\label{Z-l as integral}
\IP[Z_{-l}=z] &=& \int\int\IP\bigg[\chi_l^1=x, \chi_l^2=x',
Z=z+\frac{1}{\sqrt{M}}\Psi_l(x,x')\bigg]dxdx'\\
\label{first term Z-l as integral}
&=&
\int\int\IP[\chi_l^1=x, \chi_l^2=x', Z=z]dxdx'
\\
\label{second term Z-l as integral}
&&
+\frac{1}{\sqrt{M}}\int\int\Psi_l(x,x')\frac{\partial}{\partial z}
\IP[\chi_l^1=x, \chi_l^2=x', Z=z]dxdx'
\\
\label{third term Z-l as integral}
&&
+\frac{1}{2M}\int\int\Psi_l(x,x')^2\frac{\partial^2}{\partial z^2}
\IP[\chi_l^1=x, \chi_l^2=x', Z=z]dxdx'
+\widehat{C}_l(z)\frac{1}{M^{3/2}}.
\end{eqnarray}
Provided that $\IP[Z=z]$ has a uniformly bounded third derivative, our
assumption that the terms that make up $\Psi_l$ are uniformly bounded allows
us to deduce that $\widehat{C}_l$ is uniformly bounded in $l$ and $z$.
Notice that the expression in~(\ref{first term Z-l as integral})
is just $\IP[Z=z]$.

Since we are not conditioning on any trait values in
the pedigree, and the ancestral population is assumed to be in
linkage equilibrium, $(\chi_l^1,\chi_l^2)$ and $Z_{-l}$ are
independent.
Combining this observation with Eq.~(\ref{Z-l as integral}), and, once again
applying Taylor's Theorem, we find
\begin{align*}
&\IP[\chi_l^1=x, \chi_l^2=x', Z=z]\\
& = 
\IP[\chi_l^1=x, \chi_l^2=x']\IP\bigg[Z_{-l}=z-\frac{1}{\sqrt{M}}\Psi_l(x,x')\bigg]
\\
&=
\IP[\chi_l^1=x, \chi_l^2=x']\int\int
\IP\bigg[\chi^1_l=y, \chi^2_l=y', Z=z-\frac{1}{\sqrt{M}}\Psi_l(x,x')
+\frac{1}{\sqrt{M}}\Psi_l(y,y')\bigg]dydy'
\\
&=
\IP[\chi_l^1=x, \chi_l^2=x']\Big\{
\IP[Z=z]+\frac{1}{\sqrt{M}}\int\int\big(\Psi(y,y')-\Psi(x,x')\big)
\\
&
\qquad\qquad
\qquad\qquad
\qquad\qquad
\qquad\qquad
\times
\frac{\partial}{\partial z}\IP[\chi^1_l=y, \chi^2_l=y', Z=z]dydy'
+\widetilde{C}_l(x,x',z)\frac{1}{M}\Big\},
\end{align*}
where the function $\widetilde{C}_l$ in the last line is
uniformly bounded independent
of $l$ and $(x,x',z)$.
(To justify this last statement, recall that we are abusing notation and
implicitly subsuming the environmental noise into the distribution of $Z$. The density
function here is actually a convolution of that of the environmental noise, which is
smooth, and the true distribution of $Z$, and is therefore smooth.)
Still assuming sufficient regularity, differentiating the previous equation
we find
\begin{align}
\label{first deriv}
 \frac{\partial}{\partial z}\IP[&\chi_l^1=x, \chi_l^2=x', Z=z] \nonumber \\
& = \IP[\chi_l^1=x, \chi_l^2=x']\bigg\{
\frac{d}{d z}\IP[Z=z]
+\frac{1}{\sqrt{M}}\int\int\Big(\Psi(y,y')-\Psi(x,x')\Big) \nonumber\\
&\qquad \qquad \qquad \qquad \times
\frac{\partial^2}{\partial z^2}\IP[\chi^1_l=y, \chi^2_l=y', Z=z]dydy'
+\frac{\partial}{\partial z}\widetilde{C}_l(x,x',z)\frac{1}{M}\bigg\},
\end{align}
and
\begin{equation}
\label{2nd deriv}
\frac{\partial^2}{\partial z^2}\IP[\chi_l^1=x, \chi_l^2=x', Z=z]
=
\IP[\chi_l^1=x, \chi_l^2=x']\left\{
\frac{d^2}{d z^2}\IP[Z=z]
+\frac{1}{\sqrt{M}}\frac{\partial^2}{\partial z^2}\widetilde{\widetilde{C}}_l(x,x',z)
\right\},
\end{equation}
with $\tfrac{\partial}{\partial z}\widetilde{C}_l$
and $\tfrac{\partial^2}{\partial z^2}\widetilde{\widetilde{C}}_l$ uniformly bounded.

Finally, substituting~(\ref{first deriv}) and~(\ref{2nd deriv})
in~(\ref{second term Z-l as integral}) and~(\ref{third term Z-l as integral}),
we obtain
\begin{eqnarray*}
\IP[Z_{-l}=z]&=&\IP[Z=z]+
\frac{1}{\sqrt{M}}\int\int\Psi_l(x,x')\IP[\chi^1_l=x, \chi^2_l=x']dxdx'
\frac{d}{d z}\IP[Z=z]
\\
&&+\frac{1}{M}\int\int\int\int\Big(\Psi_l(y,y')-\Psi_l(x,x')\Big)\Psi_l(x,x')
\\
&&\qquad\qquad\times \IP[\chi^1_l=y, \chi^2_l=y']\IP[\chi^1_l=x, \chi^2_l=x']
dydy'dxdx'\frac{d^2}{d z^2}\IP[Z=z]
\\
&&
+\frac{1}{2M}\int\int\Psi_l(x,x')^2\IP[\chi^1_l=x, \chi^2_l=x']dxdx'
\frac{d^2}{d z^2}\IP[Z=z]
+\widehat{C}_l(z)\frac{1}{M^{3/2}}
\\
&=&
\IP[Z=z]+\frac{1}{\sqrt{M}}\IE[\Psi_l(\chi_l^1,\chi_l^2)]
\frac{d}{dz}\IP[Z=z]\\
&&+\frac{1}{M}\IE[\Psi_l(\chi_l^1,\chi_l^2)]^2\frac{d^2}{dz^2}\IP[Z=z]
-\frac{1}{2M}\IE[\Psi_l(\chi_l^1,\chi_l^2)^2]\frac{d^2}{dz^2}\IP[Z=z]
+\widehat{C}_l(z)\frac{1}{M^{3/2}},
\end{eqnarray*}
as required. \hfill $\Box$

We also require an analogue of Lemma~\ref{replacement for troublesome equation}
with which to control the effect of conditioning on the trait
value on the distribution of the allelic values at pairs of loci.
We write $Z_{-l-m}=Z-\tfrac{1}{\sqrt{M}}\big(
\Psi_l(\chi^1_l,\chi^2_l)+\Psi_m(\chi^1_m,\chi^2_m)\big)$
for the trait value with the contributions from loci $l$ and $m$
removed. The following lemma follows on iterating the argument that gave
us Lemma~\ref{replacement for troublesome equation}.
\begin{lemma}
\label{Z-l-m}
In the notation above,
\begin{align*}
\IP[Z_{-l-m}=z]=& \IP[Z=z]+
\frac{1}{\sqrt{M}}\Big(\IE[\Psi_l(\chi_l^1,\chi_l^2)]+\IE[\Psi_m(\chi_m^1,\chi_m^2)]\Big)
\frac{d}{d z}\IP[Z=z] \nonumber\\
& +\Big\{\frac{1}{M}\IE[\Psi_l(\chi_l^1,\chi_l^2)]^2-\frac{1}{2M}\IE[\Psi_l(\chi_l^1,\chi_l^2)^2]
+\frac{1}{M}\IE[\Psi_l(\chi_l^1,\chi_l^2)]\IE[\Psi_m(\chi_m^1,\chi_m^2)]
\nonumber\\
&  +\frac{1}{M}\IE[\Psi_m(\chi_m^1,\chi_m^2)]^2-\frac{1}{2M}\IE[\Psi_m(\chi_m^1,\chi_m^2)^2]\Big\}
\frac{d^2}{d z^2}\IP[Z=z]+C_{l,m}(z)\frac{1}{M^{3/2}},
\end{align*}
where the functions $C_{l,m}(z)$ are uniformly bounded in $l$, $m$, $z$.
\end{lemma}
{\bf Proof}

We iterate the previous result:
\begin{multline*}
\IP[Z_{-l-m}=z]=\IP[Z_{-l}=z]
+\frac{1}{\sqrt{M}}\IE[\Psi_m(\chi_m^1,\chi_m^2)]
\frac{d}{dz}\IP[Z_{-l}=z]\\
+\frac{1}{M}\IE[\Psi_m(\chi_m^1,\chi_m^2)]^2\frac{d^2}{dz^2}\IP[Z_{-l}=z]
-\frac{1}{2M}\IE[\Psi_m(\chi_m^1,\chi_m^2)^2]\frac{d^2}{dz^2}\IP[Z_{-l}=z]
+C_m(z)\frac{1}{M^{3/2}};
\end{multline*}
now substitute for $\IP[Z_{-l}=z]$ and its derivatives.
\hfill $\Box$

\begin{remark}
\label{remark on troublesome}
Just as for Lemma~\ref{replacement for troublesome equation}, the
proof of Lemma~\ref{Z-l-m} applies in any generation as long as one
interprets the expectations as being taken conditional on the pedigree.
We have assumed that our base population is in linkage equilibrium to
write $\IE[\Psi_l(y,y')\Psi_m(x,x')]=\IE[\Psi_l(y,y')]\IE[\Psi_m(x,x')]$.
\end{remark}

We shall only be presenting the detailed proofs for individuals in
generation one. To extend to the general case requires an analogue of
Lemma~\ref{replacement for troublesome equation}
when we consider the trait values of the two parents of an individual.
For completeness, we record that lemma here.
\begin{lemma}
\label{troublesome lemma with two parents}
Let us use $\IP[z_1,z_2]$ to denote
$\IP[Z^{i[1]}=z_1,Z^{i[2]}=z_2]$.
In the following expression, all expectations should be
interpreted as taken conditional on the pedigree:
\begin{align*}
 \IP\Big[Z^{i[1]}_{-l}=z_1& , Z^{i[2]}_{-l}=z_2\Big]- \IP[z_1,z_2]
\\ = &\,\frac{1}{\sqrt{M}}\IE\big[\Psi_l(\chi_l^{i[1],1},\chi_l^{i[1],2})\big]\frac{\partial}{\partial z_1}\IP[z_1,z_2]
+\frac{1}{\sqrt{M}}\IE\big[\Psi_l(\chi_l^{i[2],1},\chi_l^{i[2],2})\big]\frac{\partial}{\partial z_2}\IP[z_1,z_2]
\\
& +\left(\frac{1}{M}
\IE\big[\Psi_l(\chi_l^{i[1],1},\chi_l^{i[1],2})\big]^2
-\frac{1}{2M}
\IE\big[\Psi_l(\chi_l^{i[1],1},\chi_l^{i[1],2})^2\big]\right)
\frac{\partial^2}{\partial z_1^2}\IP[z_1,z_2]
\\& +\left(\frac{1}{M}
\IE\big[\Psi_l(\chi_l^{i[2],1},\chi_l^{i[2],2})\big]^2
-\frac{1}{2M}
\IE\big[\Psi_l(\chi_l^{i[2],1},\chi_l^{i[2],2})^2\big]\right)
\frac{\partial^2}{\partial z_2^2}\IP[z_1,z_2]
\\ & +\Bigg(\frac{2}{M}
\IE\big[\Psi_l(\chi_l^{i[1],1},\chi_l^{i[1],2})\big]
\IE\big[\Psi_l(\chi_l^{i[2],1},\chi_l^{i[2],2})\big]
\\
& \qquad -\frac{1}{M}
\IE\big[\Psi_l(\chi_l^{i[1],1},\chi_l^{i[1],2})
\Psi_l(\chi_l^{i[2],1},\chi_l^{i[2],2})\big]\Bigg)
\frac{\partial^2}{\partial z_1\partial z_2}\IP[z_1,z_2]
+\mathcal{O}\Big(\frac{1}{M^{3/2}}\Big).
\end{align*}
\end{lemma}

\section{Mean and variance of trait values conditional on parental traits}\label{cond mean and var}

We remind the reader that Notation~\ref{noise in notation} remains in force.

We now turn to calculating the conditional distribution of the trait values,
conditional not just on the pedigree, as we did in
Appendix~\ref{QG derivations},
but also on the (observed) trait values in the parental generation.
We spell out the details in generation one.
Here already we can identify the key points, without being
overwhelmed by notation. Recall that we are implicitly conditioning not
on the exact trait values of the parents, but on the observed trait
values when environmental noise is taken into account,
so that we can assume that the distribution of parental trait
values has a smooth density.

First we calculate the conditional mean.
We distinguish the case of two
distinct parents and a family produced by selfing.
Recall that we wrote ${\cal A}^i+{\cal D}^i$ for the component shared by all
individuals in the family, with ${\cal A}^i$ and ${\cal D}^i$ defined
in~(\ref{defn of A}) and~(\ref{defn of D}).

\subsubsection*{Generation one: mean trait value, distinct parents}

Since the parents are, by
assumption, unrelated, we anticipate that
the expected value of the dominance component
is zero, and so the expected value of the shared component
${\cal A}^i+{\cal D}^i$ should be the mean value of
the parental traits. However, since we are conditioning on knowing the trait
values, we do have some information about the allelic types, and we must verify
that this does not significantly distort the expectations.

We exploit again the fact that since the parents are unrelated, their
trait values (and allelic states at locus $l$) are independent. Thus
\begin{align}
\label{independence of trait values}
&\IP\left[ Z^{i[1]}=z_1, Z^{i[2]}=z_2 \Big|
\big(\chi_l^{i[1],1},\chi_l^{i[1],2}\chi_l^{i[2],1}\chi_l^{i[2],2}\big)
=(x,x',y,y')\right]
\\
& =\!
\IP\!\left[Z_{-l}^{i[1]}=z_1-\frac{1}{\sqrt{M}}\Big(\eta_l(x)+\eta_l(x')+\phi_l(x,x')
\Big)\right]\!
\IP\!\left[Z_{-l}^{i[2]}=z_2-\frac{1}{\sqrt{M}}\Big(\eta_l(y)+\eta_l(y')+\phi_l(y,y')
\Big)\right]\!. \nonumber
\end{align}

We now use Lemma~\ref{replacement for troublesome equation}
and Taylor's Theorem to deduce that
\begin{multline*}
\IP\left[\left. (\chi_l^{i[1],1}, \chi_l^{i[1],2})=(x,x') \right|
Z^{i[1]}=z\right]=
\IP\left[(\chi_l^{i[1],1}, \chi_l^{i[1],2})=(x,x')\right]
\\
\times
\left\{1-\frac{1}{\sqrt{M}}\Big(\Psi_l(x,x')-\IE[\Psi_l]\Big)
\frac{1}{\IP[Z^{i[1]}=z]}
\frac{d}{d z}\IP[Z^{i[1]}=z]+\mathcal{O} \Big(\frac{1}{M}\Big)
\right\},
\end{multline*}
with a symmetric expression for $i[2]$.
Integrating against this expression and
using~(\ref{vanishing cross variation}),
(\ref{mean phi zero}), and~(\ref{mean eta zero}),
we find, in an obvious notation,
\begin{equation}
\label{conditional dist eta gen one}
\IE\left[\left.\eta_l(\chi_l^{i[1],1})\right|i[1]\neq i[2], Z^{i[1]},
Z^{i[2]}\right]
=-\frac{1}{\sqrt{M}}\frac{\IP'[Z^{i[1]}]}{\IP[Z^{i[1]}]}
\mathtt{Var}\left(\eta_l(\widehat{\chi}_l)\right)
+{\mathcal O}\left(\frac{1}{M}\right).
\end{equation}
Note that approximating $\IP[Z^{i[1]}]$ by a normal density and ignoring the
environmental component,
the order $1/M$ terms involves $1/(\sigma_A^2+\sigma_D^2)$
and $\big(Z^{i[1]}-\bar{z}_0\big)^2/(\sigma_A^2+\sigma_D^2)$,
and is controlled through these quantities and
our bounds on $\eta_l$ and $\phi_l$.
In particular, the approximation breaks down if the genetic variance
is too small or if the trait of the parent is too extreme.
Multiplying by $1/\sqrt{M}$ and summing over loci and parents, we arrive at
\begin{equation}
\label{mean additive bit gen one}
\IE\left[\left. {\cal A}^i\right|i[1]\neq i[2], Z^{i[1]}, Z^{i[2]}\right]
=-\left(\frac{\IP'[Z^{i[1]}]}{\IP[Z^{i[1]}]}
+\frac{\IP'[Z^{i[2]}]}{\IP[Z^{i[2]}]}\right)
\frac{1}{M}\sum_{l=1}^M
\mathtt{Var}\left(\eta_l(\widehat{\chi}_l)\right)
+{\mathcal O}\left(\frac{1}{\sqrt{M}}\right).
\end{equation}

\begin{remark}
Since we already checked that
the trait $Z^{i[1]}$ is approximately normally distributed, and the same
argument evidently gives that $Z^{i[1]}_{-l}$ is approximately
normally distributed for each $l$, the derivation
above may seem unnecessarily complex. However,
in summing the terms in~\eqref{conditional dist eta gen one} over loci,
we exploited the fact that we could pull the ratio
$\IP'[Z^{i[1]}]/\IP[Z^{i[1]}]$ outside
the sum. Only then did we approximate it by the limiting normal
distribution. We could only do this because we expressed everything in terms
of the distribution of the whole trait. If we try to approximate
the distribution of $Z^{i[1]}_{-l}$ directly by a normal distribution, and
then sum, we cannot control the error. We shall use this trick
repeatedly in what follows.
\end{remark}

Similarly,
\begin{multline*}
\IE\left[\left.\phi_l(\chi_l^{i[1],1}, \chi_l^{i[2],1})\right|
i[1]\neq i[2], Z^{i[1]}=z_1,
Z^{i[2]}=z_2\right]\\
=
\int\ldots\int \phi_l(x,y)
\Big\{1-\frac{1}{\sqrt{M}}\Big(\Psi_l(x,x')-\IE[\Psi_l]\Big)
\frac{1}{\IP[Z^{i[1]}=z_1]}\frac{d}{d z_1}\IP[Z^{i[1]}=z_1]\Big\}
\\
\Big\{1-\frac{1}{\sqrt{M}}\Big(\Psi_l(y,y')-\IE[\Psi_l]\Big)
\frac{1}{\IP[Z^{i[2]}=z_2]}\frac{d}{d z_2}\IP[Z^{i[2]}=z_2]\Big\}
\\
\widehat{\nu}_l(dx)\widehat{\nu}_l(dx')
\widehat{\nu}_l(dy)\widehat{\nu}_l(dy')
+{\mathcal O}\left(\frac{1}{M}\right).
\end{multline*}
The terms of order one and $1/\sqrt{M}$ vanish as a result of
Eq.~(\ref{vanishing conditional distribution phi}),
(\ref{vanishing cross variation}),
(\ref{mean phi zero}),
and
(\ref{mean eta zero}).
Multiplying by $1/\sqrt{M}$ and summing over loci, we find that
$\IE[{\cal D}^i]={\mathcal O}(1/\sqrt{M})$.

Recalling that the trait distribution in the ancestral population is
(almost) normally distributed with mean $\bar{z}_0$, we see that if we ignore
environmental effects, so that the variance of the trait distribution in
generation zero is $\sigma_A^2+\sigma_D^2$, then
adding $\bar{z}_0$ to the right hand side
of~(\ref{mean additive bit gen one}),
and substituting
$$\IP[Z^{i[1]}]=\frac{1}{\sqrt{2\pi(\sigma_A^2+\sigma_D^2)}}
\exp\left(-\frac{(Z^{i[1]}-\bar{z}_0)^2}{2(\sigma_A^2+\sigma_D^2)}\right),$$
we recover that up to an error of order $1/\sqrt{M}$,
the expected trait value among offspring is
$$\bar{z}_0+\frac{\sigma_A^2}{\sigma_A^2+\sigma_D^2}\left(
\frac{Z^{i[1]}+Z^{i[2]}}{2}-\bar{z}_0\right),$$
as predicted by Theorem~\ref{conditioning multivariate normals}.

\begin{remark}[The breeder's equation]
\label{remark on breeder's equation}
Suppose that
as a result of environmental noise, the
observed trait of each individual in
the ancestral population is its genetic trait plus an independent
${\cal N}(0,\sigma_E^2)$ random variable.
Then assuming normality of the ancestral trait distribution, and
using
Theorem~\ref{conditioning multivariate normals},
we find that for unrelated parents
the mean trait
in generation one is
\begin{equation}
\label{breeder's equation}
\bar{z}_0+\frac{\sigma_A^2}{\sigma_Z^2}
\left(\frac{(Z^{i[1]}+Z^{i[2]})}{2}-\bar{z}_0\right),
\end{equation}
where $\sigma_Z^2$ is the total variance of the \emph{observed} trait
in the
ancestral population; that is $\sigma_Z^2=\sigma_A^2+\sigma_D^2+\sigma_E^2$.
Eq.~\eqref{breeder's equation} is
the \emph{breeder's equation}.
\end{remark}

\subsubsection*{Mean trait value, same parent}

We now turn to the expected trait value in
a family in generation one that
is produced by selfing.  The calculation for the additive term is
unchanged,
but now we have a non-trivial contribution from the
dominance component. We denote the parent $Z^{i[1]}$.
Since $Z^{i[1]}=Z^{i[2]}$, we must
calculate $\IE[\phi_l(\chi_l^{i[1],1}, \chi_l^{i[1],1})|Z^{i[1]}]$
and
$\IE[\phi_l(\chi_l^{i[1],1}, \chi_l^{i[1],2})|Z^{i[1]}]$.

Our strategy is as before: we express each of these probabilities
in terms of the distribution of the trait value minus the contribution
from locus $l$ and we apply
Lemma~\ref{replacement for troublesome equation}.
Thus, once again using that in generation zero, before conditioning, the
two alleles at locus $l$ in $Z^{i[1]}$ are independent draws
from $\widehat{\nu}_l$,
\begin{multline*}
\IE\left[\left.\phi_l(\chi_l^{i[1],1}, \chi_l^{i[1],1})\right|
i[1]=i[2], Z^{i[1]}=z\right]
=
\int\int
\phi_l(x,x)\\
\left(1-\frac{1}{\sqrt{M}}\Big(\Psi_l(x,x')-\IE[\Psi_l]\Big)
\frac{1}{\IP[Z^{i[1]}=z]}\frac{d}{d z}\IP[Z^{i[1]}=z]\right)
\widehat{\nu}_l(dx)\widehat{\nu}_l(dx')
+{\mathcal O}\left(\frac{1}{M}\right).
\end{multline*}
Using Eq.~(\ref{vanishing cross variation}), (\ref{mean phi zero}), (\ref{mean eta zero}), we
see that on integration the only non-zero contribution comes from the term
$\eta_l(x)\phi_l(x,x)$
which can be integrated to yield
\begin{equation}
\IE\big[\phi_l(\chi_l^{i[1],1}, \chi_l^{i[1],1})|i[1]=i[2], Z^{i[1]}\big]
=
\IE[\phi_l(\widehat{\chi}_l,\widehat{\chi}_l)]
-\frac{1}{\sqrt{M}}\frac{\IP'[Z^{i[1]}]}{\IP[Z^{i[1]}]}
\IE[\eta_l(\widehat{\chi}_l)\phi_l(\widehat{\chi}_l,\widehat{\chi}_l)]
+{\mathcal O}\left(\frac{1}{M}\right).
\end{equation}
Similarly,
\begin{equation}
\IE\Big[\phi_l(\chi_l^{i[1],1}, \chi_l^{i[1],2})|i[1]=i[2], Z^{i[1]}\Big]
=
-\frac{1}{\sqrt{M}}\frac{\IP'[Z^{i[1]}]}{\IP[Z^{i[1]}]}
\IE[\phi_l(\widehat{\chi}_l,\widehat{\chi}_2)^2]
+{\mathcal O}\left(\frac{1}{M}\right).
\end{equation}

Multiplying by $1/\sqrt{M}$ and summing over loci, we find that the mean
of the term ${\cal D}^i$ in~(\ref{defn of D}), conditional on $i[1]=i[2]$ and
on knowing the trait value $Z^{i[1]}$, is
\begin{equation}
\frac{1}{2\sqrt{M}}\!\sum_{l=1}^M\IE[\phi_l(\widehat{\chi}_l,\widehat{\chi}_l)]
-\frac{\IP'[Z^{i[1]}]}{\IP[Z^{i[1]}]}
\left(\frac{1}{2M}\sum_{l=1}^M
\IE[\eta_l(\widehat{\chi}_l)\phi_l(\widehat{\chi}_l,\widehat{\chi}_l)]
+\frac{1}{2M}\sum_{l=1}^M
\IE[\phi_l(\widehat{\chi}_l,\widehat{\chi}_2)^2]\right)\!
+{\mathcal O}\!\left(\frac{1}{\sqrt{M}}\right).
\end{equation}
Adding on the additive terms that we calculated before and
restating everything in terms of the quantities in
Table~\ref{QG coefficients},
we obtain that for two identical parents
\begin{equation}
\label{mean gen one identical parents}
\IE[Z^i|i[1]=i[2], Z^{i[1]}]=\bar{z}_0+\left(\frac{1}{2}\iota
-\frac{\IP'[Z^{i[1]}]}{\IP[Z^{i[1]}]}\left(\sigma_A^2+
\frac{\sigma_D^2}{2} +\frac{\sigma_{ADI}}{4} \right)\right)
+{\mathcal O}\left(\frac{1}{\sqrt{M}}\right).
\end{equation}
Notice that the factor of $1/2$ in front of $\iota$ is the probability
of identity $F^*$ of the two genes in the offspring.

Of course there is no surprise here: $\IE[({\cal A}^i+{\cal D}^i)|i[1]=i[2]]=\iota/2$
and
\begin{align}
\label{conditioned covariances}
&\mathtt{Cov}\big({\cal A}^i+{\cal D}^i, Z^{i[1]}|i[1]=i[2]\big) \nonumber \\
& = \frac{1}{M}\sum_{l=1}^M
\IE\Bigg[\Big(\eta_l(\widehat{\chi}_l^1)+\eta_l(\widehat{\chi}_2^1)
+\frac{\phi(\widehat{\chi}_l^1,\widehat{\chi}_l^1)+
\phi(\widehat{\chi}_l^2,\widehat{\chi}_l^2)+
2\phi(\widehat{\chi}_l^1,\widehat{\chi}_l^2)}{4}\Big)
\Big(\eta(\widehat{\chi}_l^1)+\eta(\widehat{\chi}_l^2)+
\phi(\widehat{\chi}_l^1,\widehat{\chi}_l^2)\Big)\Bigg]
\nonumber\\
& =\sigma_A^2+\frac{\sigma_D^2}{2}+\frac{\sigma_{ADI}}{4}.
\end{align}
Thus, up to the error term, (\ref{mean gen one identical parents}) is just
$$\bar{z}_0+\IE[{\cal A}^i+{\cal D}^i]+\mathtt{Cov}\big({\cal A}^i+{\cal D}^i,Z^{i[1]}\big)\frac{\big(Z^{i[1]}-
\IE[Z^{i[1]}]\big)}{\mathtt{Var}\big(Z^{i[1]}\big)},$$
as we expect from the (approximately) bivariate normal
distribution of $({\cal A}^i+{\cal D}^i)$ and $Z^{i[1]}$.

\subsection*{Variance of the shared parental contribution ${\cal A}^i+{\cal D}^i$, generation
one}

We now turn to the variance of the shared parental contribution. This is
where the complications associated with incorporating dominance really
start to be felt.
In the process of calculating the conditional mean above, we established
that conditioning on the parental trait values (and whether or not
they are identical) distorts the distribution of the allelic state at
a given locus by a factor of order $1/\sqrt{M}$. This distortion is
enough to shift the mean trait (as we see in the
breeder's equation), and, as we shall see, the variance of the
sum over loci will have a contribution from linkage disequilibrium.

\subsection*{Conditional variance $({\cal A}^i+{\cal D}^i)$, generation one, same parent}

First we consider the case in which the parents are the same.
We need to calculate the expectation of $({\cal A}^i+{\cal D}^i)^2$ conditional
upon the parental trait. We begin with the `diagonal' terms, corresponding
to a single locus.
We take these in three parts.
First, proceeding as before,
\begin{align}
\label{diagonal terms behave well}
& \IE\left[\eta_l(\chi_l^{i[1],1})^2\Big|i[1]=i[2], Z^{i[1]}=z\right] \nonumber\\
& =\int\int \eta_l(x)^2 \left(1-\frac{1}{\sqrt{M}}\Big(\Psi_l(x,x')-\IE[\Psi_l]\Big)
\frac{1}{\IP[Z^{i[1]}=z]}\frac{d}{dz}\IP[Z^{i[1]}=z]\right)
\widehat{\nu}_l(dx)\widehat{\nu}_l(dx')
+{\mathcal O}\left(\frac{1}{M}\right)
\nonumber\\
& =
\IE[\eta_l(\widehat{\chi}_l)^2]
+{\mathcal O}\left(\frac{1}{\sqrt{M}}\right).
\end{align}
Notice that the term arising from the Taylor expansion is already of order
$1/\sqrt{M}$, and, since we multiply each of the terms in the sum by $1/M$,
we have no need to develop the expansion further. Indeed, all terms in the
expression for the variance will be multiplied by $1/M$ and so for the
`diagonal' terms in the square of the sum, we only
need an expression to leading order.
\begin{remark}
The error that we are making in discarding the terms arising from
the Taylor expansion is $1/\sqrt{M}$ multiplied by a term that depends
on
$\IP'[Z^{i[1]}]/\IP[Z^{i[1]}]=-(Z^{i[1]}-\IE[Z^{i[1]}])/\mathtt{Var}(Z^{i[1]})$.
As usual, the approximation will be poor if the trait value of the parent
is too extreme, or the variance is too small.
\end{remark}
As a result, for these terms we can calculate with respect to the
distribution in the ancestral population and we find
\begin{eqnarray*}
\frac{1}{M}\IE\left[\left.\sum_{l=1}^M\left(\eta_l(\chi_l^{i[1],1})+
\eta_l(\chi_l^{i[1],2})\right)^2\right|i[1]=i[2], Z^{i[1]}\right]
&=&\frac{2}{M}\sum_{l=1}^M\IE\left[\eta_l(\widehat{\chi}_l)^2\right]
+{\mathcal O}\left(\frac{1}{\sqrt{M}}\right)
\\
&=&
\sigma_A^2+{\mathcal O}\left(\frac{1}{\sqrt{M}}\right).
\end{eqnarray*}
Similarly, recalling that we are still considering the case of
identical parents,
\begin{eqnarray*}
&&\frac{1}{16M}\IE\left[\left.\sum_{l=1}^M\left(
\phi_l(\chi_l^{i[1],1}, \chi_l^{i[1],1})
+2\phi_l(\chi_l^{i[1],1}, \chi_l^{i[1],2})
+\phi_l(\chi_l^{i[1],2}, \chi_l^{i[1],2})\right)^2\right|
i[1]=i[2], Z^{i[1]}\right]
\\
&=&
\frac{1}{16 M}\sum_{l=1}^M\left(2\IE[\phi_l(\widehat{\chi}_l,
\widehat{\chi}_l)^2]
+4\IE[\phi_l(\widehat{\chi}^1_l,\widehat{\chi}^2_l)^2]\right)
+{\mathcal O}\left(\frac{1}{\sqrt{M}}\right)
\\
&=&\frac{1}{8}\left(\sigma_{DI}^2+\iota^*\right)+\frac{1}{4}\sigma_D^2
+{\mathcal O}\left(\frac{1}{\sqrt{M}}\right),
\end{eqnarray*}
and
\begin{eqnarray*}
&&\frac{1}{2M}\IE\Bigg[\sum_{l=1}^M\Bigg(\eta_l(\chi_l^{i[1],1})+
\eta_l(\chi_l^{i[1],2})\Bigg)\\
&&\times
\Bigg(\phi_l(\chi_l^{i[1],1}, \chi_l^{i[1],1})
+2\phi_l(\chi_l^{i[1],1}, \chi_l^{i[1],2})
+\phi_l(\chi_l^{i[1],2}, \chi_l^{i[1],2})\Bigg)\Bigg|
i[1]=i[2], Z^{i[1]}\Bigg]
\\
&=&
\frac{1}{2M}\IE\Bigg[\sum_{l=1}^M\Bigg(\eta_l(\chi_l^{i[1],1})+
\eta_l(\chi_l^{i[1],2})\Bigg)\\
&&\times \Bigg(
\phi_l(\chi_l^{i[1],1}, \chi_l^{i[1],1})
+2\phi_l(\chi_l^{i[1],1}, \chi_l^{i[1],2})
+\phi_l(\chi_l^{i[1],2}, \chi_l^{i[1],2})\Bigg)\Bigg]
+{\mathcal O}\left(\frac{1}{\sqrt{M}}\right)
\\
&=&
\frac{1}{2M}\sum_{l=1}^M2\IE\left[\eta_l(\widehat{\chi}_l)
\phi_l(\widehat{\chi}_l,\widehat{\chi}_l)\right]
+{\mathcal O}\left(\frac{1}{\sqrt{M}}\right)
\\
&=& \frac{\sigma_{ADI}}{2}
+{\mathcal O}\left(\frac{1}{\sqrt{M}}\right).
\end{eqnarray*}
Combining all these terms we find that if the parents are identical, then
the contribution to $\IE[({\cal A}^i+{\cal D}^i)^2|i[1]=i[2], Z^{i[1]}]$
from the `diagonal' terms is
\begin{equation}
\label{diagonal terms A+D squared same parent}
\sigma_A^2+\frac{1}{8}\left(\sigma_{DI}^2+\iota^*\right)
+\frac{1}{4}\sigma_D^2+\frac{1}{2}\sigma_{ADI}
+{\mathcal O}\left(\frac{1}{\sqrt{M}}\right).
\end{equation}

We must now turn to the contribution from correlations across loci.
For this we must compute
\begin{eqnarray}
\label{correlations 1}
&&
\frac{1}{M}\IE\Bigg[\sum_{l\neq m}\Big(\eta_l(\chi_l^{i[1],1})
+\eta_l(\chi_l^{i[1],2})\Big)\Big(\eta_m(\chi_m^{i[1],1})+
\eta_m(\chi_m^{i[1],2})\Big)\Bigg|i[1]=i[2], Z^{i[1]}\Bigg]
\\
\label{correlations 2}
&&+
\frac{1}{2M}\IE\Bigg[\sum_{l\neq m}\Big(\eta_l(\chi_l^{i[1],1})
+\eta_l(\chi_l^{i[1],2})\Big)
\\
\nonumber
&&\qquad\qquad\qquad\times
\Big(\phi_m(\chi_m^{i[1],1},\chi_m^{i[1],1})+
\phi_m(\chi_m^{i[1],2},\chi_m^{i[1],2})+
2\phi_m(\chi_m^{i[1],1},\chi_m^{i[1],2})\Big)\Bigg|i[1]=i[2], Z^{i[1]}\Bigg]
\\
\label{correlations 3}
&&+
\frac{1}{16M}\IE\Bigg[\sum_{l\neq m}
\Big(\phi_l(\chi_l^{i[1],1},\chi_l^{i[1],1})+
\phi_l(\chi_l^{i[1],2},\chi_l^{i[1],2})+
2\phi_l(\chi_l^{i[1],1},\chi_l^{i[1],2})\Big)
\\
\nonumber
&&\qquad\qquad\qquad\times
\Big(\phi_m(\chi_m^{i[1],1},\chi_m^{i[1],1})+
\phi_m(\chi_m^{i[1],2},\chi_m^{i[1],2})+
2\phi_m(\chi_m^{i[1],1},\chi_m^{i[1],2})\Big)
\Bigg|i[1]=i[2], Z^{i[1]}\Bigg].
\end{eqnarray}

This time we use Lemma~\ref{Z-l-m}.
\begin{align}
\label{two loci}
& \frac{1}{\IP[Z^{i[1]}=z]}
\IP\left[\left. Z^{i[1]}=z\right| \Big(\chi_l^{i[1],1},
\chi_l^{i[1],2},\chi_m^{i[1],1},\chi_m^{i[1],2}\Big)=(x,x',y,y')\right]
\nonumber\\
& = \frac{1}{\IP[Z^{i[1]}=z]}
\IP\left[ Z_{-l-m}^{i[1]}=z-\frac{1}{\sqrt{M}}\Big(
\eta_l(x)+\eta_l(x')+\phi_l(x,x')
+\eta_m(y)+\eta_m(y')+\phi_m(y,y')\Big)
\right]\nonumber \\
& =1-\frac{1}{\sqrt{M}}\Big(\Psi_l(x,x')+\Psi_m(y,y')-\IE[\Psi_l+\Psi_m]
\Big)\frac{1}{\IP[Z^{i[1]}=z]}\frac{d}{dz}\IP[Z^{i[1]}=z] \nonumber
\\& \qquad +\frac{1}{2M}\Bigg(\big(\Psi_l(x,x')+\Psi_m(y,y')\big)^2
-\IE[(\Psi_l+\Psi_m)^2]\Bigg)
\frac{1}{\IP[Z^{i[1]}=z]}\frac{d^2}{dz^2}\IP[Z^{i[1]}=z] \nonumber\\
& \qquad -\frac{1}{M}\Bigg(\Big(\Psi_l(x,x')+\Psi_m(y,y')-\IE[(\Psi_l+\Psi_m)]\Big)
\IE[\Psi_l+\Psi_m]\Bigg)
\frac{1}{\IP[Z^{i[1]}=z]}\frac{d^2}{dz^2}\IP[Z^{i[1]}=z] \nonumber \\
& \qquad +\mathcal{O}\Big(\frac{1}{M^{3/2}}\Big).
\end{align}
Using that in the ancestral population we are at linkage
equilibrium with $x,x'$ and $y,y'$ sampled independently from
$\widehat{\nu}_l$ and $\widehat{\nu}_m$ respectively, multiplying by
$\eta_l(x)\eta_m(y)$ and integrating
against
$\widehat{\nu}(dx)\widehat{\nu}(dy)$,
the only non-zero term corresponds to the
term $\eta_l(x)\eta_m(y)$ in $(\Psi_l(x,x')+\Psi_m(y,y'))^2$, so
that
\begin{align}
\nonumber
&\frac{1}{M}\IE\Bigg[\sum_{l\neq m}\Big(\eta_l(\chi_l^{i[1],1})
+\eta_l(\chi_l^{i[1],2})\Big)\Big(\eta_m(\chi_m^{i[1],1})+
\eta_m(\chi_m^{i[1],2})\Big)\Bigg|i[1]=i[2], Z^{i[1]}\Bigg]
\\
\nonumber
&=
\frac{4}{M^2}\frac{\IP''[Z^{i[1]}]}{\IP[Z^{i[1]}]}
\sum_{l\neq m}\IE[\eta_l(\widehat{\chi}_l)^2]\IE[\eta_m(\widehat{\chi}_m)^2]
+{\mathcal O}\Big(\frac{1}{\sqrt{M}}\Big)\\
&=
\frac{\IP''[Z^{i[1]}]}{\IP[Z^{i[1]}]} (\sigma_A^2)^2
+{\mathcal O}\Big(\frac{1}{\sqrt{M}}\Big).
\label{eta by eta}
\end{align}
(The factor of 4 corresponds to the 4 possible ways of choosing the
parents at the two loci.)
Similarly, to calculate
$$\IE\left[\left.\eta_l(\chi_l^{i[1],1})\phi_m(\chi_m^{i[1],1},\chi_m^{i[1],1})
\right|i[1]=i[2], Z^{i[1]}\right]$$
we multiply~(\ref{two loci}) by $\eta_l(x)\phi_m(y,y)$ and integrate.
Once again, using
Eq.~(\ref{vanishing cross variation})-(\ref{mean eta zero}),
we find that
most of the terms vanish, leaving only
\begin{multline}
\label{eta by phi one one}
-\frac{1}{\sqrt{M}}\frac{\IP'[Z^{i[1]}]}{\IP[Z^{i[1]}]}\IE[\eta_l(\widehat{\chi}_l)^2]\IE[\phi_m(\widehat{\chi}_m,\widehat{\chi}_m)]\\
+\frac{1}{2M}
\frac{\IP''[Z^{i[1]}]}{\IP[Z^{i[1]}]}\Big\{
\IE[\eta_l(\widehat{\chi}_l)^3]\IE[\phi_m(\widehat{\chi}_m,\widehat{\chi}_m)]
+\IE[\eta_l(\widehat{\chi}_l)\phi_l(\widehat{\chi}_l^1, \widehat{\chi}_l^2)^2]
\IE[\phi_m(\widehat{\chi}_m,\widehat{\chi}_m)]
\\
+
2\IE[\eta_l(\widehat{\chi}_l)^2]
\IE[\eta_m(\widehat{\chi}_m)\phi_m(\widehat{\chi}_m, \widehat{\chi}_m)]
\Big\}.
\end{multline}
Multiplying by $1/(2M)$ and summing over loci, in the notation
of Table~\ref{QG coefficients}, the first term yields
$$-\iota\sigma_A^2\frac{\IP'[Z^{i[1]}]}{\IP[Z^{i[1]}]}.$$
(There are four terms of this form in~(\ref{correlations 2}) and we have taken account of
all of them.)
The last term gives
$$\frac{\sigma_A^2\sigma_{ADI}}{2}\frac{\IP''[Z^{i[1]}]}{\IP[Z^{i[1]}]}
$$
(again counting the contribution from all
four terms of this form
in~(\ref{correlations 2})).

Now observe that
\begin{equation*}
\frac{1}{M}\sum_{m=1}^M\IE[\phi_m(\widehat{\chi}_m,\widehat{\chi}_m)]
=\frac{\iota}{\sqrt{M}},
\end{equation*}
so that summing over loci, the contribution from the first two terms
multiplying the second derivative will be ${\mathcal O}(1/\sqrt{M})$.
\begin{remark}
\label{where we use inbreeding}
Up to this point, it has been possible to
neglect the error terms under the assumption that the within-family variance
is not too small and we
are not too far out into the tails of the distribution of $Z^{i[1]}$;
the more extreme the trait of the parent, the worse the approximation will be.
Now things change.
In order for $\IE[{\cal A}^i+{\cal D}^i]$ to be finite, we required that the
inbreeding depression~$\iota$ be well-defined; here we see that it also enters into
the error terms.
\end{remark}

In the same way we calculate
$$
\IE\left[\left.\eta_l(\chi_l^{i[1],1})\phi_m(\chi_m^{i[1],1},\chi_m^{i[1],2})
\right|i[1]=i[2], Z^{i[1]}\right]$$
by multipling~(\ref{two loci}) by $\eta_l(x)\phi_m(y,y')$ and integrating.
The only term to survive integration is
\begin{equation}
\label{eta by phi one two}
\frac{1}{2M}\frac{\IP''[Z^{i[1]}]}{\IP[Z^{i[1]}]}
\IE[2\eta_l(\widehat{\chi}_l)^2]
\IE[\phi_m(\widehat{\chi}_m^1,\widehat{\chi}_m^2)^2].
\end{equation}
There are four terms of this form in~(\ref{correlations 2}),
each of which is weighted by $1/(2M)$ and
so, summing over loci,
we arrive at an overall contribution of $\sigma_A^2\sigma_D^2\IP''[Z^{i[1]}]/\IP[Z^{i[1]}]$.
Eq.~(\ref{eta by phi one one}) and (\ref{eta by phi one two}) yield
that~(\ref{correlations 2}) equals
\begin{align}
\label{eta by phi}
& \frac{1}{2M}\IE\Bigg[\sum_{l\neq m}\Big(\eta_l(\chi_l^{i[1],1})
+\eta_l(\chi_l^{i[1],2})\Big) \nonumber \\
& \qquad \qquad \times \Big(\phi_m(\chi_m^{i[1],1},\chi_m^{i[1],1})+
\phi_m(\chi_m^{i[1],2},\chi_m^{i[1],2})+
2\phi_m(\chi_m^{i[1],1},\chi_m^{i[1],2})\Big)\Bigg|i[1]=i[2], Z^{i[1]}\Bigg]
 \nonumber\\
& =
-\frac{\IP'[Z^{i[1]}]}{\IP[Z^{i[1]}]}\iota \sigma_A^2
+
\frac{\IP''[Z^{i[1]}]}{\IP[Z^{i[1]}]}\Big\{\frac{\sigma_A^2\sigma_{ADI}}{2}
+\sigma_A^2\sigma_D^2
\Big\}
+{\mathcal O}\Big(\frac{1}{\sqrt{M}}\Big).
\end{align}

Continuing in this way,
$$\IE\left[\left.\phi_l(\chi_l^{i[1],1}, \chi_l^{i[1],1})
\phi_m(\chi_m^{i[1],1},\chi_m^{i[1],1})
\right|i[1]=i[2], Z^{i[1]}\right]$$
is obtained by multiplying~(\ref{two loci}) by $\phi_l(x,x)\phi_m(y,y)$
and integrating.
When we sum the `constant' term over loci we will obtain $\iota^2/M$ which tends
to zero. The remaining non-zero terms are
\begin{multline}
\label{phi one one by phi one one}
-\frac{1}{\sqrt{M}}
\frac{\IP'[Z^{i[1]}]}{\IP[Z^{i[1]}]}\Big\{
\IE[\eta_l(\widehat{\chi}_l) \phi_l(\widehat{\chi}_l,\widehat{\chi}_l)]
\IE[\phi_m(\widehat{\chi}_m,\widehat{\chi}_m)]
+
\IE[\phi_l(\widehat{\chi}_l,\widehat{\chi}_l)]
\IE[\eta_m(\widehat{\chi}_m)
\phi_m(\widehat{\chi}_m,\widehat{\chi}_m)]\Big\}
\\
+
\frac{1}{2M}\frac{\IP''[Z^{i[1]}]}{\IP[Z^{i[1]}]}
\Bigg\{
\IE\left[\phi_l(\widehat{\chi}_l,\widehat{\chi}_l)
\phi_m(\widehat{\chi}_m,\widehat{\chi}_m)\Big(
\eta_l(\widehat{\chi}_l)^2
+\eta_m(\widehat{\chi}_m)^2+
2\eta_l(\widehat{\chi}_l)
\eta_m(\widehat{\chi}_m)\Big)\right]
\\
+
\IE\left[\phi_l(\widehat{\chi}_l^1,\widehat{\chi}_l^1)
\phi_m(\widehat{\chi}_m^1,\widehat{\chi}_m^1)\Big(
\eta_l(\widehat{\chi}_l^2)^2+
\eta_m(\widehat{\chi}_m^2)^2\Big)\right]\Bigg\}.
\end{multline}
The terms in the last line will contribute ${\mathcal O}(1/\sqrt{M})$
when we sum, as will the first two terms in the middle line.
There are four terms of this form in~(\ref{correlations 3}) and we
are multiplying by $1/(16M)$ and summing over loci, so the top line
contributes $-\iota\sigma_{ADI}\IP'[Z^{i[1]}]/(4\IP[Z^{i[1]}])$,
similarly the second line will
contribute $\sigma_{ADI}^2\IP''[Z^{i[1]}]/(16\IP[Z^{i[1]}])$.

Now, again using~(\ref{two loci}),
\begin{align*}
&\IE\left[\phi_l(\chi_l^{i[1],1}, \chi_l^{i[1],1})
\phi_m(\chi_m^{i[1],1},\chi_m^{i[1],2})
\Big|i[1]=i[2], Z^{i[1]}\right] \nonumber \\
& = \int\ldots\int \phi_l(x,x)\phi_m(y,y')
\Bigg[ 1-\frac{1}{\sqrt{M}}\Big(\Psi_l(x,x')+\Psi_m(y,y')-\IE[\Psi_l+\Psi_m]
\Big)\frac{\IP'[Z^{i[1]}]}{\IP[Z^{i[1]}]} \\
&\qquad +\frac{1}{2M}\Bigg(\big(\Psi_l(x,x')+\Psi_m(y,y')\big)^2
-\IE[(\Psi_l+\Psi_m)^2]\Bigg)
\frac{\IP''[Z^{i[1]}]}{\IP[Z^{i[1]}]} \\
&\qquad -\frac{1}{M}\Bigg(\Big(\Psi_l(x,x')+\Psi_m(y,y')-\IE[(\Psi_l+\Psi_m)]\Big)
\IE[\Psi_l+\Psi_m]\Bigg)
\frac{\IP''[Z^{i[1]}]}{\IP[Z^{i[1]}]}\Bigg]
\widehat{\nu}_l(dx)
\widehat{\nu}_l(dx')
\widehat{\nu}_m(dy)
\widehat{\nu}_m(dy') \nonumber \\
& \qquad +{\mathcal O}\big(\frac{1}{M^{3/2}}\big).
\end{align*}
There are eight terms of this form in~(\ref{correlations 3}), and we are multiplying by $1/(16M)$ and
summing over loci, so the first term will correspond to a contribution of
$$-\frac{\IP'[Z^{i[1]}]}{\IP[Z^{i[1]}]}\frac{\iota\sigma_D^2}{2}.$$
As usual, terms multiplying the second derivative that involve the locus
$l$ only through $\phi_l(x,x)$ will contribute ${\mathcal O}(1/\sqrt{M})$
to the sum and we find that the nontrivial contributions will be
\begin{multline*}
-\frac{1}{\sqrt{M}}\frac{\IP'[Z^{i[1]}]}{\IP[Z^{i[1]}]}
\IE[\phi_l(\widehat{\chi}_l,\widehat{\chi}_l)]
\IE[\phi_m(\widehat{\chi}_m^1,\widehat{\chi}_m^2)^2]
+
\frac{1}{2M}\frac{\IP''[Z^{i[1]}]}{\IP[Z^{i[1]}]}
\IE[2\eta_l(\widehat{\chi}_l)\phi_l(\widehat{\chi}_l,\widehat{\chi}_l)]
\IE[\phi_m(\widehat{\chi}_m^1,\widehat{\chi}_m^2)^2].
\end{multline*}
There are eight terms of this form in~(\ref{correlations 3}), so multiplying by $1/(16M)$ and
summing over loci gives
\begin{equation}
\label{phi one one by phi one two}
-\frac{\iota \sigma_D^2}{2}\frac{\IP'[Z^{i[1]}]}{\IP[Z^{i[1]}]}+
\frac{\sigma_{ADI}\sigma_D^2}{4}\frac{\IP''[Z^{i[1]}]}{\IP[Z^{i[1]}]}.
\end{equation}
Finally,
when we scale and sum over loci,
the only nontrivial term in our expression for
$$
\IE\left[\left.\phi_l(\chi_l^{i[1],1}, \chi_l^{i[1],2})
\phi_m(\chi_m^{i[1],1},\chi_m^{i[1],2})
\right|i[1]=i[2], Z^{i[1]}\right]
$$
is
\begin{equation*}
+\frac{1}{2M}\frac{\IP''[Z^{i[1]}]}{\IP[Z^{i[1]}]}
\IE[2 \phi_l(\widehat{\chi}_l^1,\widehat{\chi}_l^2)^2]
\IE[\phi_m(\widehat{\chi}_m^1,\widehat{\chi}_m^2)^2].
\end{equation*}
There are four terms of this form, and so multiplying by $1/(16M)$ and
summing gives
\begin{equation}
\label{phi one two by phi one two}
\frac{(\sigma_D^2)^2}{4}\frac{\IP''[Z^{i[1]}]}{\IP[Z^{i[1]}]}.
\end{equation}

Combining \eqref{phi one one by phi one one},
\eqref{phi one one by phi one two},
and \eqref{phi one two by phi one two}, we find that~(\ref{correlations 3}) is
\begin{align}
\label{phi by phi}
& \frac{1}{16M}\IE\Bigg[\sum_{l\neq m}
\Big(\phi_l(\chi_l^{i[1],1},\chi_l^{i[1],1})+
(\phi_l(\chi_l^{i[1],2},\chi_l^{i[1],2})+
2\phi_l(\chi_l^{i[1],1},\chi_l^{i[1],2})\Big) \nonumber\\
& \qquad \times
\Big(\phi_m(\chi_m^{i[1],1},\chi_m^{i[1],1})+
(\phi_m(\chi_m^{i[1],2},\chi_m^{i[1],2})+
2\phi_m(\chi_m^{i[1],1},\chi_m^{i[1],2})\Big)
\bigg|i[1]=i[2], Z^{i[1]}\Bigg] \nonumber \\
& = -\frac{\IP'[Z^{i[1],1}]}{\IP[Z^{i[1],1}]}\Big(
\frac{\iota\sigma_{ADI}}{4}
+\frac{\iota\sigma_D^2}{2}\Big)
+\frac{\IP''[Z^{i[1],1}]}{\IP[Z^{i[1],1}]}\Big(\frac{\sigma_{ADI}^2}{16}
+\frac{\sigma_{ADI}\sigma_D^2}{4}+\frac{(\sigma_D^2)^2}{4}\Big)
+{\mathcal O}\Big(\frac{1}{\sqrt{M}}\Big).
\end{align}

Adding~(\ref{diagonal terms A+D squared same parent}), (\ref{eta by eta}),
(\ref{eta by phi}), and~(\ref{phi by phi}) yields $\IE[({\cal A}^i+{\cal D}^i)^2]$,
and subtracting the square
of~(\ref{mean gen one identical parents}), we obtain
\begin{align}
\mathtt{Var}\big(&{\cal A}^i+{\cal D}^i|i[1]=i[2], Z^{i[1]}\big)\nonumber\\
&=
\sigma_A^2+\frac{1}{8}\left(\sigma_{DI}^2+\iota^*\right)
+\frac{1}{4}\sigma_D^2+\frac{1}{2}\sigma_{ADI}
-\frac{\IP'[Z^{i[1]}]}{\IP[Z^{i[1]}]}
\left\{\iota\sigma_A^2+\frac{\iota\sigma_{ADI}}{4}+\frac{\iota\sigma_D^2}{2}
\right\} \nonumber \\
&\quad  + \frac{\IP''[Z^{i[1]}]}{\IP[Z^{i[1]}]}
\left\{\big(\sigma_A^2\big)^2+\frac{\sigma_A^2\sigma_{ADI}}{2}
+\sigma_A^2\sigma_D^2 
+\frac{\big(\sigma_D^2\big)^2}{4}
+\frac{\sigma_{ADI}^2}{16}+\frac{\sigma_D^2\sigma_{ADI}}{4}\right\} \nonumber
\\
& \quad -\left(\frac{\iota}{2}-
\frac{\IP'[Z^{i[1]}]}{\IP[Z^{i[1]}]}\left(\sigma_A^2+\frac{\sigma_D^2}{2}
+\frac{\sigma_{ADI}}{4}\right)\right)^2
+{\mathcal O}\left(\frac{1}{\sqrt{M}}\right).
\end{align}

Now if we substitute the Gaussian density for $Z^{i[1]}$, observing that
$$
\frac{\IP''[Z^{i[1]}]}{\IP[Z^{i[1]}]}
-\left(\frac{\IP'[Z^{i[1]}]}{\IP[Z^{i[1]}]}\right)^2=-\frac{1}{\sigma_A^2+\sigma_D^2},$$
we see that the variance reduces to
\begin{equation}
\label{cond variance gen one same parent}
-\frac{\iota^2}{4}+\frac{1}{\sigma_A^2+\sigma_D^2}\left(\sigma_A^2+
\frac{\sigma_D^2}{2}+\frac{\sigma_{ADI}}{4}\right)^2
+\sigma_A^2+\frac{1}{8}\left(\sigma_{DI}^2+\iota^*\right)
+\frac{1}{4}\sigma_D^2+\frac{1}{2}\sigma_{ADI}
+{\mathcal O}\left(\frac{1}{\sqrt{M}}\right).
\end{equation}

Again, that was a lot of work to recover exactly the expression that we
expected from conditioning the multivariate normal random variable
$\big(({\cal A}^i+{\cal D}^i), Z^{i[1]}\big)$ on its second argument. However, in the
process, we have identified where the normal approximation to the
conditioned process will break down.
The bounds that we have obtained will be poor if the trait value of either
parent is too extreme, or if the pedigree is too inbred (as a
result of which
the variance of trait values will be small and
inbreeding depression may be high).

Of course, we have not proved that the conditional distribution of
$({\cal A}^i+{\cal D}^i)$ converges to a normal, we have just checked that the first two
moments are asymptotically
what we would expect. We defer the proof of normality until we
have calculated the conditional variance of $({\cal A}^i+{\cal D}^i)$ in the (much simpler)
case of two distinct parents.

\subsection*{Conditional variance $({\cal A}^i+{\cal D}^i)$, generation one, distinct parents}

If the parents are distinct, then the expressions are much simpler.
First
\begin{align*}
& \frac{1}{4M}\IE\left[\left.\sum_{l=1}^M\left(\eta_l(\chi_l^{i[1],1})+
\eta_l(\chi_l^{i[1],2}) +\eta_l(\chi_l^{i[2],1}) +\eta_l(\chi_l^{i[2],2})
\right)^2\right|i[1]\neq i[2], Z^{i[1]}, Z^{i[2]}\right]
\\
&=\frac{1}{M}\sum_{l=1}^M\IE\left[\eta_l(\widehat{\chi}_l)^2\right]
+{\mathcal O}\left(\frac{1}{\sqrt{M}}\right)
= \frac{\sigma_A^2}{2}
+{\mathcal O}\left(\frac{1}{\sqrt{M}}\right).
\end{align*}
Next
\begin{align*}
&\frac{1}{16M}\IE\Bigg[\left.\sum_{l=1}^M\left(
\phi_l(\chi_l^{i[1],1}, \chi_l^{i[2],1})
+\phi_l(\chi_l^{i[1],1}, \chi_l^{i[2],2})
+\phi_l(\chi_l^{i[1],2}, \chi_l^{i[2],1})
+\phi_l(\chi_l^{i[1],2}, \chi_l^{i[2],2})\right)^2\right| \\
& \qquad \qquad \qquad \qquad i[1]\neq i[2], Z^{i[1]}, Z^{i[2]}\Bigg] \\
&=
\frac{1}{4M}\sum_{l=1}^M\IE\left[\phi_l(\widehat{\chi}^1_l,
\widehat{\chi}^2_l)^2\right]
+{\mathcal O}\left(\frac{1}{\sqrt{M}}\right)\\
& =\frac{1}{4}\sigma_D^2
+{\mathcal O}\left(\frac{1}{\sqrt{M}}\right).
\end{align*}
Finally,
\begin{align*}
&\frac{1}{4M}\IE\Bigg[\sum_{l=1}^M
\left(\eta_l(\chi_l^{i[1],1})+
\eta_l(\chi_l^{i{1},2}) +\eta_l(\chi_l^{i[2],1}) +\eta_l(\chi_l^{i[2],2})
\right)
\\
& \times
\left(
\phi_l(\chi_l^{i[1],1}, \chi_l^{i[2],1})
+\phi_l(\chi_l^{i[1],1}, \chi_l^{i[2],2})
+\phi_l(\chi_l^{i[1],2}, \chi_l^{i[2],1})
\phi_l(\chi_l^{i[1],2}, \chi_l^{i[2],2})\right)\Bigg|
i[1]\neq i[2], Z^{i[1]}, Z^{i[2]}
\Bigg] \\
& =
{\mathcal O}\left(\frac{1}{\sqrt{M}}\right).
\end{align*}

We now turn to the off-diagonal terms. We need to be able to calculate
the conditional expectation of
\begin{multline}
\label{off-diag term}
\Bigg[\frac{1}{2}
\left(\eta_l(\chi_l^{i[1],1})+
\eta_l(\chi_l^{i[1],2}) +\eta_l(\chi_l^{i[2],1}) +\eta_l(\chi_l^{i[2],2})
\right)
\\
+\frac{1}{4}
\left(
\phi_l(\chi_l^{i[1],1}, \chi_l^{i[2],1})
+\phi_l(\chi_l^{i[1],1}, \chi_l^{i[2],2})
+\phi_l(\chi_l^{i[1],2}, \chi_l^{i[2],1})
+\phi_l(\chi_l^{i[1],2}, \chi_l^{i[2],2})\right)\Bigg]
\\
\times
\Bigg[\frac{1}{2}
\left(\eta_m(\chi_m^{i[1],1})+
\eta_m(\chi_m^{i[1],2}) +\eta_m(\chi_m^{i[2],1}) +\eta_m(\chi_m^{i[2],2})
\right)
\\ +\frac{1}{4}
\left(
\phi_m(\chi_m^{i[1],1}, \chi_m^{i[2],1})
+\phi_m(\chi_m^{i[1],1}, \chi_m^{i[2],2})
+\phi_m(\chi_m^{i[1],2}, \chi_m^{i[2],1})
+\phi_m(\chi_m^{i[1],2}, \chi_m^{i[2],2})\right)\Bigg]
\end{multline}
given the trait values in the (unrelated) parents $i[1]$ and $i[2]$.
Because the parents are distinct, and
they are in generation zero, as
in~(\ref{independence of trait values})
in our calculation of the
conditional mean, we can exploit the fact that
the trait values $Z^{i[1]}$ and $Z^{i[2]}$ are
independent so that
the joint probability that
$$\big(\chi_l^{i[1],1}, \chi_l^{i[1],2}, \chi_m^{i[1],1},
\chi_m^{i[1],2}\big)=(x,x',y,y'),$$
conditional on $Z^{i[1]}, Z^{i[2]}$
is just the same as if we only
condition on $Z^{i[1]}$. 
Recalling~(\ref{independence of trait values}),
we can calculate the conditional expectation of~(\ref{off-diag term})
using~(\ref{two loci}).
None of the genes at either locus are identical
by descent, and so integrating against the term of order $1/\sqrt{M}$ in the
Taylor expansion in~(\ref{two loci}) gives zero, but since we are calculating the conditional
expectation of ${\cal O}(M^2)$
terms, each of which is of order $1/M$,
we can expect to see a contribution from the term of order $1/M$.
All the terms involving the dominance components vanish, as do those
terms involving only one copy of the additive component at one of the loci.
In total we find that the conditional expectation of~\eqref{off-diag term}
is
$$\frac{1}{M}\IE[\eta_l(\widehat{\chi}_l)^2]\IE[\eta_m(\widehat{\chi}_m)^2]
\left\{\frac{\IP''[Z^{i[1]}]}{\IP[Z^{i[1]}]}+
\frac{\IP''[Z^{i[2]}]}{\IP[Z^{i[2]}]}
+2\frac{\IP'[Z^{i[1]}]}{\IP[Z^{i[1]}]}\frac{\IP'[Z^{i[2]}]}{\IP[Z^{i[2]}]}
\right\}.$$
Summing over loci (and noting that we may include the diagonal terms and
only incur an error of order $1/M$), we find that,
in the case of different parents, the variance of the shared terms ${\cal A}^i+
{\cal D}^i$,
conditional on the trait values of the parent is
\begin{multline}
\frac{1}{2}\sigma_A^2
+\frac{1}{4}\sigma_D^2
-\left(\left(\frac{\IP'(Z^{i[1]})}{\IP(Z^{i[1]})}
+\frac{\IP'(Z^{i[2]})}{\IP(Z^{i[2]})}\right)\frac{\sigma_A^2}{2}\right)^2
\\+
\left\{\frac{\IP''[Z^{i[1]}]}{\IP[Z^{i[1]}]}+
\frac{\IP''[Z^{i[2]}]}{\IP[Z^{i[2]}]}
+2\frac{\IP'[Z^{i[1]}]}{\IP[Z^{i[1]}]}\frac{\IP'[Z^{i[2]}]}{\IP[Z^{i[2]}]}
\right\}\Big(\frac{\sigma_A^2}{2}\Big)^2
+{\mathcal O}\left(\frac{1}{\sqrt{M}}\right).
\end{multline}

Once again we see that if we approximate the distribution of $Z^{i[1]}$
and $Z^{i[2]}$ by that of independent normal random variables with mean
$\bar{z}_0$ and variance $\sigma_A^2+\sigma_D^2$, most of these terms
cancel and we are left with
$$\frac{\sigma_A^2}{2}+\frac{\sigma_D^2}{4}
-\frac{\sigma_A^4}{2(\sigma_A^2+\sigma_D^2)},$$
exactly as predicted by Theorem~\ref{conditioning multivariate normals}.

\subsubsection*{The general case}

So far we have only dealt with generation one, where expressions
are simplified by the fact that $Z^{i[1]}$, $Z^{i[2]}$ are
either identical or independent. More generally, we can perform
entirely analogous calculations using
Lemma~\ref{troublesome lemma with two parents}
in place of Lemma~\ref{replacement for troublesome equation}.
In the interests of sanity, we omit the details.

\section{Convergence to normal of $({\cal A}^i+{\cal D}^i)$ conditional on parental traits}
\label{convergence to normal}

\begin{notn}
\label{notation F1}
We remind the reader that Notation~\ref{noise in notation} remains in force.
Moreover, since the environmental noise $E^i$ is assumed to be shared by
all offspring of the couple $i[1]$, $i[2]$, with this convention we can also
assume that the distribution of
${\cal A}^i+{\cal D}^i$ has a smooth density.
\end{notn}

We have verified that the first two moments of the conditional
distribution converge to the limits that we would expect if the limit
of $({\cal A}^i+{\cal D}^i)$ were multivariate normal, but this is not sufficient.
To prove that the conditional distribution is indeed asymptotically normal, we appeal
to Proposition~\ref{chen propn}, or rather Corollary~\ref{chen corollary}.
We perform the calculation in the case of identical parents, the case
of distinct parents being analogous (and less surprising). For
definiteness, we consider only generation one.
The same argument will work in any generation, but the calculations
become considerably more involved,
\emph{c.f.}~Lemma~\ref{troublesome lemma with two parents}.

Recall that ${\cal A}^i+{\cal D}^i=\sum_{l=1}^M\Phi(l)/\sqrt{M}$ with $\Phi$ defined
in~(\ref{defn of Phil}). Since we are considering the case of a single parent, $Z^{i[1]}=Z^{i[2]}$.
We shall write $\Phi_l(\chi_l^1,\chi_l^2)$
when we need to specify the alleles at locus $l$ in $Z^{i[1]}$ on which
this is evaluated.

Writing $W=\sum_{l=1}^M\Phi(l)/\sqrt{M}$ (as a shorthand for ${\cal A}^i+
{\cal D}^i$),
we write
$$\widehat{W}=\frac{1}{\sqrt{M}}\sum_{l=1}^M \Phi(l)\Big|
i[1]=i[2], Z^{i[1]};$$
that is $\widehat{W}$ is the random variable $W$ in the $i$th individual,
conditional on it being produced by selfing and on the parental trait value.
This is the quantity that we should like to prove is normally distributed.
The first step is to find a suitable exchangeable pair.
We write $\widehat{\Phi}(l)$ for the conditioned version of $\Phi(l)$.

For each $l\in\{1,\ldots ,M\}$, let $\widehat{\Phi}^*(l)$ be
an independent draw from the conditional distribution
of $\widehat{\Phi}(l)$ given the sum of $\widehat{\Phi}(m)$ over all $m\neq l$;
that is, in an obvious notation,
$\widehat{\Phi}^*(l)$ has the same distribution as
$$\Phi(l)\Big|\sum_{m\neq l}\widehat{\Phi}(m), i[1]=i[2], Z^{i[1]}.$$
Now let $L$ be a uniform random variable on $\{1,\ldots ,M\}$ and
define
$$\widehat{W}'=\widehat{W}-
\frac{\big(\widehat{\Phi}(L)-\widehat{\Phi}^*(L)\big)}{\sqrt{M}}.$$
Then $(\widehat{W},\widehat{W}')$ is an exchangeable pair.

Observe that
\begin{eqnarray}
\nonumber
\IE\big[\widehat{W}-\widehat{W}'|\widehat{W}\big]&=&
\IE\left[\left.\frac{1}{\sqrt{M}}\frac{1}{M}\sum_{l=1}^M
\big(\widehat{\Phi}(l)-\widehat{\Phi}^*(l)\big)
\right|\widehat{W}\right]
\\
\nonumber
&=&\frac{1}{M}\widehat{W}-\frac{1}{\sqrt{M}}\frac{1}{M}\sum_{l=1}^M
\IE\big[\widehat{\Phi}^*(l)\big|\widehat{W}\big]
\\
\label{defn of first remainder}
&:=&\frac{1}{M}\widehat{W}-
T(\widehat{W}).
\end{eqnarray}
\begin{remark}
We wish to apply Corollary~\ref{chen corollary}.
Our first instinct is to write $\IE[\widehat{W}'|\widehat{W}]=\widehat{W}(1-1/M)+T(\widehat{W})$
and take $\lambda =1/M$ in~(\ref{exchangeable pair requirement}). This will not suffice, as, with
this choice, the first term on the right of~(\ref{unnormalised difference})
will be too big. As we shall see, the resolution is to take a larger value of $\lambda$ which
captures the dependence of $\widehat{W}'$ on $\widehat{W}$.
\end{remark}
Before we can apply Corollary~\ref{chen corollary},
we need to investigate $T$. The first step is to establish the distribution of
$\chi_l^1, \chi_l^2$ conditional on $i[1]=i[2]$, $Z^{i[1]}$ (which we
shall for the rest of this section abbreviate to $Z$) and $W_{-l}$.

Keeping in mind Notation~\ref{notation F1}, and recalling that $(Z,W)$ is
shorthand for
$(Z^{i[1]}, {\cal A}^i+{\cal D}^i)$, we write $\IP[z,w]$ for the
density function of $(Z,W)$ evaluated at $(z,w)$ and $\IP_z[z,w]$, $\IP_w[z,w]$, and so on,
for the corresponding partial derivatives.
The proof of the following lemma mirrors those of Appendix~\ref{key lemmas}.
\begin{lemma}
\label{Z-l and W-l}
The (unconditional) distribution of $(Z_{-l}, W_{-l})$ can be
written as
\begin{align*}
\IP[&Z_{-l}=z, W_{-l}=w] \nonumber\\
& =\IP[z,w]+\frac{1}{\sqrt{M}}\IE[\Psi_l]\IP_z[z,w]
+\frac{1}{\sqrt{M}}\IE[\Phi_l]\IP_w[z,w] +\frac{1}{M}\Big(\IE[\Psi_l]^2-\IE[\Psi_l^2]\Big)\IP_{zz}[z,w] \nonumber
\\
& \quad +\frac{2}{M}\Big(\IE[\Phi_l]\IE[\Psi_l]-\IE[\Phi_l\Psi_l]\Big)\IP_{zw}[z,w]
+\frac{1}{M}\Big(\IE[\Phi_l]^2-\IE[\Phi_l^2]\Big)\IP_{ww}[z,w] \nonumber\\
& \quad +\frac{1}{2M}\IE[\Psi_l^2]\IP_{zz}[z,w]
+\frac{1}{M}\IE[\Phi_l\Psi_l]\IP_{zw}[z,w]
+\frac{1}{2M}\IE[\Phi_l^2]\IP_{ww}[z,w]
+\mathcal{O}\Big(\frac{1}{M^{3/2}}\Big).
\end{align*}
\end{lemma}
\noindent

\newpage
{\bf Proof}

The key, as usual, is Taylor's Theorem.
\begin{align*}
& \IP[Z_{-l}=z, W_{-l}=w] \\
&=\int\int\IP\bigg[\chi_l^1=x, \chi_l^2=x',
Z=z+\frac{1}{\sqrt{M}}\Psi_l(x,x'), W=w+\frac{1}{\sqrt{M}}\Phi_l(x,x')\bigg]dxdx'\\
&=
\int\int \IP[\chi_l^1=x,\chi_l^2=x', Z=z, W=w]dxdx'
\\
& \quad +\frac{1}{\sqrt{M}}\int\int\Psi_l(x,x')\frac{\partial}{\partial z}
\IP[\chi_l^1=x,\chi_l^2=x', Z=z, W=w]dxdx'
\\
& \quad+\frac{1}{\sqrt{M}}\int\int\Phi_l(x,x')\frac{\partial}{\partial w}
\IP[\chi_l^1=x,\chi_l^2=x', Z=z, W=w]dxdx'
\\
& \quad+\frac{1}{2M}\int\int\Psi_l^2(x,x')\frac{\partial^2}{\partial z^2}
\IP[\chi_l^1=x,\chi_l^2=x', Z=z, W=w]dxdx'
\\
& \quad+\frac{1}{M}\int\int\Psi_l(x,x')\Phi_l(x,x')\frac{\partial^2}{\partial z\partial w}
\IP[\chi_l^1=x,\chi_l^2=x', Z=z, W=w]dxdx'
\\
& \quad+\frac{1}{2M}\int\int\Phi_l^2(x,x')\frac{\partial^2}{\partial w^2}
\IP[\chi_l^1=x,\chi_l^2=x', Z=z, W=w]dxdx'
+\mathcal{O}\Big(\frac{1}{M^{3/2}}\Big).
\end{align*}
Now write
\begin{multline*}
\IP[\chi_l^1=x,\chi_l^2=x', Z=z, W=w]=
\IP[\chi_l^1=x,\chi_l^2=x']\\
\times
\IP\left[Z_{-l}=z-\frac{1}{\sqrt{M}}\Psi_l(x,x'),
W_{-l}=w-\frac{1}{\sqrt{M}}\Phi_l(x,x')\right].
\end{multline*}
Using the notation $\IP[x,x',z,w]:=
\IP[\chi_l^1=x,\chi_l^2=x', Z=z, W=w]$
we substitute from above and apply Taylor's Theorem to obtain,
\begin{multline*}
\IP[x, x',z, w]
=
\IP[\chi_l^1=x,\chi_l^2=x']
\\
\times
\int\int\IP\Big[y, y',
z-\frac{1}{\sqrt{M}}\Psi_l(x,x')+\frac{1}{\sqrt{M}}\Psi_l(y,y'),
w-\frac{1}{\sqrt{M}}\Phi_l(x,x')+\frac{1}{\sqrt{M}}\Phi_l(y,y')\Big]dydy'
\\
=
\IP[\chi_l^1=x,\chi_l^2=x']
\Bigg\{
\IP[Z=z, W=w]
+\frac{1}{\sqrt{M}}\int\int\Big(\Psi_l(y,y')-\Psi_l(x,x')\Big)
\frac{\partial}{\partial z}\IP[y,y',z,w]dydy'
\\
+\frac{1}{\sqrt{M}}\int\int\Big(\Phi_l(y,y')-\Phi_l(x,x')\Big)
\frac{\partial}{\partial w}\IP[y,y',z,w]dydy'
+\mathcal{O}\Big(\frac{1}{M}\Big)\Bigg\}.
\end{multline*}
Differentiating with respect to $z$ (and assuming sufficient regularity),
\begin{multline*}
\frac{\partial}{\partial z}
\IP[\chi_l^1=x,\chi_l^2=x', Z=z, W=w]
=
\IP[\chi_l^1=x,\chi_l^2=x']
\\
\times
\Bigg\{
\frac{\partial}{\partial z}\IP[Z=z, W=w]
+\frac{1}{\sqrt{M}}\int\int\Big(\Psi_l(y,y')-\Psi_l(x,x')\Big)
\frac{\partial^2}{\partial z^2}\IP[y,y',z,w]dydy'
\\
+\frac{1}{\sqrt{M}}\int\int\Big(\Phi_l(y,y')-\Phi_l(x,x')\Big)
\frac{\partial^2}{\partial z\partial w}\IP[y,y',z,w]dydy'
+\mathcal{O}\Big(\frac{1}{M}\Big)\Bigg\},
\end{multline*}
and similarly
\begin{multline*}
\frac{\partial}{\partial w}
\IP[\chi_l^1=x,\chi_l^2=x', Z=z, W=w]
=
\IP[\chi_l^1=x,\chi_l^2=x']
\\
\times
\Bigg\{
\frac{\partial}{\partial w}\IP[Z=z, W=w]
+\frac{1}{\sqrt{M}}\int\int\Big(\Psi_l(y,y')-\Psi_l(x,x')\Big)
\frac{\partial^2}{\partial z\partial w}\IP[y,y',z,w]dydy'
\\
+\frac{1}{\sqrt{M}}\int\int\Big(\Phi_l(y,y')-\Phi_l(x,x')\Big)
\frac{\partial^2}{\partial w^2}\IP[y,y',z,w]dydy'
+\mathcal{O}\Big(\frac{1}{M}\Big)\Bigg\}.
\end{multline*}
As in the proof of Lemma~\ref{replacement for troublesome equation}
we only require the second derivatives to leading order
$$\frac{\partial^2}{\partial z^2}
\IP[\chi_l^1=x,\chi_l^2=x', Z=z, W=w]
=
\IP[\chi_l^1=x,\chi_l^2=x']
\frac{\partial^2}{\partial z^2}\IP[Z=z, W=w]
+\mathcal{O}\Big(\frac{1}{\sqrt{M}}\Big),$$
with similar expressions for the other second partial derivatives.
Substituting back into the first display yields the result.
\hfill $\Box$

\begin{lemma}
The conditional distribution of $\chi_l^1, \chi_l^2$ given $Z$ and $W_{-l}$
is given by
\begin{multline}
\label{distbn alleles in Phistar}
\IP[\chi_l^1=x,\chi_l^2=x'|Z=z, W_{-l}=w_{-l}]
\\
=\IP[\chi_l^1=x,\chi_l^2=x']
\Bigg\{
1-\frac{\Psi_l(x,x')}{\sqrt{M}\IP[z,w_{-l}]} 
\IP_z[z, w_{-l}]
+\frac{\IE[\Psi_l]}{\sqrt{M}\IP[z, w_{-l}]} 
\IP_z[z, w_{-l}]
\Bigg\}
+\mathcal{O}\Big(\frac{1}{M}\Big).
\end{multline}
\end{lemma}
\noindent
{\bf Proof}

This is just an application of Bayes' rule:
\begin{align}
\IP[\chi_l^1=x,\chi_l^2=x'|Z=z, W_{-l}=w_{-l}]
&=
\frac{\IP[Z=z, W_{-l}=w_{-l}| \chi_l^1=x,\chi_l^2=x']}{
\IP[Z=z, W_{-l}=w_{-l}]}
\IP[\chi_l^1=x,\chi_l^2=x']
\nonumber
\\
&=
\frac{\IP[Z_{-l}=z-\frac{\Psi_l(x,x')}{\sqrt{M}}, W_{-l}=w_{-l}]}{
\IP[Z=z, W_{-l}=w_{-l}]}
\IP[\chi_l^1=x,\chi_l^2=x']
\label{cond dist alleles}
\end{align}
Using Lemma~\ref{Z-l and W-l} and Taylor's Theorem,
\begin{align*}
&\IP\bigg[Z_{-l}=z-\frac{\Psi_l(x,x')}{\sqrt{M}}, W_{-l}=w_{-l}\bigg]
\\
&=\IP\bigg[Z=z-\frac{\Psi_l(x,x')}{\sqrt{M}}, W_{-l}=w_{-l}\bigg]
+\frac{1}{\sqrt{M}}\IE[\Psi_l]\IP_z\bigg[z-\frac{\Psi_l(x,x')}{\sqrt{M}}, w_{-l}\bigg] \\
& \qquad \qquad
+\frac{1}{\sqrt{M}}\IE[\Phi_l]\IP_w\bigg[z-\frac{\Psi_l(x,x')}{\sqrt{M}}, w_{-l}\bigg]
\\
&=
\IP[Z=z, W=w_{-l}]-\frac{\Psi_l(x,x')}{\sqrt{M}}
\IP_z[z,w_{-l}]
+\frac{\IE[\Psi_l]}{\sqrt{M}}
\IP_z[z,w_{-l}]
+\frac{\IE[\Phi_l]}{\sqrt{M}}
\IP_w[z,w_{-l}] +
\mathcal{O}\Big(\frac{1}{M}\Big).
\end{align*}
When we integrate this expression with respect to $x$ and $x'$, to
calculate the denominator in~(\ref{cond dist alleles})
we recover
$\IP_z[z,w_{-l}]
+\frac{\IE[\Phi_l]}{\sqrt{M}} \IP_w[z,w_{-l}] +
\mathcal{O}\big(\frac{1}{M}\big)$
(since the
expectation of $\Psi_l$ cancels).
Expanding the ratio in~(\ref{cond dist alleles}) in powers of $1/\sqrt{M}$, the
terms involving $\IE[\Phi_l]$ cancel, and the result follows. \hfill $\Box$

Finally we are in a position to calculate the quantity
$T(\widehat{W})$ that was defined in~(\ref{defn of first remainder}).
Recall that $\widehat{\Phi}_l^*$ is an independent draw from the
conditional distribution of $\widehat{\Phi}_l$ given $W_{-l}$ and $Z$,
so using~(\ref{distbn alleles in Phistar}),
\begin{equation}
\label{cond expn Phistar}
\IE[\widehat{\Phi}_l^*|\widehat{W}]
=\IE[\Phi_l]-\Big(\IE[\Phi_l\Psi_l]-\IE[\Phi_l]\IE[\Psi_l]\Big)
\frac{1}{\sqrt{M}}\IE\left[\left.\frac{1}{\IP[Z,W_{-l}]}
\frac{\partial}{\partial z}\IP[Z,W_{-l}]\right| W=\widehat{W}\right] +
\mathcal{O}\Big(\frac{1}{M}\Big).
\end{equation}

Conditioning only on $i[1]=i[2]$,
using the calculations in Appendix~\ref{QG derivations}
and Eq.~(\ref{conditioned covariances}),
by an
application of Theorem~\ref{rinott clt},
(up to an error of order $1/\sqrt{M}$)
the joint distribution of $({\cal A}^i+{\cal D}^i, Z^{i[1]})$ is approximately
that of a bivariate
normal. 

We will need that for a bivariate normal distribution with mean vector
$(\mu_Z,\mu_W)$ and covariance matrix
$$\begin{pmatrix}
\sigma_{Z}^2 & \mathtt{Cov}(Z,W)\\
\mathtt{Cov}(Z,W) & \sigma_W^2
\end{pmatrix}$$
the density function takes the form
$$p(z,w)\!=\!\frac{1}{2\pi\sigma_Z\sigma_W\sqrt{1-\rho^2}}
\exp\!\bigg\{\!-\frac{1}{2(1-\rho^2)}\left(\frac{(z-\mu_Z)^2}{\sigma_Z^2}
-\frac{2\rho (z-\mu_Z)(w-\mu_W)}{\sigma_Z\sigma_W}+
\frac{(w-\mu_W)^2}{\sigma_W^2}\right)\!\bigg\},$$
where $\rho=\mathtt{Cov}(Z,W)/(\sigma_Z\sigma_W)$. Differentiating, we find
\begin{equation}
\label{bivariate normal calculation}
\frac{1}{p(z,w)}\frac{\partial}{\partial z}p(z,w)=
\frac{1}{(1-\rho^2)}\left\{\frac{\rho(w-\mu_W)}{\sigma_Z\sigma_W}
-\frac{(z-\mu_Z)}{\sigma_Z^2}\right\}.
\end{equation}

Recall the definition of $T$ from~(\ref{defn of first remainder}).
Multiplying~(\ref{cond expn Phistar}) by $1/\sqrt{M}$,
observing that $\IE[W_{-l}|W,Z]=W+\mathcal{O}(1/\sqrt{M})$ (and since $\Phi_l$ is uniformly bounded
independent of $l$, the
error is bounded independent of $l$),
and then averaging
out over $l$ as in the definition of $T(\widehat{W})$, on
substituting~(\ref{bivariate normal calculation}) and $\mathtt{Cov}(Z,W)=\rho\sigma_Z\sigma_W$,
we find
\begin{align}
\label{TWHat}
T(\widehat{W})& =
\frac{1}{M}\IE[W]+\frac{\mathtt{Cov}(Z,W)}{M}
\left\{\frac{Z-\IE[Z]}{\sigma_Z^2(1-\rho^2)}-\frac{\rho}{1-\rho^2}
\frac{\widehat{W}-\IE[W]}{\sigma_Z\sigma_W}\right\}
+\mathcal{O}\Big(\frac{1}{M^{3/2}}\Big) \nonumber\\
& =
\frac{1}{M}\IE[W]+\frac{\rho\sigma_Z\sigma_W}{M}
\left\{\frac{Z-\IE[Z]}{\sigma_Z^2(1-\rho^2)}-\frac{\rho}{1-\rho^2}
\frac{\widehat{W}-\IE[W]}{\sigma_Z\sigma_W}\right\}
+\mathcal{O}\Big(\frac{1}{M^{3/2}}\Big) \nonumber
\\
& =\frac{1}{M}\IE[W]+\frac{1}{M}\rho\frac{\sigma_W}{\sigma_Z}
\frac{Z-\IE[Z]}{1-\rho^2}-\frac{1}{M}\frac{\rho^2}{1-\rho^2}(\widehat{W}-\IE[W])
+\mathcal{O}\Big(\frac{1}{M^{3/2}}\Big).
\end{align}
Using the approximation for the conditional distribution
of $(\chi_l^1, \chi_l^2)$ given $Z$ obtained in Appendix~\ref{key lemmas},
$$\IE[\widehat{W}]=\IE[W|Z]=\IE[W]+\rho\frac{\sigma_w}{\sigma_z}\big(Z-\IE[Z]\big)
+\mathcal{O}\Big(\frac{1}{\sqrt{M}}\Big),$$
so we can rewrite~(\ref{TWHat}) as
$$T(\widehat{W})=\frac{1}{M}\frac{1}{(1-\rho^2)}\IE[\widehat{W}]-
\frac{1}{M}\frac{\rho^2}{1-\rho^2}\widehat{W}
+\mathcal{O}\Big(\frac{1}{M^{3/2}}\Big).
$$
Substituting in~(\ref{defn of first remainder}),
\begin{equation}
\label{choice of lambda}
\IE[\widehat{W}-\widehat{W}'|\widehat{W}]=\frac{1}{M}\frac{1}{1-\rho^2}
\Big(\widehat{W}-\IE[\widehat{W}]\Big)
+\mathcal{O}\Big(\frac{1}{M^{3/2}}\Big).
\end{equation}

We are going to apply Corollary~\ref{chen corollary} to
$(\widehat{W},\widehat{W}')$ with ${\cal F}=\sigma(\widehat{W})$.
We set $\lambda =1/(M(1-\rho^2))$ and
observe from~(\ref{choice of lambda})
that we may take a remainder term $R$ with $\IE[|R|]$
of order $1/M^{1/2}$ in~(\ref{unnormalised difference}).
Moreover,
$$\widehat{K}_2=\frac{1}{2\lambda}\frac{|\Delta|^3}{2},$$
and so, since by construction $|\Delta|<C/\sqrt{M}$,
$\IE[\widehat{K}_2]$ is also order at most $1/M^{1/2}$.

Since with these definitions
$$\widehat{K}_1=\frac{M(1-\rho^2)}{2} 
\IE\Big[(\widehat{W}-\widehat{W}')^2|\widehat{W}\Big],$$
it remains to control
\begin{equation}
\label{difference from one control}
\IE\left[\left| \sigma_{\widehat{W}}^2-\frac{M(1-\rho^2)}{2}
\IE\Big[(\widehat{W}-\widehat{W}')^2|\widehat{W}\Big]
\right|
\right].
\end{equation}

Again using the results of Appendix~\ref{key lemmas},
$$\sigma_{\widehat{W}}^2=(1-\rho^2)\sigma_W^2
+\mathcal{O}\bigg(\frac{1}{\sqrt{M}}\bigg),$$
(the first term being the conditional variance if the
random variables were distributed exactly as a bivariate normal),
whereas
$$\IE\Big[\IE[(\widehat{W}-\widehat{W}')^2 |\widehat{W}]\Big]
=\IE[(\widehat{W}-\widehat{W}')^2]
=\IE\Big[\frac{1}{M^2}\sum_{l=1}^M (\widehat{\Phi}_l-\widehat{\Phi}_l^*)^2\Big]
=2\frac{1}{M}\sigma_W^2
+\mathcal{O}\bigg(\frac{1}{M^{3/2}}\bigg).$$
(Note that we see the unconditioned $\sigma_W^2$ in this second expression
since it involves only diagonal terms.)

To control~(\ref{difference from one control}),
observing that, by Cauchy-Schwarz inequality,
\begin{equation}
\IE\left[\left|\IE[M(\widehat{W}-\widehat{W}')^2]-\IE[M(\widehat{W}-\widehat{W}')^2|\widehat{W}]\right|\right]
\leq \mathtt{Var}\left(\IE[M(\widehat{W}-\widehat{W}]')^2|\widehat{W}]\right)^{1/2},
\end{equation}
it suffices
to control
$$\mathtt{Var}\left(\IE[M(\widehat{W}-\widehat{W}')^2|\widehat{W}]\right).$$
In particular, we should like to show that this expression is of order
$\mathcal{O}(1/M)$.

Now we use the standard decomposition of conditional expectations:
for two random variables $X$ and $F$,
\begin{eqnarray*}
\mathtt{Var}(X)&=&
\IE\left[\IE[X^2|F]-\big(\IE[X|F]\big)^2
+\big(\IE[X|F]\big)^2\right]-\IE\left[\IE[X|F]\right]^2\\
&=&\IE\big[\mathtt{Var}(X|F)\big] +
\mathtt{Var}\big(\IE[X|F]\big).
\end{eqnarray*}
So
$$\mathtt{Var}(\IE[X|F])=\mathtt{Var}(X)-\IE[\mathtt{Var}(X|F)].$$
For us, $X=M(\widehat{W}-\widehat{W}')^2=(\Phi_L-\Phi_L^*)^2$, and
$F=\widehat{W}$,
so
$$\mathtt{Var}(X)=\frac{1}{M}\sum_{l=1}^M\IE[(\Phi_l-\Phi_l^*)^4]
-\left(\frac{1}{M}\sum_{l=1}^M\IE\big[(\Phi_l-\Phi_l^*)^2\big]\right)^2,$$
and we seek
\begin{multline*}
\frac{1}{M}\sum_{l=1}^M\IE[(\Phi_l-\Phi_l^*)^4]
-\left(\frac{1}{M}\sum_{l=1}^M\IE[(\Phi_l-\Phi_l^*)^2]\right)^2
\\
-\frac{1}{M}\sum_{l=1}^M\IE\Big[\IE[(\Phi_l-\Phi_l^*)^4] |\widehat{W}\Big]
+\IE\Bigg[\left(\frac{1}{M}\sum_{l=1}^M\IE[(\Phi_l-\Phi_l^*)^2|\widehat{W}]
\right)^2\Bigg],
\end{multline*}
where the expectation is with respect to the distribution of $\widehat{W}$.
By the tower property, the terms involving $(\Phi_l-\Phi_l^*)^4$ cancel,
leaving
\begin{equation}
\label{controlling the variance}
\frac{1}{M^2}\sum_{l=1}^M\sum_{m=1}^M\Bigg\{
\IE\Bigg[\IE\big[(\Phi_l-\Phi_l^*)^2|\widehat{W}\big]
\IE\big[(\Phi_m-\Phi_m^*)^2|\widehat{W}\big]\Bigg]
-
\IE[(\Phi_l-\Phi_l^*)^2]
\IE[(\Phi_m-\Phi_m^*)^2]\Bigg\}.
\end{equation}
Expanding
$\IE\big[(\Phi_l-\Phi_l^*)^2|\widehat{W}\big]
\IE\big[(\Phi_m-\Phi_m^*)^2|\widehat{W}\big]$
in an entirely analogous way to~(\ref{cond expn Phistar}),
when we take expectations, using the tower property of conditional
expectations,
the part of the product that is an affine function of $\widehat{W}$
will cancel in~(\ref{controlling the variance}), leaving quadratic (and higher
order) terms, each of which is of order $\mathcal{O}(1/M)$ in the summand.
Overall then~(\ref{controlling the variance}) is $\mathcal{O}(1/M)$,
and applying Corollary~\ref{chen corollary}, the proof that ${\cal A}^i+{\cal D}^i$ is
normal with an error of order $1/\sqrt{M}$ is complete.

\section*{The residuals, generation one}

Proving that $R_A^i+R_D^i$ is normal is much simpler. Since Mendelian inheritance
is independent across loci we are able to use Theorem~\ref{rinott clt}
in much the same way as in generation zero. A combination of
Lemma~\ref{replacement for troublesome equation}
and Bayes' rule suffices to show that the variance is
not affected by conditioning on parental trait values, after which
the proof proceeds essentially as in the additive case and so is
omitted.

\section{Generation $t$: accumulation of information}
\label{appendix: accumulation of information}

If we wanted to prove a strict analogue of the results of
Barton et al.~(2017) in the
\nocite{barton/etheridge/veber:2017}
additive case, then we would want to condition not just on the
trait values of the parents, but on the trait values of an
arbitrary collection of individuals in the pedigree. Such a proof
can follow essentially the same line as that above, although the
calculations are considerably longer to write out. The only thing that
must be checked is that we do not accumulate too much
information from knowing those trait values; it is this
that controls for how long the infinitesimal approximation will
remain accurate.
This requires more care
than the additive case of Barton et al.~(2017),
\nocite{barton/etheridge/veber:2017}
so we present the argument here.

Recall that we write ${\cal P}(t)$ for the pedigree up to and
including generation $t$ and $Z(t)$ for the
corresponding vector of trait values of all individuals in ${\cal P}(t)$.
We would like to understand the distribution of the allelic types
$\chi_l^1(j^*),\chi_l^2(j^*)$ at locus $l$ of an individual $j^*$ in generation $t$,
conditional on knowing the trait values of all individuals in the pedigree
up to generation $t-1$.
That is, we would like to estimate
\begin{multline}
\label{thing to bound}
\IP\Big[\left. (\chi_l^1(j^*),\chi_l^2(j^*))=(x,x') \right| {\cal P}(t),
Z(t-1)=\big(z^j\big)_{j\in {\cal P}(t-1)}\Big]
\\
=\frac{\IP\big[Z(t-1)=\big(z^j\big)_{j\in {\cal P}(t-1)}\big|
(\chi_l^1(j^*),\chi_l^2(j^*))=(x,x'), {\cal P}(t)\Big]}
{\IP[Z(t-1) =\big(z^j\big)_{j\in {\cal P}(t-1)}
\big|{\cal P}(t)]}\IP\Big[(\chi_l^1(j^*),\chi_l^2(j^*))=(x,x')\big|
{\cal P}(t)\Big].
\end{multline}
To estimate the numerator in the fraction,
we partition over the possible patterns of
identity at locus~$l$ in the pedigree, conditional on that pedigree; that is we condition on
the values of the Bernoulli random variables that determine Mendelian inheritance at locus $l$
across the pedigree.
We denote this $M_l(t)$ and abuse notation by writing $(M_l^1(j), M_l^2(j))$ for the allelic
states at locus $l$ in individual $j\in {\cal P}(t-1)$ conditional on $M_l(t)$.
More precisely, if $\chi_l^1(j^*)=x$ and $\chi_l^2(j^*)=x'$,
$(M_l^1(j), M_l^2(j))
=(y,y'), (y,x'), (x,y'), (x,x')$ according to whether $j$ is identical
by descent with the chosen individual $j^*$ on neither chromosome, one
chromosome or both chromosomes. We use $\IE_{M_l}$ when we wish to emphasize that we are
taking the expectation
with respect to this quantity.
We proceed as in Lemma~\ref{troublesome lemma with two parents}:
\begin{align*}
 \IP\Big[&Z(t-1)=\big(z^j\big)_{j\in {\cal P}(t-1)}
\Big| (\chi_l^1(j^*), \chi_l^2(j^*))=(x,x'),
{\cal P}(t), M_l(t)\Big]
\\
& =\IP\left[Z_{-l}^j =z^j-\frac{1}{\sqrt{M}}
\Psi_l\big(M_l^1(j),M_l^2(j)\big), \ \forall j\in {\cal P}(t-1)\Big|{\cal P}(t)\right]
\\
& = \IE\bigg[\IP\left[Z^j=z^j-\frac{1}{\sqrt{M}}
\Psi_l\big(M_l^1(j),M_l^2(j)\big)+\frac{1}{\sqrt{M}}\Psi_l\big(\chi_l^1(j),\chi_l^2(j)\big),\ \forall j\in {\cal P}(t-1) \Big|{\cal P}(t)\right]\bigg],
\end{align*}
where in the last line the expectation is taken with respect to the unconditional law of the random family $\{(\chi_l^1(j),\chi_l^2(j)),\, j\in {\cal P}(t-1)\}$.

Substituting in~(\ref{thing to bound}), in an obvious notation,
\begin{align*}
& \IP\Big[(\chi_l^1(j^*),\chi_l^2(j^*))=(x,x') \big| {\cal P}(t), Z(t-1)=\mathbf{z}\Big]\\
& = \IP\Big[\left. (\chi_l^1(j^*),\chi_l^2(j^*))=(x,x') \right| {\cal P}(t)\Big]
\\
& \, \times \!
\left(1 -\!\! \sum_{j\in{\cal P}(t-1)}\!\frac{1}{\sqrt{M}}\bigg\{\IE\Big[\Psi_l\big(M_l^1(j),M_l^2(j)\big)\big| {\cal P}(t) \Big]-\IE\Big[\Psi_l\big(\chi_l^1(j),\chi_l^2(j)\big)|{\cal P}(t)\Big]
\bigg\}\frac{\IP_{Z^j}[\mathbf{z}]}{\IP [\mathbf{z}]}\right)\!
+\!\mathcal{O}\bigg(\frac{1}{M}\bigg).
\end{align*}
In particular, the summand will vanish if $j$ and $j^*$ are not identical
by descent in at least one copy at locus $l$, since then the allelic states at locus $l$ in individuals $j$ and $j^*$ are independent.
Furthermore, the more distant the relationship between $j$ and $j^*$ (that is, the smaller the probability of their being identical by descent), the less information we glean about the allelic states in $j^*$ from
observing the trait value of individual $j$, resulting in a small contribution of the $j$-th term to the difference between the conditional and unconditional laws of $(\chi_l^1(j^*),\chi_l^2(j^*))$. The infinitesimal model
can be expected to break down for an individual if we know that one
of its close relatives had a particularly extreme trait value, or if
the pedigree is particularly inbred (so that there is little variation
between offspring).

\section{Supplementary Material and Codes}\label{appendix:SM}
The following supplementary material can be found in the public repository \cite{barton:2023}:
\begin{itemize}
\item The \emph{Mathematica} notebook \texttt{Algorithm for calculating identities.nb}, comprising a set of codes to compute the identity coefficients of Section~\ref{s:identity} and Appendix~\ref{appendix:id coef}.
\item The \emph{Mathematica} notebook \texttt{Infinitesimal with dominance.nb}, accompanying and complementing the simulations and figures presented in the paper.
\item The different datasets allowing to reproduce the numerical experiments presented in the paper.
\end{itemize}

\end{appendix}
\bibliographystyle{apalike}
\bibliography{structure}
\end{document}